\documentclass[11pt]{article}

\usepackage[margin=1in]{geometry}
\usepackage{amsmath,amssymb,amsthm,mathtools,bm,bbm}
\usepackage{booktabs,longtable,array,tabularx}
\usepackage{graphicx,float,enumitem,multirow}
\usepackage[square]{natbib}
\usepackage[colorlinks=true,linkcolor=black,citecolor=black,urlcolor=black]{hyperref}
\usepackage{caption}
\usepackage{ifthen}
\usepackage{algorithm}
\usepackage{algpseudocode}
\usepackage[section]{placeins}
\usepackage{authblk}

\usepackage{xcolor}
\title{Generalized Hierarchical Conformal Prediction}

\date{August 15, 2026}

\author{Soham Mallick\thanks{Correspondence to: \texttt{kcillam@wharton.upenn.edu}.}}
\author{Eric Tchetgen Tchetgen}
\author{Edgar Dobriban}
\author{Yonghoon Lee}

\affil{University of Pennsylvania}

\newtheorem{theorem}{Theorem}[section]
\newtheorem{proposition}[theorem]{Proposition}
\newtheorem{corollary}[theorem]{Corollary}
\newtheorem{lemma}[theorem]{Lemma}

\newtheorem{remark}[theorem]{Remark}
\newtheorem{assumption}{Assumption}

\newcommand{\assumptionprefix}{A}
\newcommand{\setassumptionprefix}[1]{%
  \renewcommand{\assumptionprefix}{#1}%
  \setcounter{assumption}{0}%
}

\newcommand{\PP}{\mathbb{P}}
\newcommand{\EE}{\mathbb{E}}
\newcommand{\R}{\mathbb{R}}

\newcommand{\cX}{\mathcal{X}}
\newcommand{\cY}{\mathcal{Y}}

\newcommand{\cW}{\mathcal{W}}

\newcommand{\1}{\mathbf{1}}

\newcommand{\iid}{\stackrel{\mathrm{i.i.d.}}{\sim}}
\DeclareMathOperator{\length}{length}
\DeclareMathOperator{\Unif}{Unif}

\newcommand{\quant}[2]{Q_{#1}\!\left(#2\right)}

\begin{document}

\maketitle

\begin{abstract}
Many prediction problems arise with data collected in groups. In this setting, hierarchical conformal prediction (HCP) \citep{lee2023distribution} provides distribution-free prediction sets for a new observation from a previously unseen group under hierarchical exchangeability. In many
applications, however, prediction is conducted only after a few observations from the group of interest have already been collected. Standard HCP cannot leverage these observations, as its required symmetry conditions do not hold in this setting. At the same time, the initial sample may still be too small for standard conformal prediction applied within the test group to be informative.

We develop predictive inference methods for this setting. Our proposed method, Generalized HCP (GHCP), restores the relevant symmetry needed for conformal inference by assigning the test group a randomly ``donated'' reference group size. GHCP further leverages the initial test group observations to improve the quality of the nonconformity scores for prediction within that group. To improve efficiency, we introduce a variant that restricts the set of eligible donors. We demonstrate the performance of the proposed method through simulations and an illustration on the American Community Survey dataset.
\end{abstract}

\tableofcontents

\section{Introduction}

In many settings, data are collected from multiple groups or sources, such as patients treated at different hospitals, students from different schools, or households from different geographic areas. We consider predictive inference problems with such hierarchical data structures, where the goal is to predict the response of a new unit from a designated \emph{test group} based on data from other groups, which we call \emph{reference groups}. While conformal prediction~\citep{vovk1999machine,saunders1999transduction} provides distribution-free prediction sets for i.i.d. or exchangeable data, it does not directly provide valid inference in hierarchical settings where individual-level exchangeability is violated. Intuitively, this is because observations within the same group tend to be more similar to one another than those across different groups.

Hierarchical conformal prediction (HCP), introduced in \citet{lee2023distribution}, addresses this problem under the weaker assumption of hierarchical exchangeability: the groups are exchangeable, and the observations within each group are exchangeable. 
For instance, schools are considered irrespective of their order, and within schools the students are also considered in an arbitrary order. 
Under this assumption, HCP provides finite-sample valid prediction sets for a new observation from a previously unseen group (such as the first student from a school of interest), without requiring additional distributional assumptions.

In many applications, however, prediction for the test group (e.g., school of interest) is made after a few observations from that group have already been collected (e.g., the school already has some students). We call these observations the \emph{initial test group sample}. For example, a survey agency using data from many geographic areas may need to predict responses in a newly sampled area while fieldwork is still ongoing. In surveys such as the American Community Survey, data are collected continuously over the field period \citep{salvo2006moving,uscensus2022acsdesign}, and survey operations are often monitored and adjusted using information observed during collection \citep{groves2006responsive,west2023deriving}. Thus, after a number of completed interviews in a new area, the analyst may want a prediction interval for the next respondent. 

Although the initial observations are informative about the test group, they may be too few for ordinary split conformal prediction applied within that group alone to yield informative prediction sets \citep{vovk2005algorithmic}. Standard HCP can instead ``borrow strength" from the reference groups, but only by discarding the initial test group sample in this setting. The aim of this paper is to combine the reference groups with the initial test group sample, thereby adapting the prediction set to the group of interest. The resulting procedure aims to produce tighter prediction sets while retaining finite-sample distribution-free validity under the hierarchical sampling structure.

\subsection{Problem setup}
\label{subsec:problem-setup}

\begin{figure}
\centering
\includegraphics[width=0.9\linewidth]{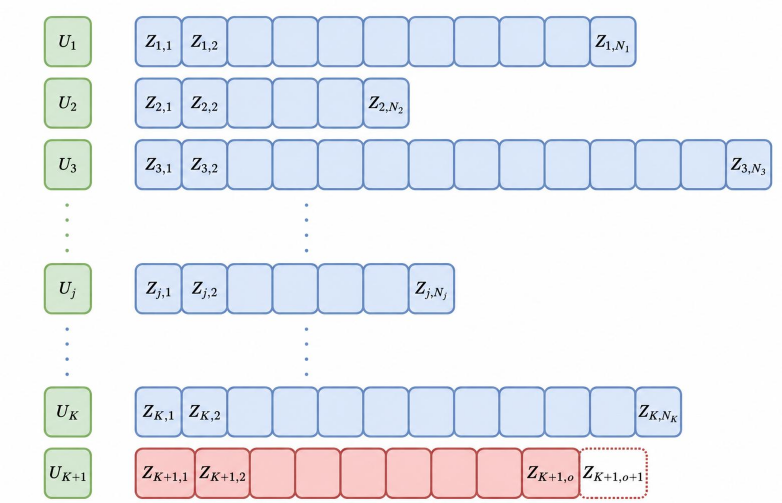}
\caption{\footnotesize A representation of our setup. Blue rows correspond to reference groups, and the red row corresponds to the test group. The goal is to predict the last observation of the test group (shown dashed).}
\label{fig:represent-diagram}
\end{figure}

Suppose that we observe $K>0$ groups, referred to as reference groups, together with a test group for which some initial observations are available prior to prediction. We allow each group to have group-level features $U_j \in \mathcal U$, which can be omitted when no such feature is available. We denote the $i$-th observation in the $j$-th group by
\(
Z_{j,i}=(X_{j,i},Y_{j,i}) \in \mathcal X \times \mathcal Y\), \(W_{j,i}=(U_{j}, Z_{j,i})\in \mathcal U \times \mathcal X \times \mathcal Y = \mathcal{W}.
\)

For the reference groups, let $N_j$ denote the size of group $j$, for $j=1,\ldots,K$. Denote the $(K+1)$-th group as the test group with a given, fixed number $o$ of observations available at the time of prediction. 
The data available take the form
\[
\widetilde W_j = (W_{j,1}, W_{j,2}, \ldots, W_{j,N_j}), ~j\in[K],~\text{and}~
\widetilde W_{K+1}^\ast = (W_{K+1,1}, W_{K+1,2}, \ldots, W_{K+1,o}).
\] 

Given a new feature vector $(U_{\mathrm{test}},X_{\mathrm{test}})$ from the test group, we aim to construct a prediction set for the corresponding unobserved outcome $Y_{\mathrm{test}}$. Under the sampling assumptions introduced below, the test pair can be viewed as the $(o+1)$-th observation from the test group. 
We therefore write $(U_{\mathrm{test}},X_{\mathrm{test}},Y_{\mathrm{test}})=(U_{K+1},X_{K+1,o+1},Y_{K+1,o+1})$, using the two notations interchangeably. See Figure~\ref{fig:represent-diagram} for an illustration of the setup.

\paragraph{Hierarchical i.i.d. sampling.}
\label{subsubsec:hier-iid-sampling}
 In order to introduce some of the ideas, let us keep the following model in mind as a starting point. 
 We generalize this model below. 
Let each group $j$ have its own feature $U_j$: 
$U_1,\dots,U_{K+1}\iid P_U$, 
and sample sizes 
$N_1,\dots,N_K\iid P_N$.
 Given these, 
 the observations within each group are i.i.d.:
\begin{equation}
\label{eq:hier-sampling}
\begin{split}
&Z_{j,1},Z_{j,2},\ldots,Z_{j,N_j}\mid U_j, N_j \iid P_{Z\mid U_j},
\qquad j=1,\dots,K,\\
&Z_{K+1,1}, Z_{K+1,2}, \ldots, Z_{K+1,o+1} \mid U_{K+1} \iid P_{Z \mid U_{K+1}}.
\end{split}
\end{equation}

\paragraph{Review of HCP.}
\label{subsubsec:hcp-review-main}
Hierarchical conformal prediction (HCP) \citep{lee2023distribution} 
begins by splitting the groups into two sets. The first $K_0$ groups are used to construct a nonconformity score function $s:\cW\to\R$. 
For instance, one may fit a regression model $\widehat\mu$ on these groups and take the absolute residual $s(u,x,y)=|y-\widehat\mu(u,x)|$. The remaining $K_1=K-K_0$ groups are used for calibration. Given new feature inputs $U_{K+1}, X_{K+1,1}$ from a new group, HCP 
constructs the following prediction set for the corresponding outcome $Y_{K+1,1}$:
\begin{equation}
\label{eq:hcp-measure-main}
\widehat C_{\mathrm{HCP}}(U_{K+1},X_{K+1,1})
=
\Bigl\{y\in\cY:s(U_{K+1},X_{K+1,1},y)\le
\quant{1-\alpha}{\nu_{\mathrm{HCP}}}\Bigr\},
\end{equation}
where  $\nu_{\mathrm{HCP}}$  is a weighted empirical distribution over the calibration scores:
\begin{equation*}
\nu_{\mathrm{HCP}}
=
\sum_{j=K_0+1}^{K}\sum_{i=1}^{N_j}
\frac{1}{(K_1+1)N_j}\delta_{s(W_{j,i})}
+
\frac{1}{K_1+1}\delta_{\infty}.
\end{equation*}
A weight of $1/(K_1+1)$ is assigned to each calibration group and distributed equally among its observations, while the remaining weight $1/(K_1+1)$ is assigned to $+\infty$, serving as a placeholder for the test group, whose scores are unobserved. 
Two features of HCP are relevant to our setting.
\begin{enumerate}[label=(\roman*)]
    \item HCP is designed to predict the first observation from a previously unseen group and therefore cannot use the initial test group observations $W_{K+1,1},\ldots,W_{K+1,o}$. These observations also cannot simply be appended to the calibration data, as the test group has a fixed number $o$ of observations, whereas the reference group sizes $N_1,\ldots,N_K$ may arise from a different  mechanism.
    Thus, the resulting $K+1$ groups need not satisfy the hierarchical exchangeability (Assumption~\ref{assump:HCP-group-exch} in Appendix~\ref{appsubsec:hcp-review})  required by HCP; and thus the required coverage may not hold.

    \item HCP can be conservative when the number of calibration groups $K_1$ is small. Since $\nu_{\mathrm{HCP}}$ assigns weight $1/(K_1+1)$ to $+\infty$, the $(1-\alpha)$ quantile corresponds to the quantile level
    $(1-\alpha)(K_1+1)/K_1$ among the finite calibration scores. Thus, HCP calibrates at a level $(1-\alpha)/K_1$ above the nominal level $1-\alpha$, resulting in a more extreme calibration score when there are only a few calibration groups.
\end{enumerate}

\paragraph{Our setting: a few initial observations available in the test group.}
\label{subsubsec:setting}
We consider the setting in which a small number of initial observations from the test group are available prior to prediction. Such a setting arises naturally in several common scenarios. 

\begin{enumerate}

\item \textbf{Prediction in a new environment.}
Some prior data is available from several environments, while prediction is required in a newly encountered environment after only a limited amount of local data has become available. This setting arises in multi-environment predictive inference, where information from the test environment can be used to improve prediction for subsequent observations \citep{duchi2025predictive}.

\item \textbf{Partial availability during data collection.}
A latent full test-group size $N_{K+1}$ may be generated in the same way as the reference-group sizes, only the first $o$ observations of which are available when prediction is required. This is natural in continuous surveys and other field operations in which observations become available over time \citep{uscensus2022acsdesign}. Here, $o$ reflects the stage of data collection at which prediction is made, rather than the eventual size $N_{K+1}$ of the group.

\item \textbf{Prediction after a fixed initial sample.}
An analyst may decide in advance to make predictions after a fixed number of observations from the test group have been collected. More generally, sequential data-collection designs often make inferential or operational decisions at intermediate stages of data collection \citep{groves2006responsive,west2023deriving}. In this case, $o$ is a design parameter rather than a realization of the random group-size mechanism.
\end{enumerate}

These scenarios correspond to different mechanisms by which the observed test-group size $o$ may arise. Our goal is to combine information from the initial test-group sample and the reference groups while accommodating the resulting size mismatch, without explicitly modeling any of these mechanisms.

\paragraph{Hierarchical exchangeability.}
\label{subsubsec:dhcp-assumptions}
The hierarchical i.i.d. sampling model in \eqref{eq:hier-sampling} can be generalized as follows. 
At a high level, we assume that \emph{(i) groups are exchangeable at the level of their group features and unobserved laws, (ii) the reference group sizes are ignorable relative to these group features and laws, and (iii) conditional on the group type, within-group observations are i.i.d.} We formalize each component below. 

\begin{assumption}[Group-level exchangeability]
\label{assump:group-exch}
There exist random group-level probability measures
$\mu_1,\ldots,\mu_{K+1}$ such that the pairs
$\{(U_j,\mu_j):j\in[K+1]\}$ are exchangeable. Equivalently, for every
$\sigma\in\mathcal S_{K+1}$, where $\mathcal S_{K+1}$ is the set of
permutations of $[K+1]$,
\[
\bigl((U_1,\mu_1),\ldots,(U_{K+1},\mu_{K+1})\bigr)
\overset{d}{=}
\bigl((U_{\sigma(1)},\mu_{\sigma(1)}),\ldots,
      (U_{\sigma(K+1)},\mu_{\sigma(K+1)})\bigr).
\]
\end{assumption}
Above, $A\overset{d}{=}B$ denotes that the two random objects $A,B$ have the same distribution. 

\begin{assumption}[Group-size ignorability]
\label{assump:independence}
The reference group sizes $N_{1:K}:=(N_1,\ldots,N_K)$ are exchangeable and independent of
$
(U_{1:K+1},\mu_{1:K+1})
:=
\big((U_1,\mu_1),\ldots,(U_{K+1},\mu_{K+1})\big),
$
i.e.,
\(N_{1:K} \perp\!\!\!\perp (U_{1:K+1},\mu_{1:K+1}).
\)
\end{assumption}

\begin{assumption}[Within-group i.i.d.]
\label{assump:within-group-exch}
Conditional on $(U_{1:K+1},\mu_{1:K+1},N_{1:K})$, the within-group observations in the $j$-th group are i.i.d.~from $\mu_j$ and are independent of observations from other groups, i.e.,
\begin{equation*}
\begin{split}
&Z_{j,1},\ldots,Z_{j,N_j}\mid (U_{1:K+1},\mu_{1:K+1},N_{1:K})
\ \sim\ \mu_j^{\otimes N_j}, ~ j\in[K], \\
&Z_{K+1,1}, Z_{K+1,2}\ldots \mid (U_{1:K+1},\mu_{1:K+1},N_{1:K})
\ \iid\ \mu_{K+1}.
\end{split}
\end{equation*}
\end{assumption}

These assumptions formalize the idea that the groups are a priori exchangeable draws from a common population of groups, and that the number of observations in a group carries no information about that group's law. Assumption~\ref{assump:group-exch} states that group identities are uninformative to distinguish the test group. Assumption~\ref{assump:independence} states that the reference group sample sizes are determined by mechanisms that are unrelated to the group-level data-generating laws. 
The sample sizes may be random and even dependent across groups. 
Assumption~\ref{assump:within-group-exch} states that conditional on the generating measure, each group contains i.i.d. samples. See Appendix~\ref{appsec:background} for a detailed comparison with the assumptions used for HCP in \citet{lee2023distribution}.

\subsection{Main contributions}

Our contributions are as follows.

\begin{enumerate}

\item \textbf{Generalized HCP.}
We introduce Generalized HCP (GHCP), a predictive inference procedure for hierarchical data settings in which an initial sample from the test group is available. GHCP improves upon standard HCP, which discards the initial sample, as well as ordinary split conformal prediction applied within the test group, which relies on a small sample and can produce trivial prediction sets when the test-group size is small.

\item \textbf{Test-group adaptation and extensions.}
%Since the prediction target comes from the test group,
GHCP leverages part of the initial test-group sample to adapt the score function to the group-specific structure, while keeping the remaining observations for calibration. This allows the resulting prediction set to be more specifically tailored to the group for which prediction is required. To improve efficiency, we further introduce a restricted-donor variant of GHCP that adapts the donor pool to the observed group sizes.

\item \textbf{Empirical evaluation.}
We evaluate GHCP through simulations and a real-data application on the ACS income dataset~\citep{ding2021folktables}, illustrating its
validity and practical advantages over baselines.

\end{enumerate}

\subsection{Related work}
\label{subsec:related-work}

Distribution-free predictive inference has been studied extensively in statistics and machine learning. Under exchangeability, conformal prediction provides finite-sample valid prediction sets; see \citet{vovk2005algorithmic}, \citet{fontana2023conformal}, and \citet{angelopoulos2024theoretical} for overviews. A widely used computational variant is split conformal prediction \citep{papadopoulos2002inductive}.

A substantial literature extends conformal prediction beyond exchangeable data, including covariate shift \citep{tibshirani2019conformal} and label shift \citep{podkopaev2021distribution}. 
Grouped and multilevel data have long been modeled using random-effects and mixed-effects models \citep{laird1982random,liang1986longitudinal}, as well as small-area estimation and empirical Bayes methods \citep{fay1979estimates,battese1988error,jiang2006mixed}. These methods typically impose parametric or semiparametric assumptions on the between- and within-group distributions.

Distribution-free prediction for hierarchical data is studied by \citet{dunn2022distribution}, who consider both prediction for a new group and prediction of a subsequent observation from an observed group, and by \citet{lee2023distribution}, who introduce hierarchical conformal prediction for a new observation from a previously unseen group. \citet{dobriban2025symmpi} develop a more general symmetry-based framework that includes prediction of unobserved outcomes within a partially observed branch of a hierarchical model. \citet{duchi2025predictive} study predictive inference across multiple environments, and primarily consider a notion of batch coverage. 
Our setting allows the number of initially observed test group observations to be determined separately from the reference group size mechanism, while targeting marginal coverage for a new observation from the test group.

\section{Main results}
\label{sec:donor-method}

\subsection{Proposed method: Generalized HCP (GHCP)}
\label{subsec:G-HCP}

In this section, we introduce our main procedure, Generalized HCP (GHCP). The overall steps of the proposed procedure are as follows: 
We first select the reference groups whose sizes are \emph{compatible} with the test group.
We then recover hierarchical exchangeability through a \emph{donation strategy}.
Finally, we construct the nonconformity score and the prediction set.
The detailed steps are as follows.

\paragraph{Step 1: Selection.}

We first choose the reference groups whose sizes are compatible with the test group---i.e., groups with more than $o$ observations. We denote the resulting subset of the reference set as $S:=\bigl\{j\in[K]:N_j>o\bigr\}$.
As this selection is based only on the group sizes,
Assumption~\ref{assump:independence} ensures that the sampling distribution of the group-level distributions remains the same when conditioned on $S$.

\paragraph{Step 2: Donation.}
We obtain a surrogate size for the test group that is exchangeable with the selected reference group sizes by randomly choosing a donor group-index $J_0\mid S\sim\Unif(S)$. The size
$N_{J_0}$
of this donor group is then assigned to the test group for the
purpose of calibration; 
 we set $N_{K+1} = N_{J_0}$. 

The test group is treated as having ``total size" $N_{J_0}$, of which only the first $o$ observations are observed before prediction. The donor group is then removed from the reference calibration set, so the retained reference groups are indexed by $S_{\mathrm{cal}}:=S\setminus\{J_0\}$. The remaining inferential steps are carried out based on reference groups in $S_{\mathrm{cal}}$, together with the initial samples $(U_{K+1},Z_{K+1,1},\ldots,Z_{K+1,o})$. If $S=\varnothing$, we skip the donation step; see Remark \ref{rem:no-donors}.

\paragraph{Step 3: Global and local training.}
The unselected  reference groups with indices in $S_{\mathrm{train}}=[K]\setminus S$ are used to train a global predictor $\widehat\mu^{\mathrm{glob}}$. Next, we set aside a small within-group training block of size $\lfloor o/2\rfloor$---i.e., for each group $j\in S_{\mathrm{cal}}\cup \{K+1\}$, define the within-group predictor as the group mean
\(\widehat\mu_j^{\mathrm{loc}}(U_j, x) = \frac{1}{\lfloor o/2\rfloor} \sum_{i=1}^{\lfloor o/2\rfloor} Y_{j, i} \). 
We then merge the global and local fits\footnote{If $S_{\mathrm{train}}=\varnothing$, set
$\widehat\mu^{\mathrm{glob}}\equiv0$. If additionally, $\lfloor o/2\rfloor=0$, set $\widehat\mu_j^{\mathrm{loc}}\equiv0$ and $\lambda=0$.} 
as
\begin{equation}
\label{eq:weighted-merge}
\widetilde\mu_j(U_j,x)
:=
(1-\lambda)\widehat\mu^{\mathrm{glob}}(U_j,x)+\lambda\widehat\mu_j^{\mathrm{loc}}(U_j,x),
\qquad
\lambda:=\frac{\lfloor o/2\rfloor}{|S_{\mathrm{train}}|+\lfloor o/2\rfloor}.
\end{equation}
Other approaches are presented in Appendix \ref{appsubsec:more-mergers}.
The calibration score for a held-out point in group $j$ is then $s_j(U_j,X,Y):=\bigl|Y-\widetilde\mu_j(U_j,X)\bigr|$.

\paragraph{Step 4: Constructing the prediction set.}
Let $L_j:=N_j-\lfloor o/2\rfloor$ denote the number of samples in group $j$ that are used for inference for each $j\in S_{\mathrm{cal}}\cup \{K+1\}$. For $j\in S_{\mathrm{cal}}$, all $L_j$ held-out scores are observed. For the test group, only $o-\lfloor o/2\rfloor$ held-out scores are available. 
The Generalized HCP (GHCP) prediction set is defined as\footnote{$\quant{\beta}{\nu}$ is the $\beta$-quantile of the probability measure $\nu$, defined as $\quant{\beta}{\nu}
:=\inf\{t\in\overline{\R}:\nu((-\infty,t])\ge\beta\}$.}
\begin{equation}
\label{eq:donor-set}
\widehat C_{\mathrm{GHCP}}(U_{K+1},X_{K+1,o+1})
=
\Bigl\{y\in\cY:s_{K+1}(U_{K+1},X_{K+1,o+1},y)\le \quant{1-\alpha}{\nu_{\mathrm{donor}}}\Bigr\},
\end{equation}
where
\begin{align}
\label{eq:donor-measure}
\nu_{\mathrm{donor}}
={}&
\sum_{j\in S_{\mathrm{cal}}}\sum_{i=\lfloor o/2\rfloor+1}^{N_j}
\frac{1}{(|S_{\mathrm{cal}}|+1)L_j}\delta_{s_j(U_j,Z_{j,i})} \nonumber\\
&+\sum_{i=\lfloor o/2\rfloor+1}^{o}
\frac{1}{(|S_{\mathrm{cal}}|+1)L_{K+1}}\delta_{s_{K+1}(U_{K+1},Z_{K+1,i})}
+\frac{N_{K+1}-o}{(|S_{\mathrm{cal}}|+1)L_{K+1}}\cdot\delta_{\infty}.
\end{align}

\subsection{Coverage guarantee}
To simplify the upper bound coverage results that follow, we impose the following no-ties condition.
\begin{assumption}[No ties]
\label{assump:no-ties}
Let $s_{j, i}= s_j(U_j, Z_{j, i})$ denote the held-out score for $i$-th observation in the $j$-th calibration group. We assume that the finite held-out scores are distinct almost surely within and across groups, i.e., $s_{j, i}\neq s_{j', i'}$ a.s. for any $(j, i)\neq (j', i')$.
\end{assumption}

\begin{theorem}[Finite-sample coverage of GHCP]
\label{thm:donor-valid}
If Assumptions~\ref{assump:group-exch}--\ref{assump:within-group-exch} hold, then the prediction set $\widehat C_{\mathrm{GHCP}}$ from \eqref{eq:donor-set} satisfies
\[
\PP\bigl\{
Y_{K+1,o+1}\in
\widehat C_{\mathrm{GHCP}}(U_{K+1},X_{K+1,o+1})
\bigr\}
\ge 1-\alpha.
\]
If, in addition, Assumption~\ref{assump:no-ties} holds, then,
\[
\PP\!\left\{Y_{K+1,o+1}\in
\widehat C_{\mathrm{GHCP}}(U_{K+1},X_{K+1,o+1}) \mid |S|>0 \!\right\} \le 1-\alpha 
+ \EE\!\left[\frac1{|S|}\left(1+  \rho_o\left(\max_{j\in S} N_j\right) \right) \,\middle|\, |S|>0 \right].
\]
Here, $\rho_o(n) = (n-o)/(n- \lfloor o/2\rfloor)$ is the proportion of test group calibration scores that are unobserved at time of prediction relative to the total number of test group calibration scores if the test group had size $n$.
\end{theorem}

The upper bound slack involving $\rho_o\left(\max_{j\in S} N_j\right)$ corresponds to the candidate donor size for which the proportion of unobserved test group calibration scores is largest. 

\begin{remark}[No available donors]
\label{rem:no-donors}
When $S=\varnothing$, set $S_{\mathrm{cal}}=\varnothing$ and assign the
surrogate size $N_{K+1}=o+1$. GHCP then reduces to split conformal
prediction within the test group, with the first $\lfloor o/2\rfloor$
observations used for local training. Consequently,
\[
1-\alpha
\le
\PP\!\left\{
Y_{K+1,o+1}\in
\widehat C_{\mathrm{GHCP}}(U_{K+1},X_{K+1,o+1})
\,\middle|\,|S|=0
\right\}
\le
1-\alpha+
\frac{1}{o+1-\lfloor o/2\rfloor},
\]
where the upper bound holds under Assumption~\ref{assump:no-ties}.
\end{remark}

Under Assumption~\ref{assump:no-ties}, Theorem~1 of
\citet{lee2023distribution} proves the HCP coverage upper bound
\begin{equation}
\label{eq:hcp-coverage-ub}
\PP\!\left\{Y_{\mathrm{test}}\in
\widehat C_{\mathrm{HCP}}(U_{\mathrm{test}},X_{\mathrm{test}})\right\}
\le
1-\alpha+\frac{2}{K_1+1},
\end{equation}
where $K_1$ is the number of HCP calibration groups. We next compare this quantity with the GHCP bound in
Theorem~\ref{thm:donor-valid}. When
$\PP\{N_1>o\}=1$, all $K$ reference groups are eligible donors. Hence, writing
$
N_{\max}:=\max_{1\le j\le K}N_j,
$
the GHCP upper-bound slack in Theorem~\ref{thm:donor-valid} reduces to
\begin{equation*}
%\label{eq:ghcp-all-donors-slack}
\Delta_{\mathrm{GHCP}}(o)
:=
\frac{1+\EE\{\rho_o(N_{\max})\}}{K}.
\end{equation*}
The following result describes how this bound changes with the amount of
test-group information and compares it with the HCP bound.

\begin{corollary}[Comparison with the HCP coverage bound]
\label{cor:all-donors}
Suppose the conditions of Theorem~\ref{thm:donor-valid} hold. For any integer $o\ge1$ such that $\PP\{N_1>o\}=1$,
$
\Delta_{\mathrm{GHCP}}(o)
<
\Delta_{\mathrm{GHCP}}(o-1).
$
Moreover, whenever $\PP\{N_1>o\}=1$, the GHCP upper-bound slack is no larger than the HCP slack in \eqref{eq:hcp-coverage-ub} for any value of $K_1$.
%HCP split that uses at least one reference group for training.
\end{corollary}

Corollary~\ref{cor:all-donors} applies when all reference groups remain eligible for donation. When some groups may be ineligible, the number of available donors must also be taken into account. Combining Theorem~\ref{thm:donor-valid} with Remark~\ref{rem:no-donors}, and applying $\rho_o(n)\leq 1$, we have the following marginal upper bound for coverage:
\begin{equation}
\label{eq:ghcp-donor-count-upper}
\PP\left\{
Y_{K+1,o+1}\in
\widehat C_{\mathrm{GHCP}}(U_{K+1},X_{K+1,o+1})
\right\}
\le
1-\alpha+
\frac{\PP\{|S|=0\}}{o+1-\lfloor o/2\rfloor}
+
2\,\EE\left[
\frac{\1\{|S|>0\}}{|S|}
\right].
\end{equation}

The following corollary provides a simple sufficient condition under which the marginal GHCP upper bound is no larger than the HCP upper bound in \eqref{eq:hcp-coverage-ub}, when eligible donors are sufficiently likely to be available. A proof is provided in Appendix \ref{app:proof-cor:sufficient-donors}.

\begin{corollary}
\label{cor:sufficient-donors}
Suppose the conditions of Theorem~\ref{thm:donor-valid} hold and suppose $N_1,\ldots,N_K$ are i.i.d. If
\(
\PP\{N_1>o\} \ge \sqrt{\frac{K_1+1}{K}},
\)
then the GHCP coverage upper bound in \eqref{eq:ghcp-donor-count-upper} is no larger than the HCP coverage upper bound in \eqref{eq:hcp-coverage-ub}.

\end{corollary}

\begin{remark}
The estimator in \eqref{eq:weighted-merge} is a heuristic merger of global
and local information. When $o$ is small, the global fit receives more weight; as $o$ grows, the group-specific fit contributes more strongly.
For validity, the same group-wise merger must be applied to every selected group. Further discussion of more general scores and merger constructions is provided in Appendix~\ref{appsubsec:more-mergers}. Experiments on the sensitivity of these weights is provided in Appendix \ref{appsubsec:weight-sensitivity}.
\end{remark}

\subsection{Extension: adaptive restriction of the donor pool}
\label{subsec:restricted-selection}

The default donor pool $S=\{j\in[K]:N_j>o\}$ can be undesirable for two
reasons. \begin{enumerate}[label=(\roman*)]
    \item When $o$ is small, $S$ may contain nearly all reference groups, leaving few groups in $S_{\mathrm{train}}$ for fitting the global predictor. At the same time, the within-group predictor is estimated from only a few observations, so neither component may be estimated accurately.
    \item If the donated size $N_{J_0}$ is much larger than $o$, then the weight $(N_{J_0}-o)/(|S|L_{K+1})$ placed at $+\infty$ in \eqref{eq:donor-measure} becomes large, reducing the benefit of the observed test group data and potentially leading to a trivial prediction set. This motivates favoring compatible donors whose sizes are closer to $o$.
\end{enumerate} 

Theorem~\ref{thm:donor-valid} does not require every compatible
reference group to be eligible as a donor. Let
$S(n,o)\subseteq[K]$ be a possibly randomized selection rule for a
deterministic size vector $n\in\mathbb N^K$. We call the rule
permutation-equivariant if, for every $n\in\mathbb N^K$ and every
permutation $\pi$ of $[K]$,
\[
S(\pi n,o)\overset{d_R}{=}\pi\{S(n,o)\},
\qquad
(\pi n)_j:=n_{\pi^{-1}(j)},
\]
where equality in distribution is over auxiliary randomization that is
independent of the data, and
$\pi(A):=\{\pi(j):j\in A\}$. Thus relabeling the reference groups before
selection merely relabels the selected set. Any such rule based only on
$N_{1:K}$ and $o$ preserves the coverage argument.

We use this flexibility to restrict the number of compatible donors, thereby addressing the two issues described above. Fix $\eta\in[0,1)$ and let
\(
m_\eta=\min\left\{|S|,\left\lceil(1-\eta)K\right\rceil\right\}.
\)
We define
\begin{equation}
\label{eq:restricted-region}
S_\eta
=
\left\{
j\in S:
N_j \text{ is among the } m_\eta
\text{ smallest values of } \{N_\ell:\ell\in S\}
\right\}.
\end{equation}
Thus, when sufficiently many compatible groups are available,
$S_\eta$ retains the $\lceil(1-\eta)K\rceil$ groups whose sizes are
closest to $o$; otherwise, all compatible groups are retained. Ties at
the cutoff are broken at random in a permutation-equivariant manner; see
Appendix~\ref{appsubsec:restriction} for details.
We use $\eta=0.5$ in the experiments; this guarantees
$|S_{\mathrm{train}}|\ge\lfloor K/2\rfloor$.

\begin{algorithm}
\caption{\footnotesize GHCP with a Restricted Donor Pool}
\label{alg:dhcp}
\begin{algorithmic}[1]
\State \textbf{Input}: Reference data $\{\widetilde W_j\}_{j=1}^K$,
initial test group sample $\widetilde W^*_{K+1}$,
target level $\alpha\in(0,1)$, and restriction parameter $\eta\in[0,1)$.
\State Construct the restricted donor pool $S_\eta$ as in
\eqref{eq:restricted-region}, and set
$S_{\mathrm{train}}:=[K]\setminus S_\eta$.
\State Fit the global predictor $\widehat\mu^{\mathrm{glob}}$ using the
groups in $S_{\mathrm{train}}$.
\If{$S_\eta\neq\varnothing$}
    \State Draw $J_0\sim\Unif(S_\eta)$, set
    $S_{\mathrm{cal}}:=S_\eta\setminus\{J_0\}$, and set $N_{K+1}:=N_{J_0}$.
\Else
    \State Set $S_{\mathrm{cal}}:=\varnothing$ and set
    $N_{K+1}:=o+1$.
\EndIf
\State For each $j\in S_{\mathrm{cal}}\cup\{K+1\}$, construct
$\widetilde\mu_j$ as in \eqref{eq:weighted-merge} and compute the held-out
scores using
$s_j(U_j,X,Y)=|Y-\widetilde\mu_j(U_j,X)|$.
\State Construct $\nu_{\mathrm{donor}}$ from \eqref{eq:donor-measure} and output $\widehat C_{\mathrm{GHCP},\eta}
(U_{K+1},X_{K+1,o+1})$ as in \eqref{eq:donor-set}.
\end{algorithmic}
\end{algorithm}

\begin{corollary}[Finite-sample coverage of GHCP with a restricted donor pool]
\label{cor:restricted-valid}
Suppose Assumptions~\ref{assump:group-exch}--\ref{assump:within-group-exch}
hold, and let $S_\eta$ be the restricted donor pool in
\eqref{eq:restricted-region}. Then the prediction set from Algorithm~\ref{alg:dhcp}
satisfies
\(
\PP\bigl\{
Y_{K+1,o+1}\in
\widehat C_{\mathrm{GHCP},\eta}(U_{K+1},X_{K+1,o+1})
\bigr\}
\ge 1-\alpha.
\)
Suppose, in addition, that Assumption~\ref{assump:no-ties} holds.
\[
\begin{aligned}
\PP\!\left\{Y_{K+1,o+1}\in \widehat C_{\mathrm{GHCP},\eta}(U_{K+1},X_{K+1,o+1})\right\} 
\le 1-\alpha
+
\frac{\PP\{|S_\eta|=0\}}{o+1-\lfloor o/2\rfloor}
+
2\,\EE\!\left[\frac{\1\{|S_\eta|>0\}}{|S_\eta|}\right].
\end{aligned}
\]
\end{corollary}
Thus, restricting the donor pool preserves finite-sample validity, while the
upper bound depends on the resulting number of eligible donors through
$|S_\eta|$.

\section{Experiments}
\label{sec:experiments}

We now illustrate the performance of the proposed GHCP procedure compared to HCP through experiments\footnote{Code to reproduce these experiments is available at \href{https://github.com/soham-penn/hierarchical_CP}{\texttt{https://github.com/soham-penn/hierarchical\_CP}}.}. Further comparisons are provided in Appendix~\ref{appsec:additional-experiments}.

\subsection{Simulations}
\label{subsec:main-simulations}

\paragraph{Data generation.}
We fix the number of reference groups as $K=20$ and use group sizes $N_j$, which may be kept fixed or generated from a distribution. For each $j\in[K+1]$,  we draw the group-covariate $U_j \in \R^d$ as
\(U_j \iid \mathrm{Unif}([1,5]^d)\),
\(U_j=(U_{j1},\dots,U_{jd})^\top,~d=5.\)
We then draw a latent group feature $B_j \iid N(0,\gamma^2), \gamma=5$.
We generate the individual samples as
\[
Z_{j,1},\dots,Z_{j,N_j}\mid U_j,B_j \iid \mathcal N_d\left(\mu(U_j)+B_j e_d,~\Sigma(U_j)\right),
\qquad j\in[K],
\]
and for the test group,
\[
Z_{K+1,1},\dots,Z_{K+1,o+1}\mid U_{K+1},B_{K+1} \iid \mathcal N_d\left(\mu(U_{K+1})+B_{K+1} e_d,~\Sigma(U_{K+1})\right).
\]
Here $e_d=(0, 0, \ldots, 0, 1)^\top$ is the $d$-th canonical unit vector in $\R^d$. 
The latent group feature $B_j$ 
induces heterogeneity between the groups, so within-group adaptation can help significantly. Each  $Z_{j,i}=(X_{j,i},Y_{j,i})\in\R^{d-1}\times\R$, meaning that the last coordinate represents the response.  The test group sample length is varied over $o\in\{0,5,10,15,20\}$,
assuming that the prediction is for $Y_{K+1,o+1}$. The mean and covariance functions are set as
\(
\mu(U_j)=(U_{j1}^2,\dots,U_{jd}^2)^\top\),
\(\Sigma(U_j)=(1-\rho)\,\mathrm{diag}(U_{j1},\dots,U_{jd})+\rho\,\mathbf 1_d\mathbf 1_d^\top\),
where \(\rho=0.5.
\)

\paragraph{Evaluation.} We use random forest (RF) regression (50 trees with a minimum leaf size of five) as the global learner. We use the absolute residual score $s(u, x,y)=|y-\hat{\mu}(u, x)|$, and the sample mean $\frac{1}{\lfloor o/2\rfloor}\sum_{i=1}^{\lfloor o/2\rfloor}Y_{j,i}$ as the within-group local predictor, with the convention that this quantity is set to zero whenever $\lfloor o/2\rfloor=0$. We then apply~\eqref{eq:weighted-merge} to construct the merged predictor. We evaluate marginal coverage by regenerating the full dataset with $K=20$ reference groups and one test group on each of $n_\text{rep}=1000$ independent replicates. For each replicate, we record the coverage indicator and interval width for every method. We report replicate-level boxplots and summary tables.

\subsubsection{GHCP vs HCP}
Here, we compare the GHCP from Algorithm \ref{alg:dhcp} with $\eta=0.5$ with HCP, which ignores the observed test group samples.
    
\paragraph{Avoiding trivial prediction sets.}
We first consider equal reference group sizes $N_j\equiv 21$, $j\in[K]$, at test level $\alpha=0.05$. The results are shown in Table~\ref{tab:dhcp-hcp-alpha005}. In this setting, HCP returns trivial (infinite-width) prediction sets, as the weight $1/(K_1+1)=1/11$ that \eqref{eq:hcp-measure-main} places at $+\infty$ exceeds $\alpha=0.05$. In addition, the standard split conformal prediction applied within the test group also leads to trivial prediction sets for $o\le 20$, once half of the initial sample is reserved for training.

By contrast, GHCP can return non-trivial prediction sets in such settings. Indeed, according to the weight at $+\infty$ in \eqref{eq:donor-measure}, the GHCP prediction set becomes non-trivial only when $(N_{J_0}-o)/(|S_\eta|L_{K+1})\le\alpha$. In our setting where $N_{J_0}=21$, $|S_\eta|=10$, this condition reduces to $(21-o)/\{10(21-\lfloor o/2\rfloor)\}\le 0.05$, which holds for $o\ge 14$. 
We show infinite intervals in  Table~\ref{tab:dhcp-hcp-alpha005}, but omit these cases on the plots.

\begin{table}
\centering
\caption{\footnotesize Empirical coverage and mean interval width for GHCP across test group size $o$, together with HCP, for fixed $N_j=21$, $\gamma=5$, and $\alpha=0.05$, using a random forest global predictor. Entries are means across $1000$ simulation repetitions, with standard errors in parentheses. Width entries equal to $+\infty$ indicate that all reported intervals were trivial.}
\label{tab:dhcp-hcp-alpha005}
\small
\setlength{\tabcolsep}{6pt}
\begin{tabular}{lcccccc}
\toprule
& \multicolumn{5}{c}{GHCP} & HCP \\
\cmidrule(lr){2-6}
Value of $o$ & 0 & 5 & 10 & 15 & 20 & -- \\
\midrule
Coverage ($\alpha=0.05$)
& 1.00 (0.0)
& 1.00 (0.0)
& 1.00 (0.0)
& 0.98 (0.004)
& 0.95 (0.007)
& 1.00 (0.0) \\
Mean Width
& $+\infty$
& $+\infty$
& $+\infty$
& 21.28 (0.149)
& 15.83 (0.113)
& $+\infty$ \\
\bottomrule
\end{tabular}
\end{table}

\begin{figure}
    \centering
    \includegraphics[width=0.9\linewidth]{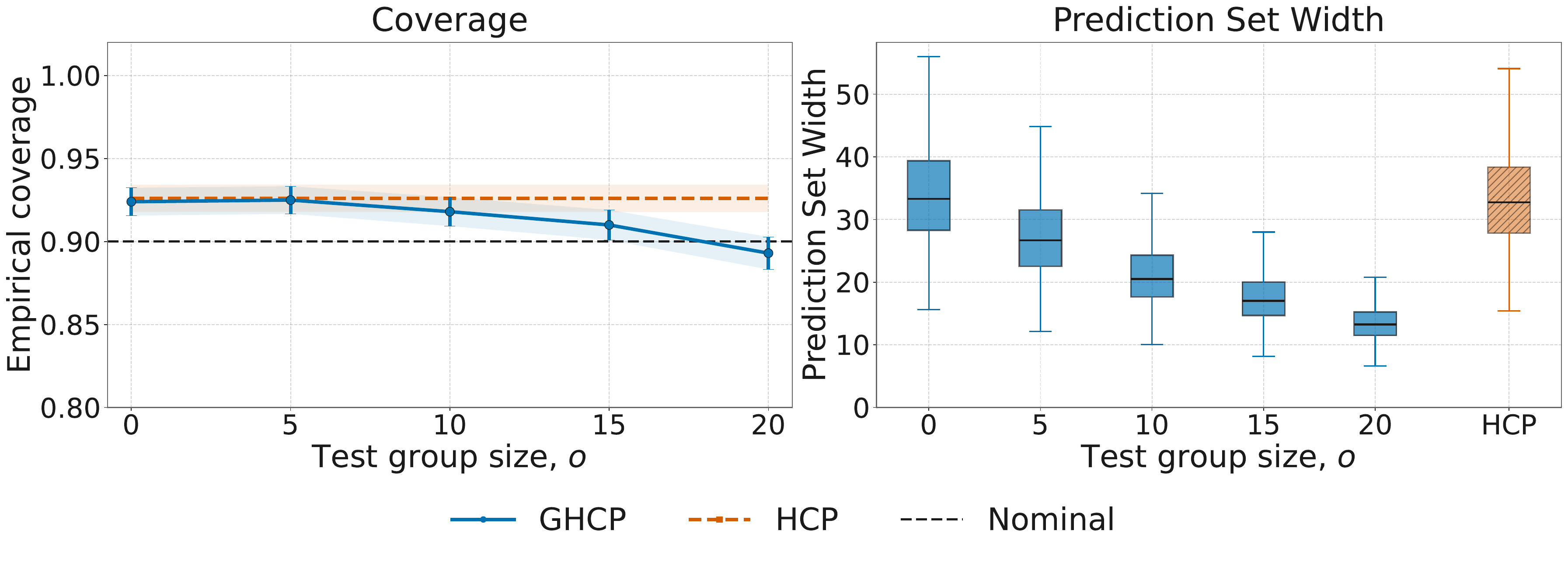}
    \caption{\footnotesize Empirical coverage and width of HCP and GHCP with $o\in\{0,5,10,15,20\}$, using a random forest global predictor. Here $K=20$, $N_j\equiv 21$, and $\alpha=0.1$.}
    \label{fig:N21-cov-width-20}
\end{figure}

\paragraph{Width improvements when all prediction sets are nontrivial.}
Next, we repeat the simulations at a higher level $\alpha=0.1$, where HCP also produces nontrivial intervals with finite widths. Figure~\ref{fig:N21-cov-width-20} and Table~\ref{tab:dhcp-hcp-alpha01} show the coverage and width of the GHCP prediction set across different values of $o$.

\begin{table}
\centering
\caption{\footnotesize Empirical coverage and mean interval width for GHCP across test group size $o$, together with HCP, for fixed $N_j=21$, $\gamma=5$, and $\alpha=0.1$, using a random forest global predictor. Entries are means across $1000$ simulation repetitions, with standard errors in parentheses.}
\label{tab:dhcp-hcp-alpha01}
\small
\setlength{\tabcolsep}{6pt}
\begin{tabular}{lcccccc}
\toprule
& \multicolumn{5}{c}{GHCP} & HCP \\
\cmidrule(lr){2-6}
Value of $o$ & 0 & 5 & 10 & 15 & 20 & -- \\
\midrule
Coverage ($\alpha=0.1$)
& 0.92 (0.008)
& 0.93 (0.008)
& 0.92 (0.009)
& 0.91 (0.009)
& 0.89 (0.010)
& 0.93 (0.008) \\
Mean Width
& 34.15 (0.27)
& 27.52 (0.21)
& 21.22 (0.16)
& 17.54 (0.13)
& 13.57 (0.09)
& 33.54 (0.26) \\
\bottomrule
\end{tabular}
\end{table}

When $o=0$, GHCP nearly coincides with HCP, since there are no
test group observations to exploit. As discussed in
Section~\ref{subsubsec:hcp-review-main}, HCP calibrates above the nominal level $1-\alpha$ because of the weight placed at $+\infty$. In our setting, with $K_1=10$ and $\alpha=0.1$, the effective calibration level among the
finite HCP scores is $0.9+0.9/10=0.99$. Thus, HCP uses an unusually high calibration quantile, which explains the conservative coverage and wider prediction sets in Fig.~\ref{fig:N21-cov-width-20}. As $o$ increases, GHCP incorporates the initial test group observations and its calibration becomes substantially less conservative, leading to the reduction in prediction set width seen in the figure. At $o=20$, the mean width is reduced by $60\%$, and the standard error of the width is reduced by $65\%$, relative to $o=0$. HCP, by contrast, does not use the test group samples and therefore remains unchanged across values of $o$.

\begin{figure}[ht]
    \centering
    \includegraphics[width=0.9\linewidth]{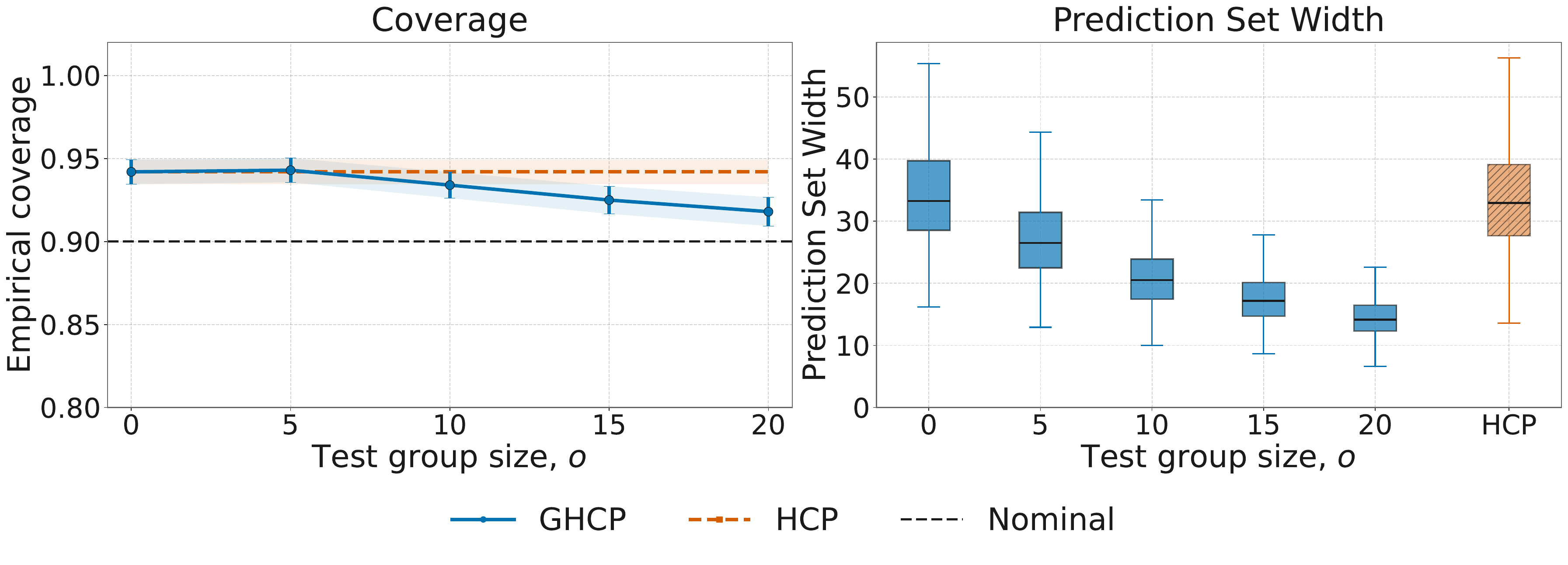}
    \caption{\footnotesize Empirical coverage and width of HCP and GHCP with $o\in\{0,5,10,15,20\}$, using a random forest global predictor. Here $K=20$, $N_j\overset{\text{iid}}{\sim} \text{Poi}(25)$, and $\alpha=0.1$.}
    \label{fig:poi-cov-width-20}
\end{figure}

\paragraph{Experiment with random group sizes.}
Here, we consider heterogeneous reference group sizes drawn as $N_j\overset{\mathrm{i.i.d.}}{\sim}\mathrm{Poisson}(25), j\in[K]$,
again at test level $\alpha=0.1$; the sample sizes are positive with overwhelming probability. The donor group size $N_{J_0}$ also becomes random in this setting. The plots of the coverage and width in Table~\ref{tab:dhcp-hcp-poisson-alpha01} and Figure~\ref{fig:poi-cov-width-20} show similar results as in the fixed-size setting: GHCP continues to attain valid coverage while its width decreases as $o$ increases.

\begin{table}[t]
\centering
\caption{\footnotesize Empirical coverage and mean interval width for GHCP across test group size $o$, together with HCP, for $N_j\overset{\mathrm{i.i.d.}}{\sim}\mathrm{Poi}(25)$, $\gamma=5$, and $\alpha=0.1$, using a random forest global predictor. Entries are means across $1000$ simulation repetitions, with standard errors in parentheses.}
\label{tab:dhcp-hcp-poisson-alpha01}
\small
\setlength{\tabcolsep}{6pt}
\begin{tabular}{lcccccc}
\toprule
& \multicolumn{5}{c}{GHCP} & HCP \\
\cmidrule(lr){2-6}
Value of $o$ & 0 & 5 & 10 & 15 & 20 & -- \\
\midrule
Coverage ($\alpha=0.1$)
& 0.94 (0.007)
& 0.94 (0.007)
& 0.93 (0.008)
& 0.93 (0.008)
& 0.92 (0.009)
& 0.94 (0.007) \\
Mean Width
& 34.29 (0.26)
& 27.23 (0.20)
& 21.02 (0.15)
& 17.60 (0.13)
& 14.52 (0.10)
& 33.60 (0.26) \\
\bottomrule
\end{tabular}
\end{table}

\subsubsection{Benefit of within-group training}

We compare GHCP without within-group training, 
where all the observations in the selected groups are used for inference with GHCP with within-group training, using
$\lfloor o/2\rfloor$ samples; both using $\eta=0.5$.

\begin{figure}[ht]
    \centering
    \includegraphics[width=0.9\linewidth]{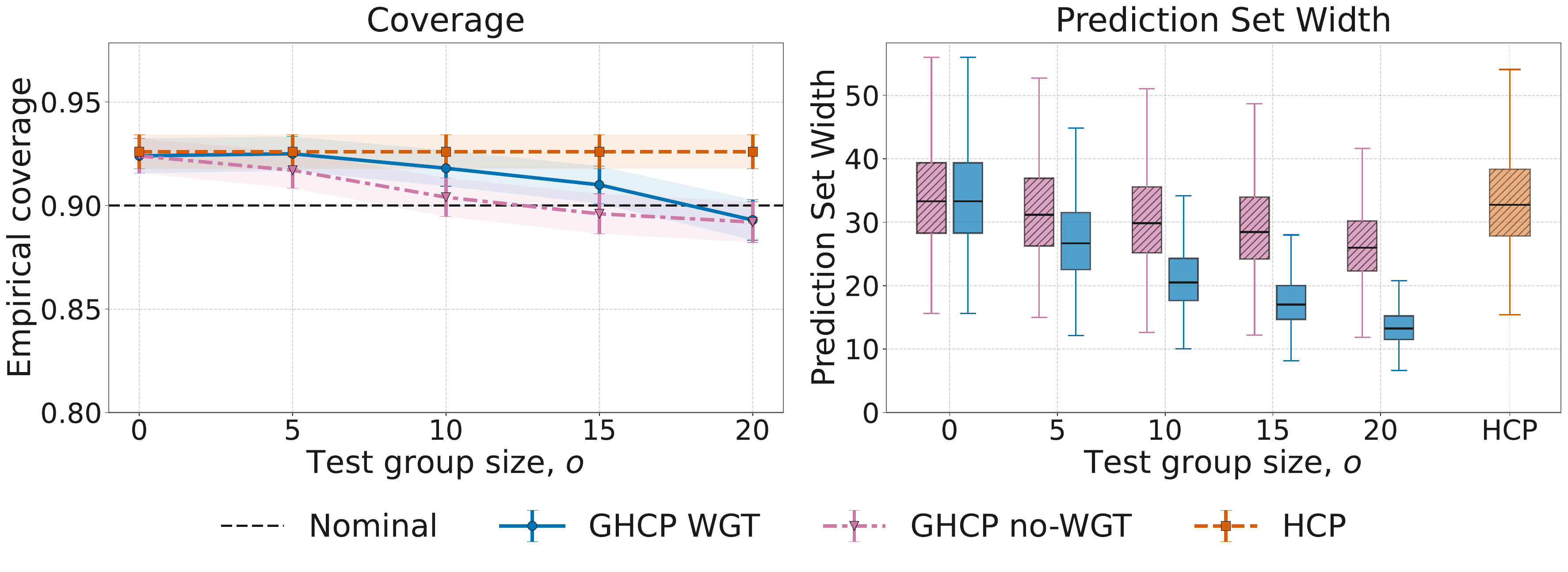}
    \caption{\footnotesize Improvement due to within-group training for $o\in\{0,5, 10, 15, 20\}$, using a random forest global predictor. Here $K=20$, $N_j\equiv 21$, and $\alpha=0.1$.}
    \label{fig:N21-wgt-vs-no-wgt}
\end{figure}

The coverage and width of the prediction sets for different values of $o$ are shown in Figure~\ref{fig:N21-wgt-vs-no-wgt} and Table \ref{tab:sim-with-vs-without-training}. We observe that within-group training helps significantly as $o$ increases. When $o=0$, GHCP with and without WGT coincide, since there are essentially no test group samples to exploit. For $o=20$, WGT yields a $49\%$ reduction in mean width relative to no-WGT, while maintaining coverage closer to the nominal level.

\begin{table}[t]
\centering
\caption{\footnotesize Empirical coverage and mean prediction interval width of GHCP with and without within-group training as functions of the test group size $o$, together with HCP as a baseline, under $N_j=21$, $\gamma=5$, and $\alpha=0.1$, using a random forest global predictor. Standard errors are reported in parentheses.}
\label{tab:sim-with-vs-without-training}
\small
\setlength{\tabcolsep}{5pt}
\begin{tabular}{llccccc}
\toprule
Metric & Method & 0 & 5 & 10 & 15 & 20 \\
\midrule
\multirow{3}{*}{Coverage}
& GHCP (no WGT) & 0.924 (0.008) & 0.917 (0.009) & 0.904 (0.009) & 0.896 (0.010) & 0.892 (0.010) \\
& GHCP (WGT)    & 0.924 (0.008) & 0.925 (0.008) & 0.918 (0.009) & 0.910 (0.009) & 0.893 (0.010) \\
& HCP           & 0.926 (0.008) & 0.926 (0.008) & 0.926 (0.008) & 0.926 (0.008) & 0.926 (0.008) \\
\addlinespace
\multirow{3}{*}{Mean width}
& GHCP (no WGT) & 34.15 (0.267) & 32.16 (0.259) & 30.80 (0.251) & 29.47 (0.241) & 26.68 (0.200) \\
& GHCP (WGT)    & 34.15 (0.267) & 27.52 (0.214) & 21.22 (0.161) & 17.54 (0.131) & 13.57 (0.094) \\
& HCP           & 33.54 (0.260) & 33.54 (0.260) & 33.54 (0.260) & 33.54 (0.260) & 33.54 (0.260) \\
\bottomrule
\end{tabular}
\end{table}

\subsection{Illustration to ACS PUMS income data}
\label{subsec:acs-real}

As an empirical illustration, we show results on the American Community Survey Public Use Microdata Sample (ACS PUMS 2018) dataset (\citet{ding2021folktables}) using Public Use Microdata Areas (PUMAs) as groups. PUMAs are non-overlapping geographic units within a single state, each defined to contain at least 100,000 residents. Each observation corresponds to an adult individual, and the response variable $Y$ is annual income (in U.S.\ dollars). We focus on California PUMAs and study the income of recent foreign-born individuals. 

Specifically, we restrict the analysis to foreign-born adults aged $25$-$54$ who report at least $40$ usual hours worked per week and who entered the United States in 2000 (YOEP) or later. The age restriction focuses on the prime working-age population, while the hours worked and year-of-entry restrictions yield a more comparable labor-market population across PUMAs.  Under this pre-processing, the filtered dataset contains 12,285 individuals across 265 California PUMAs. We retain only PUMAs with at least $21$ observations after filtering, so that an initial sample of size up to $20$ and a subsequent prediction target are available in every eligible PUMA. After this filtering, 212 PUMAs remain eligible for the experiment. Among eligible PUMAs, group sizes $N_j$ range from 21 to 252 with median 46.

Restricting the analysis to California further removes state-level differences in survey geography and socioeconomic factors that could otherwise induce stronger associations between group size and the group-specific data-generating law. 
Although group-size ignorability (Assumption~\ref{assump:independence}) is only an approximation, we find that GHCP continues to attain near-nominal empirical coverage while substantially reducing interval width, illustrating that the method can remain useful beyond the exact conditions of Theorem~\ref{thm:donor-valid}.

\paragraph{Overview of experiment.}
For each test PUMA, we fix the individual at position 21 as the prediction target and vary the number of preceding observations treated as the initial
test group sample. Specifically, for each
\(
o\in\{0,5,10,15,20\},
\)
we use the first $o$ observations as the initial sample and predict the
outcome of the individual at position 21. We evaluate marginal coverage and
prediction set width at nominal coverage level $1-\alpha=0.9$.

\paragraph{Covariates and score construction.}
There are no additional group-level covariates, and thus we construct the scores using the individual observations $(X_{j,i},Y_{j,i})$. For each individual, the covariate vector consists of the continuous variables \texttt{age} and \texttt{hours worked}, together with indicators for \texttt{marital status}, \texttt{sex}, \texttt{education}, and \texttt{English proficiency}. We apply random forest regression (50 trees, minimum leaf size 5) to construct the global predictor. For GHCP, the within-group training step uses \eqref{eq:weighted-merge}. We use the absolute residual score.

\paragraph{Evaluation protocol.}
In each of $B=1000$ trials, we draw 20 reference PUMAs and one test PUMA uniformly without replacement, reflecting the group-level symmetry in Assumption \ref{assump:group-exch}. Since ACS records have no meaningful within-PUMA ordering, we independently apply a uniform random permutation to the individuals within each selected PUMA. 

\paragraph{Results.}

\begin{figure}[ht]
    \centering
    \includegraphics[width=0.9\linewidth]{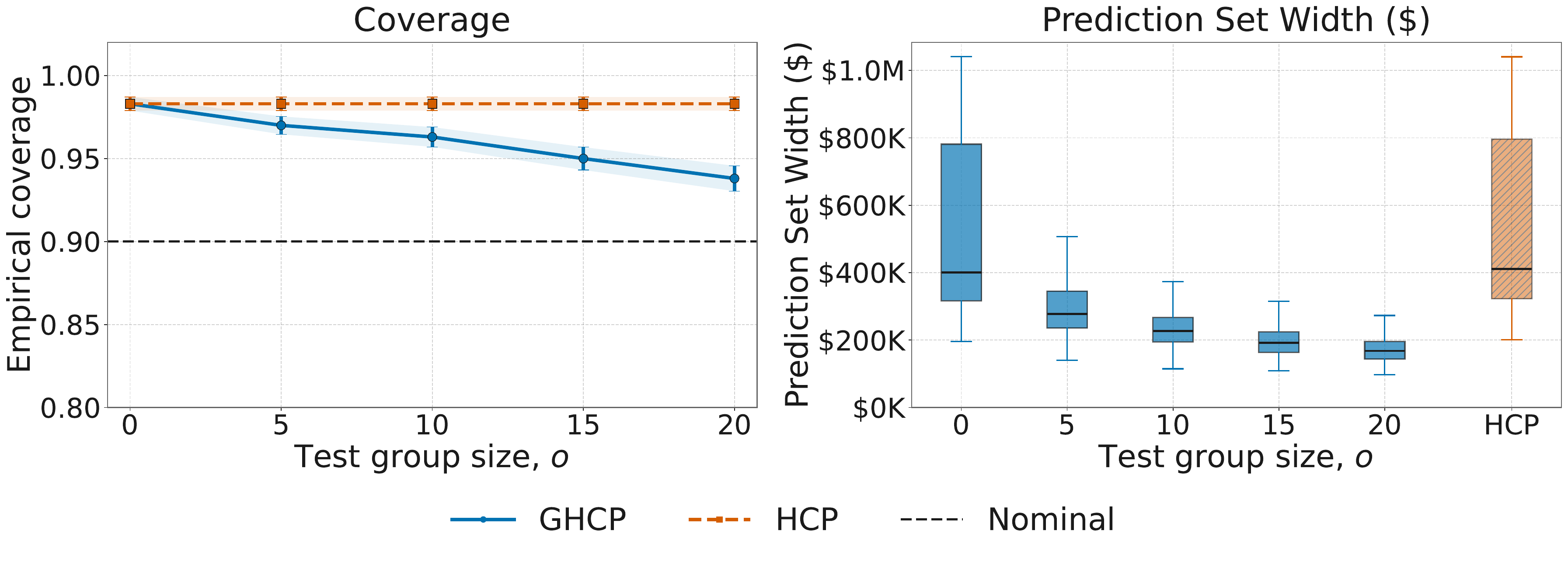}
    \caption{\footnotesize Empirical coverage and width on the ACS PUMS data for HCP and GHCP, both with an RF global predictor; GHCP additionally uses local within-group training.}
    \label{fig:acs-main-coverage}
\end{figure}

\begin{table}[t]
\centering
\caption{\footnotesize Empirical coverage and mean interval width for
GHCP across initial test-group sample sizes $o$, together with HCP, for
the ACS experiment with $\alpha=0.1$. Entries are means across
$B=1000$ trials, with standard errors in parentheses. HCP does not use
the initial test-group sample and hence does not vary with $o$.}
\label{tab:acs-main-results}
\small
\setlength{\tabcolsep}{6pt}
\begin{tabular}{lccccc}
\toprule
& \multicolumn{5}{c}{Initial test-group sample size $o$} \\
\cmidrule(lr){2-6}
Method & 0 & 5 & 10 & 15 & 20 \\
\midrule
\multicolumn{6}{l}{\textit{Empirical coverage ($1-\alpha=0.90$)}} \\[2pt]
GHCP
& 0.983 (0.004)
& 0.970 (0.005)
& 0.963 (0.006)
& 0.950 (0.007)
& 0.938 (0.008) \\
HCP
& 0.983 (0.004) & & & & \\
\addlinespace[4pt]
\multicolumn{6}{l}{\textit{Mean interval width}} \\[2pt]
GHCP
& 523{,}195 (7{,}685)
& 315{,}773 (4{,}285)
& 247{,}714 (3{,}027)
& 200{,}557 (1{,}845)
& 174{,}906 (1{,}595) \\
HCP
& 531{,}277 (7{,}722) & & & & \\
\bottomrule
\end{tabular}
\end{table}

Fig.\ \ref{fig:acs-main-coverage} and Table~\ref{tab:acs-main-results} show the benefit of using test group information. At $o=0$, GHCP has similar mean width as HCP. However, as the number of test group observations $o$ increases, the width of GHCP prediction set shrinks steadily and by $o=20$, it decreases to $174{,}906$, which is a $67.08\%$ reduction relative to HCP. 

The variability of interval widths also decreases with $o$. The standard error of replicate-level GHCP widths drops from $7{,}722$ for HCP to $1{,}595$ at $o=20$ ($79.3\%$ reduction). Coverage decreases mildly with $o$, remaining above the nominal $0.90$ level throughout.

GHCP also compares favorably to a naive test group standard conformal prediction (Std-CP) which uses the first half of the test group data for training and remaining half for calibration. 
To further correct for potential heterogeneity, we 
use 
Std-CP with a \emph{studentized} score: a local random forest is fit on half of the $o$ observations in the target PUMA, a second forest estimates a local scale $\widehat\sigma$, and nonconformity uses $s(u, x,y) = |y - \hat{\mu}(u, x)|/\widehat{\sigma}(u, x)$ on the held-out half. Under this construction, Std-CP yields infinite mean widths for all the settings where $o<20$, at $\alpha=0.1$. At $o=20$, GHCP’s mean width is smaller ($174{,}906$ versus $211{,}223$)
and less volatile
($1{,}595$  versus $6{,}916$)
than Std-CP’s. 

\section{Discussion}
\label{sec:discussion}

We proposed GHCP for hierarchical predictive inference when initial observations from the test group are available. The key idea is to
assign the test group a size donated by a reference group, thereby restoring the symmetry needed for conformal calibration without specifying
a probabilistic model for the test-group size.

The cost of avoiding such a size model is the group-size ignorability condition in Assumption~\ref{assump:independence}, which rules out systematic dependence between reference-group sizes and their data-generating laws. Our sensitivity experiments (Appendix \ref{appsubsec:size-ignorability-sensitivity}) suggest that moderate departures from this condition need not have a large empirical effect, but the finite-sample guarantee no longer applies once the assumption is violated. Developing procedures that remain valid under dependence between group size and group law is therefore an important direction for future work.

Several questions concerning efficiency also remain open. The restricted donor rule used here depends only on group sizes; allowing donor selection
to incorporate group-level covariates while preserving the required symmetry could further improve efficiency. Likewise, the global-local
merger used in our experiments is deliberately simple, and Appendix \ref{appsubsec:weight-sensitivity} shows that the preferred weighting depends on the information carried by the global and local predictors. Developing optimal data-adaptive
mergers with finite-sample validity is a natural next step. Finally, while Appendix \ref{appsec:derandomization} gives a derandomized version of GHCP, its generic $p$-value
merger can be conservative, motivating less conservative approaches to removing donor randomization.

\section*{Acknowledgments}

This work was supported in part by 
NIH R01-AG065276, R01-GM139926, NSF 2210662, P01-AG041710, R01-CA222147, as well as the ARO, ONR, and the Sloan Foundation.

{\small
\setlength{\bibsep}{0.2pt plus 0.3ex}
\bibliographystyle{plainnat-abbrev}
\bibliography{ref}
}

\appendix

\addcontentsline{toc}{section}{Appendix}
\begin{center} \section*{Appendix} \end{center}

\section{Background}
\label{appsec:background}

\subsection{Split conformal prediction}
\label{appsubsec:split-cp}

We begin with the standard setting of exchangeable data. Let
$Z_1,\ldots,Z_{n+1}$ be exchangeable, where
$Z_i=(X_i,Y_i)\in\cX\times\cY$. The prediction task is to construct,
given the new feature input $X_{n+1}$, a prediction set for $Y_{n+1}$.
In split conformal prediction, the first $n$ observations are partitioned
into a training set $(Z_i)_{1\le i\le n_0}$ and a calibration set
$(Z_i)_{n_0+1\le i\le n}$, where $n_0+n_1=n$. The training data are used
to fit a prediction rule and thereby construct a nonconformity score
$s:\cX\times\cY\to\R$. The split conformal prediction set is
\begin{equation*}
%\label{eq:split-cp}
\widehat C_{\mathrm{split}}(X_{n+1})
=
\left\{y\in\cY:
 s(X_{n+1},y)
 \le
 \quant{1-\alpha}{
 \frac{1}{n_1+1}\sum_{i=n_0+1}^{n}\delta_{s(Z_i)}
 +\frac{1}{n_1+1}\delta_{\infty}}
\right\}.
\end{equation*}
Conditional on the training data, the $n_1$ calibration scores and the
test score are exchangeable. Consequently,
\(
\PP\{Y_{n+1}\in\widehat C_{\mathrm{split}}(X_{n+1})\}\ge1-\alpha
\)
\citep{vovk2005algorithmic}.

\subsection{Hierarchical exchangeability in HCP} \label{appsubsec:hcp-review} Following the notation in Section~\ref{subsec:problem-setup}, let $\widetilde{\mathcal W}$ be the set of all finite sequences with entries in $\mathcal W$, 
$\widetilde{\mathcal W}=\bigcup_{m\ge 1}\mathcal W^m.$ For a group $\widetilde W_j\in\widetilde{\mathcal W}$, let $\length(\widetilde W_j)$ denote its length. As before, the $(K+1)$-th group is the test group. The HCP procedure \citep{lee2023distribution}, reviewed in Section~\ref{subsec:problem-setup}, assumes hierarchical exchangeability of the calibration and test groups. Formally, this can be stated as follows.

\setassumptionprefix{HCP-A} 

\begin{assumption}[Across-group exchangeability] \label{assump:HCP-group-exch} 
The random vectors $(\widetilde W_1,\ldots,\widetilde W_{K+1})$ are exchangeable, i.e., for any permutation $\sigma\in\mathcal S_{K+1}$, 
\( (\widetilde W_1,\ldots,\widetilde W_{K+1}) 
 \overset{d}{=} (\widetilde W_{\sigma(1)},\ldots,\widetilde W_{\sigma(K+1)}). 
\) 
\end{assumption} 

\begin{assumption}[Within-group exchangeability]
\label{assump:HCP-within-group-exch}
For each $j\in[K+1]$, each $m\geq 1$ such that
$\PP\{\operatorname{length}(\widetilde W_j)=m\}>0$, and every
$\sigma\in \mathcal{S}_m$,
\[
\begin{aligned}
&(\widetilde W_1,\ldots,\widetilde W_j,\ldots,\widetilde W_{K+1}) \, \mid \, \{\operatorname{length}(\widetilde W_j)=m\}
\overset{d}{=}\
(\widetilde W_1,\ldots,\widetilde W_j^{\sigma},\ldots,\widetilde W_{K+1} )
\, \mid \,
\{\operatorname{length}(\widetilde W_j)=m\},
\end{aligned}
\]
where
\(
\widetilde W_j^{\sigma}
=
(W_{j,\sigma(1)},\ldots,W_{j,\sigma(m)}).
\)
\end{assumption}

\setassumptionprefix{A} 

We say that the hierarchical data $(\widetilde W_1,\ldots,\widetilde W_{K+1})$ satisfy hierarchical exchangeability if Assumptions~\ref{assump:HCP-group-exch} and~\ref{assump:HCP-within-group-exch} hold. This condition is weaker than full exchangeability of all individual observations.
\begin{remark}[Relation with Assumptions \ref{assump:group-exch}-\ref{assump:within-group-exch}]
\label{rem:relation-to-hcp}
Let $\Pi_m$ denote the class of exchangeable distributions on $\mathcal W^m$, and define \[ \Pi_m^{\mathrm{extend}} := \Big\{ P\in\Pi_m:\ \exists\ \text{an exchangeable distribution }P_\infty \text{ on }(\mathcal W^{\mathbb N},\mathcal B^{\otimes\mathbb N}) \text{ such that } P=P_\infty\circ\, \mathrm{pr}_{1:m}^{-1} \Big\}, \] where $\mathrm{pr}_{1:m}$ denotes projection onto the first $m$ coordinates. 

\begin{enumerate}[label=(\roman*)]
    \item Assumption~\ref{assump:HCP-within-group-exch} requires only that the observations within each group have an exchangeable distribution in $\Pi_m$. In contrast, Assumption~\ref{assump:within-group-exch} restricts each finite-dimensional conditional law to $\Pi_m^{\mathrm{extend}}$, the subclass of infinitely extendible exchangeable laws, and also imposes conditional independence across groups. Under mild measurability conditions (e.g., when $\mathcal W$ is a standard Borel space), $\Pi_m^{\mathrm{extend}}$ is the class of finite-dimensional marginals of de~Finetti mixtures. Thus Assumption~\ref{assump:within-group-exch} may be viewed as a conditionally i.i.d.\ random-measure model, together with conditional independence across groups.

    \item At the group level, HCP assumes full group exchangeability (Assumption~\ref{assump:HCP-group-exch}), which implies the exchangeability of the length vector $(N_1,\ldots,N_K,o)$ in our setting. In contrast, Assumptions~\ref{assump:group-exch}--\ref{assump:independence} do not require the variable-length observed rows to be exchangeable. These assumptions are therefore not comparable to Assumption~\ref{assump:HCP-group-exch} in general. The size-ignorability assumption, Assumption~\ref{assump:independence}, makes the group sizes non-predictive of the group-level laws and is what justifies the donor replacement step in Algorithm~\ref{alg:dhcp}.
\end{enumerate} 
\end{remark}

\subsection{Special Case: A known-size extension of HCP}
\label{appsubsec:known-size-hcp}

We first consider a special case in which the test group has a well-defined
complete size. Specifically, suppose that each group consists of a finite
collection of observations, as in Section \ref{subsec:problem-setup}, and that
the complete test group size $N_{K+1}$ is known at the time of prediction and
$N_{K+1}>o$ almost surely, although only the first $o$ observations from that
group have been observed. Such a setting may arise, for example, when the test group is a
closed cohort with known membership. This differs from our general setup, where the observed test group sample size $o$ need not be associated with any underlying complete group size.

Suppose Assumptions~\ref{assump:HCP-group-exch} and
\ref{assump:HCP-within-group-exch} hold. Let groups
$1,\ldots,K_0$ be used for training and the remaining
$K_1=K-K_0$ reference groups for calibration, and let $s$ be a
nonconformity score constructed from the training data. Since the complete
size $N_{K+1}$ is known, the $o$ observed test group scores can be included
with their usual within-group weights, while the remaining
$N_{K+1}-o$ unobserved scores are represented by weight at $+\infty$. This leads to
\begin{equation*}
%\label{eq:known-size-set}
\widehat C_{\mathrm{known}}(U_{K+1},X_{K+1,o+1})
=
\Bigl\{
y\in\cY:
s(U_{K+1},X_{K+1,o+1},y)
\le
\quant{1-\alpha}{\nu_{\mathrm{known}}}
\Bigr\},
\end{equation*}
where
\[
\nu_{\mathrm{known}}
=
\sum_{j=K_0+1}^{K}\sum_{i=1}^{N_j}
\frac{1}{(K_1+1)N_j}\delta_{s(W_{j,i})}
+
\sum_{i=1}^{o}
\frac{1}{(K_1+1)N_{K+1}}\delta_{s(W_{K+1,i})}
+
\frac{N_{K+1}-o}{(K_1+1)N_{K+1}}\delta_{\infty}.
\]

\begin{proposition}
\label{prop:known-size}
Suppose Assumptions~\ref{assump:HCP-group-exch} and
\ref{assump:HCP-within-group-exch} hold, and suppose that the complete
test group size $N_{K+1}$ is known and $N_{K+1}>o$ almost surely. Then the prediction set above satisfies
\(
\PP\bigl\{
Y_{K+1,o+1}\in
\widehat C_{\mathrm{known}}(U_{K+1},X_{K+1,o+1})
\bigr\}
\ge 1-\alpha.
\)
\end{proposition}

\begin{proof}
Condition on the groups used to construct the score function $s$. The
remaining $K_1$ calibration groups and the test group are still
hierarchically exchangeable. Relabel these groups as
$1,\ldots,M$, where $M=K_1+1$ and group $M$ is the test group. Write
$s_{j,i}=s(W_{j,i})$ and define the full-data measure
\begin{equation*}
\nu_{\mathrm{full}}
:=
\sum_{j=1}^{M}\sum_{i=1}^{N_j}
\frac{1}{MN_j}\delta_{s_{j,i}},
\qquad
q_{1-\alpha}:=\quant{1-\alpha}{\nu_{\mathrm{full}}}.
\end{equation*}
The map from the collection of score vectors to $q_{1-\alpha}$ is
invariant under permutations of the groups and under permutations within
each group. Therefore, by the within-group exchangeability,
\begin{align*}
\PP\{s_{M,o+1}\le q_{1-\alpha}\mid\text{training data}\}=
\EE\left[
\frac{1}{N_M}\sum_{i=1}^{N_M}
\1\{s_{M,i}\le q_{1-\alpha}\}
\,\middle|\,\text{training data}
\right].
\end{align*}
From the across-group exchangeability, we then have
\begin{align*}
&\PP\{s_{M,o+1}\le q_{1-\alpha}\mid\text{training data}\}\\
&\qquad=
\EE\left[
\frac{1}{M}\sum_{j=1}^{M}\frac{1}{N_j}
\sum_{i=1}^{N_j}
\1\{s_{j,i}\le q_{1-\alpha}\}
\,\middle|\,\text{training data}
\right]\\
&\qquad=
\EE\left[
\nu_{\mathrm{full}}((-\infty,q_{1-\alpha}])
\,\middle|\,\text{training data}
\right]
\ge1-\alpha.
\end{align*}
For every finite $t$, replacing the unobserved test-group scores by
$+\infty$ can only decrease the cumulative distribution function:
\begin{equation*}
\nu_{\mathrm{known}}((-\infty,t])
\le
\nu_{\mathrm{full}}((-\infty,t]).
\end{equation*}
Hence
$\quant{1-\alpha}{\nu_{\mathrm{known}}}\ge q_{1-\alpha}$, and the
claimed coverage follows after removing the conditioning on the training
data.
\end{proof}

Thus, no additional construction is needed when the complete test-group
size is both well defined and known. The difficulty addressed by GHCP
arises when such a complete size is unavailable or may not be well
defined.

\section{Implementation Details}
\label{appsec:implementation}

\subsection{General global--local mergers and score construction}
\label{appsubsec:more-mergers}

Algorithm~\ref{alg:dhcp} uses a simple score construction. First, using
the training groups, we fit a global predictor
$\widehat\mu^{\mathrm{glob}}:\mathcal U\times\mathcal X\to\mathbb R$.
This predictor may be constructed by any regression or machine-learning
method applied to the training data. For every group used in calibration,
$j\in S_{\mathrm{cal}}\cup\{K+1\}$, let
\begin{equation*}
%\label{eq:with-group-training-size}
m_j:=\tau(N_j,o),
\qquad
\tau:\mathbb N\times\mathbb N_0\to\mathbb N_0,
\qquad
0\le\tau(n,o)\le\min\{o,n-1\}.
\end{equation*}
Thus, at least one observation is left in the held-out block. In the main
procedure, every calibration group has $N_j>o$ and
$m_j=\lfloor o/2\rfloor$, which satisfies this condition.

The convex merger in~\eqref{eq:weighted-merge} is the special case. More generally, let $\mathcal A$ be a common measurable group-wise rule.
The score in group $j$ may take the form
\begin{equation*}
%\label{eq:general-groupwise-score}
s_j(u,x,y)
=
\mathcal A\!\left(
\widehat\mu^{\mathrm{glob}},U_j,
Z_{j,1},\ldots,Z_{j,m_j};u,x,y
\right),
\qquad j\in S_{\mathrm{cal}}\cup\{K+1\}.
\end{equation*}
The same rule $\mathcal A$ must be used in every participating group, and
held-out scores must be computed only from observations outside the local
training block. The use of the first $m_j$ observations is only a
convention. One may instead use a prespecified subset of positions, or a
random subset chosen independently of the group observations by the same
permutation-equivariant rule in every group. A data-dependent split needs
a separate symmetry argument and is not covered by the statement above.
Under the stated conditions, the proof of Theorem~\ref{thm:donor-valid}
is unchanged.

One useful extension treats $\widehat\mu^{\mathrm{glob}}$ as a baseline
and estimates a group-specific correction from the within-group training
block. For $i\in[m_j]$, define
$
R_{j,i}:=Y_{j,i}-\widehat\mu^{\mathrm{glob}}(U_j,X_{j,i}).
$
Let $\mathcal H$ be a class of functions on $\mathcal X$, let $\ell$ be
a loss function, and let $\mathrm{pen}:\mathcal H\to[0,\infty]$ control
complexity. For $m_j\ge1$, one may fit
\begin{equation*}
\widehat g_j
\in
\arg\min_{g\in\mathcal H}
\left\{
\frac{1}{m_j}\sum_{i=1}^{m_j}
\ell\bigl(R_{j,i}-g(X_{j,i})\bigr)
+
\kappa_j\,\mathrm{pen}(g)
\right\},
\end{equation*}
where $\kappa_j\ge0$ is chosen by the same group-wise rule in every group (for example, $\kappa_j=\kappa(N_j,o)$), and set
$\widetilde\mu_j(U_j,x)=\widehat\mu^{\mathrm{glob}}(U_j,x)+\widehat g_j(x)$.
For $m_j=0$, set $\widehat g_j\equiv0$. Here, $\widehat g_j$ estimates the portion of the conditional mean in group $j$ that is not captured by the global predictor. If $\mathcal H$ contains only constant functions, this corresponds to a
group-specific intercept correction. More flexible choices, such as
spline spaces or reproducing-kernel Hilbert spaces
\citep{hastie2009elements,wahba1990spline}, allow the correction to depend
on $x$. Regularization keeps $\widehat g_j$ close to zero when $m_j$ is
small.

The same flexibility applies to the score. Instead of
$s_j(U_j,X,Y)=|Y-\widetilde\mu_j(U_j,X)|$, one may use, for example,
$
s_j(U_j,X,Y)
=
|Y-\widetilde\mu_j(U_j,X)|/\widehat\sigma_j(U_j,X).
$
This remains covered by the same validity argument provided that
$\widehat\sigma_j$ is constructed in every group by the same group-wise
procedure using only that group's local training block. Such
studentization can be useful under heteroskedasticity.

\subsection{Selecting the restricted donor pool}
\label{appsubsec:restriction}

We give the precise construction of the restricted donor pool
$S_\eta$ in \eqref{eq:restricted-region} from Section \ref{subsec:restricted-selection} when ties occur at the cutoff.
Recall that
\(
S=\{j\in[K]:N_j>o\},
m_\eta=\min\left\{|S|,\left\lceil(1-\eta)K\right\rceil\right\}.
\)
If $m_\eta=0$, we set $S_\eta=\varnothing$. Otherwise, let $c_\eta$ denote the $m_\eta$-th smallest value among
$\{N_j:j\in S\}$, i.e.,
\(
c_\eta
:=
\min\left\{
c:\left|\{j\in S:N_j\le c\}\right|\ge m_\eta
\right\},
\)
 and define
\[
S_{<}:=\{j\in S:N_j<c_\eta\},
\qquad
S_{=}:=\{j\in S:N_j=c_\eta\}.
\]
All groups in $S_{<}$ are retained. From the groups tied at the cutoff,
we select uniformly at random a subset
$
B_\eta\subseteq S_{=},
|B_\eta|=m_\eta-|S_{<}|,
$
and set
$S_\eta=S_{<}\cup B_\eta$.
Thus, $S_\eta$ contains exactly the $m_\eta$ compatible groups with the
smallest sizes, with uniform randomization only when the cutoff size is
shared by more groups than can be retained. This construction is
equivariant under permutations of the reference group labels. The randomization used to resolve ties is separate from the subsequent
donor selection. Once $S_\eta$ has been constructed, Algorithm~\ref{alg:dhcp} draws $J_0\mid S_\eta\sim\Unif(S_\eta)$
whenever $S_\eta\neq\varnothing$.

\section{Derandomization of GHCP}
\label{appsec:derandomization}

The GHCP procedure in Section \ref{subsec:G-HCP} is randomized through the choice of the donor group. We now provide a derandomized procedure by forming a donor-specific predictive $p$-value for different realizations and combining these $p$-values using the arithmetic-mean merger of
\citet{vovk2020combining}. The construction in this section is restricted to the GHCP procedure with no within-group training. Let $S\subseteq\{j\in[K]:N_j>o\}$ be the donor pool, which for the purposes of this section is assumed to be non-empty. Note that no derandomization is necessary if the donor pool is empty. 
For a candidate response $y\in\mathcal Y$, define
\(
t(y):=s(U_{K+1},X_{K+1,o+1},y), r_i:= s(U_{K+1},X_{K+1,i},Y_{K+1,i}), i\in[o].
\)
Let $\mathcal S_K$ denote the set of all permutations of $[K]$. For each $\sigma\in\mathcal S_K$, 
define
\(
J_\sigma
:=
\sigma\!\left(
\min\{\ell\in[K]:\sigma(\ell)\in S\}
\right).
\)
Observe that each permutation $\sigma \in \mathcal S_K$
defines a donor for every realization of $S$. For $\sigma\in\mathcal S_K$, define
\begin{equation}
\label{eq:order-pvalue}
\pi_\sigma(y)
:=
\sum_{m\in S\setminus\{J_\sigma\}}
\frac{1}{|S|N_m}
\sum_{i=1}^{N_m}
\1\{s(W_{m,i})\ge t(y)\}
+
\frac{1}{|S|N_{J_\sigma}}
\sum_{i=1}^{o}
\1\{r_i\ge t(y)\}
+
\frac{N_{J_\sigma}-o}{|S|N_{J_\sigma}}.
\end{equation}
Lemma~\ref{lem:fixed-order-valid} shows that, for every fixed
$\sigma\in\mathcal S_K$ and every nonempty possible realization of $S$,
\(
\PP\left\{
\pi_\sigma(Y_{K+1,o+1})\le u
\,\middle|\,
S
\right\}
\le u,~\text{for}~ u\in(0,1),
\)
implying that $\pi_\sigma(Y_{K+1,o+1})$ is a valid $p$-value. 

Note that the random variables
$\bigl\{\pi_\sigma(Y_{K+1,o+1}):\sigma\in\mathcal S_K\bigr\}$
are computed from the same observations and are thus dependent.
We combine them using the arithmetic-mean $p$-value merger:
\begin{equation}
\label{eq:merged-pvalue}
\pi_{\mathrm{merge}}(y)
:=
\min\left\{
1,\,
\frac{2}{|\mathcal S_K|}
\sum_{\sigma\in\mathcal S_K}
\pi_\sigma(y)
\right\}.
\end{equation}
Although $\mathcal{S}_K$ contains $K!$ permutations, the average in~\eqref{eq:merged-pvalue} admits a simple closed-form expression and can therefore be computed efficiently. 

\begin{lemma}
\label{lem:donor-closed-form}
Suppose $S\neq\varnothing$, and let $M:=|S|$. Define
\(
\bar w
:=
\frac1{M^2}\sum_{j\in S}\frac1{N_j}
\)
and
\begin{equation*}
%\label{eq:bar-nu-donor}
\bar\nu_{\mathrm{donor}}
:=
\frac{M-1}{M}
\sum_{j\in S}\sum_{i=1}^{N_j}
\frac{1}{MN_j}\delta_{s(W_{j,i})}
+
\bar w\sum_{i=1}^{o}\delta_{r_i}
+
\left(
\frac1M-o\bar w
\right)\delta_\infty.
\end{equation*}
Then, for every $y\in\mathcal Y$,
\(
\frac{1}{|\mathcal S_K|}
\sum_{\sigma\in\mathcal S_K}
\pi_\sigma(y)
=
\bar\nu_{\mathrm{donor}}
\left([t(y),\infty]\right).
\)
\end{lemma}
The weights in  $\bar\nu_{\mathrm{donor}}$ are nonnegative because
$N_j>o$ for every $j\in S$, and they sum to one. Thus
$\bar\nu_{\mathrm{donor}}$ is a probability measure with finite support
on $\overline{\mathbb R}$. For any finitely supported probability measure
$\nu$, any $a\in(0,1)$, and any finite $t$,
\begin{equation*}
\nu([t,\infty])>a
\quad\Longleftrightarrow\quad
t\le\quant{1-a}{\nu}.
\end{equation*}
Combining this fact with Lemma~\ref{lem:donor-closed-form}, we define, on
$\{S\neq\varnothing\}$,
\begin{equation}
\label{eq:ddhcp-pvalue-set}
\widehat C_{\mathrm{D\text{-}GHCP}}
(U_{K+1},X_{K+1,o+1})
:=
\left\{
y\in\mathcal Y:
\pi_{\mathrm{merge}}(y)>\alpha
\right\}
=
\left\{
y\in\mathcal Y:
t(y)
\le
\quant{1-\alpha/2}{\bar\nu_{\mathrm{donor}}}
\right\}.
\end{equation}
If $S=\varnothing$, no donor averaging is needed. In that case use the ordinary split-conformal set
\begin{equation*}
\widehat C_{\mathrm{D\text{-}GHCP}}
(U_{K+1},X_{K+1,o+1})
:=
\left\{y\in\mathcal Y:
 t(y)\le\quant{1-\alpha}{\frac{1}{o+1}\sum_{i=1}^{o}\delta_{r_i}
+
\frac{1}{o+1}\delta_{\infty}}
\right\}.
\end{equation*}

\begin{theorem}
\label{thm:donor-merge}
Suppose that Assumptions~\ref{assump:group-exch}--\ref{assump:within-group-exch} hold. Let $S$ denote the donor pool, and suppose that the score function $s$ is constructed from the reference groups outside $S$. Then
\[
\PP\left\{
Y_{K+1,o+1}\in
\widehat C_{\mathrm{D\text{-}GHCP}}
(U_{K+1},X_{K+1,o+1})
\right\}
\ge
1-\alpha.
\]
\end{theorem}

A proof of Theorem~\ref{thm:donor-merge} is given in
Appendix~\ref{appsec:proofs-alt}.

\begin{remark}[Conservativeness of the merger]
The factor $2$ in \eqref{eq:merged-pvalue} can be viewed as the price for imposing no assumptions on the dependence among the donor-specific $p$-values. Consequently, the derandomized procedure can be more conservative in terms of prediction set size than randomized GHCP. 
Less conservative mergers may be available under additional information about the joint distribution of the donor-specific $p$-values.
\end{remark}

\section{Additional Experiments}
\label{appsec:additional-experiments}

In this section, we provide additional experimental results. We begin with an oracle conditional-mean calculation for the simulation DGP in Section~\ref{subsec:main-simulations}, which provides a scale for interpreting the quality of the fitted predictor. We then compare GHCP with additional baseline methods, explore the performance under heterogeneous group sizes, and provide these comparisons also for the ACS PUMS application.\footnote{Code to reproduce these experiments is available at \href{https://github.com/soham-penn/hierarchical_CP}{\texttt{https://github.com/soham-penn/hierarchical\_CP}}.}

\subsection{Comparison with the Bayes-optimal predictor}
\label{appsubsec:simulation-bayes-error}

The efficiency of the prediction sets in Section~\ref{sec:experiments} depends in part on the predictor used to construct the nonconformity scores. We therefore compare the random forest used in Section~\ref{sec:experiments} with the Bayes-optimal predictor based only on $(X,U)$ under the true data-generating model. This oracle predictor uses the exact conditional mean of $Y$ given $(X,U)$, but it cannot account for the realized latent effect $B$ of the test group. The initial test group responses contain information about this latent group effect that is unavailable from $(X,U)$ alone. By comparing GHCP under the random forest and the Bayes predictor, we assess whether the reduction in prediction set width from incorporating test group observations persists even when the global predictor already extracts all information available from $(X,U)$.

We first derive the oracle predictor. For $u\in\mathbb{R}^d$, let
\[
\mu_X(u)
=
(u_1^2,\ldots,u_{d-1}^2)^\top,
\quad
\mu_Y(u)
=
u_d^2,
\quad
\Sigma_X(u)
=
(1-\rho)\operatorname{diag}(u_1,\ldots,u_{d-1})
+
\rho\mathbf{1}_{d-1}\mathbf{1}_{d-1}^\top.
\]
Under the simulation model,
\[
\operatorname{Cov}(X,Y\mid U=u,B=b)
=
\rho\mathbf{1}_{d-1},
\quad
\operatorname{Var}(Y\mid U=u,B=b)
=
(1-\rho)u_d+\rho.
\]
Hence, by the conditional distribution of a multivariate Gaussian,
\[
Y\mid X=x,U=u,B=b
\sim
N\left(
\mu_Y(u)+b
+
\rho\mathbf{1}_{d-1}^\top
\Sigma_X(u)^{-1}
\{x-\mu_X(u)\},
\;
v(u)
\right),
\]
where
\(
v(u)
=
(1-\rho)u_d+\rho
-
\rho^2
\mathbf{1}_{d-1}^\top
\Sigma_X(u)^{-1}
\mathbf{1}_{d-1}.
\)
Since $B\sim N(0,\gamma^2)$ independently of $(X,U)$, integrating out
$B$ leads to
\[
Y\mid X=x,U=u
\sim
N\left(
\mu_Y(u)
+
\rho\mathbf{1}_{d-1}^\top
\Sigma_X(u)^{-1}
\{x-\mu_X(u)\},
\;
v(u)+\gamma^2
\right).
\]
Therefore, the Bayes-optimal predictor based only on $(X,U)$ is
\[
f^*_{X,U}(u,x)
=
u_d^2
+
\rho\mathbf{1}_{d-1}^\top
\Sigma_X(u)^{-1}
\left\{
x-(u_1^2,\ldots,u_{d-1}^2)^\top
\right\}.
\]

The term $\gamma^2$ in the conditional variance is important for this
comparison. Even the Bayes-optimal predictor based on $(X,U)$ averages
over the unobserved group effect $B$ and therefore cannot account for
its realized value in the test group. In contrast, the initial
test group responses contain information about $B_{K+1}$ and can be
used by the within-group adjustment. To examine this, we repeat the simulation with
\(
N_j\equiv 21,
K=20,
\gamma=5,
\alpha=0.1,
\)
and vary $o$ as in the main experiments. To make this comparison, we rerun the same GHCP procedure, replacing only the random forest global predictor $\hat{\mu}_{\mathrm{RF}}$ with the Bayes-optimal predictor $f^\ast_{X, U}$.
All subsequent steps of the procedure, including the within-group adjustment, the construction of the nonconformity scores, and the GHCP weighting and calibration scheme, are kept unchanged. Table~\ref{tab:oracle-comparison-predictor} reports the resulting mean prediction set widths as $o$ varies.

\begin{table}
\centering
\caption{\footnotesize Mean GHCP prediction set widths using the random forest predictor
and the Bayes-optimal predictor based on $(X,U)$. Entries are Monte Carlo means over $B=1000$ independent replicates. The two columns come from separate experiments that differ only in the global predictor (random forest versus $f^*_{X,U}=E[Y\mid X,U]$). Parentheses report standard errors $s/\sqrt{B}$.}
\label{tab:oracle-comparison-predictor}
\begin{tabular}{rcc}
\hline
$o$ & Random forest predictor & Bayes-optimal predictor \\
\hline
0  & 34.15\,(0.27) & 22.12\,(0.16) \\
5  & 27.52\,(0.21) & 17.81\,(0.13) \\
10 & 21.22\,(0.16) & 13.81\,(0.10) \\
15 & 17.54\,(0.13) & 11.44\,(0.08) \\
20 & 13.57\,(0.09) &  9.07\,(0.05) \\
\hline
\end{tabular}
\end{table}

The prediction sets become substantially narrower as more test group observations are incorporated under both choices of the global predictor. As $o$ increases from zero to $20$, the mean width decreases from $34.15$ to $13.57$ with the random forest, corresponding to a reduction of approximately $60\%$, and from $22.12$ to $9.07$ with the Bayes-optimal predictor, corresponding to a reduction of approximately
$59\%$. The improvement from incorporating test-group observations remains essentially unchanged when the fitted random forest is replaced
by the Bayes-optimal predictor based on $(X,U)$. The reduction in prediction set width is not merely due to the choice of the global predictor and reflects additional predictive information contained in the initial test group observations, in this simulation through the realized latent group effect $B$.

\subsection{Added baselines for the simulations in Section~\ref{subsec:main-simulations}}
\label{appsubsec:added-baselines}

\begin{figure}
    \centering
    \includegraphics[width=0.8\linewidth]{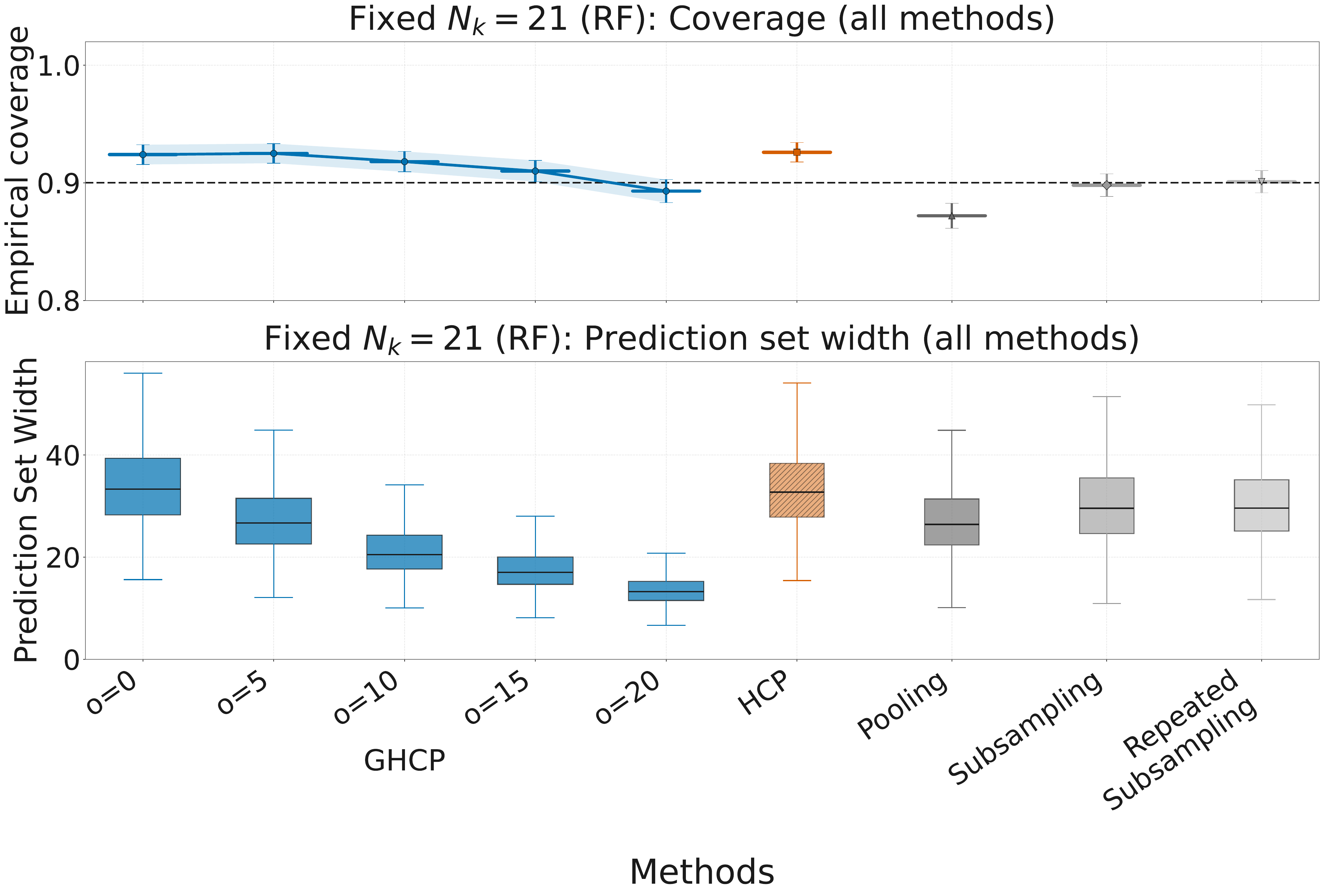}
    \caption{\footnotesize Empirical coverage and width for GHCP with
    $o\in\{0,5,10,15,20\}$ and the baseline procedures, under $K=20$,
    $N_j\equiv 21$, $\gamma=5$, and $\alpha=0.1$.}
    \label{fig:width-append-dhcp-fixed}
\end{figure}

Here, we compare GHCP with additional baseline methods under the simulation
settings of Section~\ref{subsec:main-simulations}. The data are generated
exactly as in Section~\ref{subsec:main-simulations} with $K=20$, $\gamma=5$, and $\alpha=0.1$.
We consider the fixed-size design of Figure~\ref{fig:N21-cov-width-20}
($N_j\equiv 21$) and the Poisson group-size design of
Figure~\ref{fig:poi-cov-width-20}
($N_j$ drawn from $\mathrm{Poi}(25)$ with any zero draw redrawn, so that group sizes are positive). In both cases, we compare
\begin{enumerate}[label=(\roman*)]
    \item GHCP, using Algorithm~\ref{alg:dhcp} with the restricted donor rule
    $\eta=0.5$;
    \item HCP~\citep{lee2023distribution};
    \item Pooling CDFs, Subsampling Once, and Repeated Subsampling
    \citep{dunn2022distribution}.
\end{enumerate}
None of the baseline methods make use of the test-group data.

\paragraph{Fixed group sizes.}
Figure~\ref{fig:width-append-dhcp-fixed} shows the empirical coverage and width for the fixed-size setting, and Table~\ref{tab:dhcp-hcp-alpha01-extra} shows the corresponding numerical summary. The subsampling based procedures from \citet{dunn2022distribution} are less conservative than HCP but do not use the observed test group data, so their widths remain essentially unchanged across $o$, while the pooling method undercovers.

\begin{table}
\centering
\caption{\footnotesize Empirical coverage and mean interval width for GHCP across initial test group sample sizes $o$, together with the baseline procedures, under the fixed-size simulation setting of Section \ref{subsec:main-simulations}. Values are averaged over $B=1000$ repetitions, with standard errors reported in parentheses.}
\label{tab:dhcp-hcp-alpha01-extra}
\small
\setlength{\tabcolsep}{6pt}
\begin{tabular}{lcc}
\toprule
Method & Coverage & Width \\
\midrule
GHCP, $o=0$  & 0.924 (0.008) & 34.15 (0.267) \\
GHCP, $o=5$  & 0.925 (0.008) & 27.52 (0.214) \\
GHCP, $o=10$ & 0.918 (0.009) & 21.22 (0.161) \\
GHCP, $o=15$ & 0.910 (0.009) & 17.54 (0.131) \\
GHCP, $o=20$ & 0.893 (0.010) & 13.57 (0.094) \\
HCP & 0.926 (0.008) & 33.54 (0.260) \\
Pooling CDF & 0.872 (0.011) & 27.25 (0.226) \\
Subsampling Once & 0.898 (0.010) & 30.45 (0.268) \\
Repeated Subsampling & 0.901 (0.009) & 30.40 (0.251) \\
\bottomrule
\end{tabular}
\end{table}

\begin{figure}
    \centering
    \includegraphics[width=0.8\linewidth]{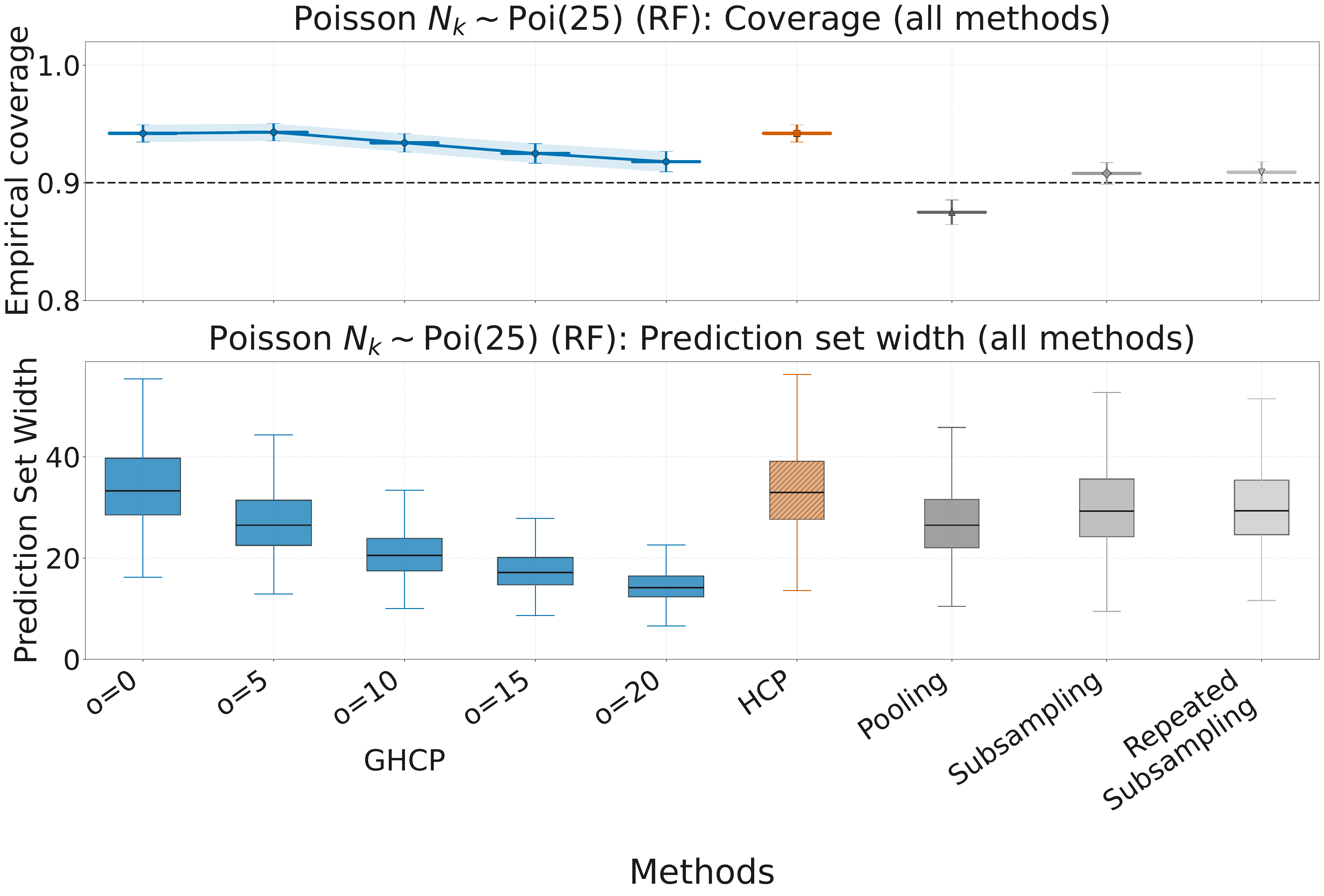}
    \caption{\footnotesize Empirical coverage and width for GHCP with
    $o\in\{0,5,10,15,20\}$ and the baseline procedures, under $K=20$,
    $N_j$ drawn independently from $\mathrm{Poi}(25)$ with zero draws redrawn, $\gamma=5$, and
    $\alpha=0.1$.}
    \label{fig:width-append-dhcp-poisson}
\end{figure}

\paragraph{Poisson group sizes.}
We repeat the comparison under the Poisson group-size design of
Figure~\ref{fig:poi-cov-width-20} in Section \ref{subsec:main-simulations}. Figure~\ref{fig:width-append-dhcp-poisson} and Table~\ref{tab:dhcp-hcp-poisson-alpha01-extra} report the corresponding coverage and width summaries. The results are consistent with the fixed group-size setting.

\begin{table}
\centering
\caption{\footnotesize Empirical coverage and mean interval width for GHCP across different values of the initial test group sample size $o$, together with the comparison methods, under the Poisson group-size setting described in Section~\ref{subsec:main-simulations}. Values are averaged over $B=1000$ trials, with standard errors reported in parentheses.}
\label{tab:dhcp-hcp-poisson-alpha01-extra}
\small
\setlength{\tabcolsep}{6pt}
\begin{tabular}{lcc}
\toprule
Method & Coverage & Width \\
\midrule
GHCP, $o=0$  & 0.94 (0.007) & 34.29 (0.256) \\
GHCP, $o=5$  & 0.94 (0.007) & 27.23 (0.204) \\
GHCP, $o=10$ & 0.93 (0.008) & 21.02 (0.155) \\
GHCP, $o=15$ & 0.93 (0.008) & 17.60 (0.128) \\
GHCP, $o=20$ & 0.92 (0.009) & 14.52 (0.101) \\
HCP & 0.94 (0.007) & 33.60 (0.259) \\
Pooling CDF & 0.88 (0.010) & 27.05 (0.226) \\
Subsampling Once & 0.91 (0.009) & 30.17 (0.266) \\
Repeated Subsampling & 0.91 (0.009) & 30.18 (0.250) \\
\bottomrule
\end{tabular}
\end{table}
\FloatBarrier

\subsection{Sensitivity to departures from group-size ignorability}
\label{appsubsec:size-ignorability-sensitivity}

Assumption~\ref{assump:independence} requires the reference-group sizes to be independent of the group-level data-generating laws. We examine the sensitivity of GHCP when this condition is violated. To isolate the
role of group-size ignorability, we modify only the dependence between
the group size and the latent group effect in the Poisson-size simulation
of Section~\ref{subsec:main-simulations}; all remaining components of
the data-generating mechanism and the implementation of GHCP are kept
unchanged.

\paragraph{Data generation.}
For each $j\in[K+1]$, generate
$
M_j\overset{\mathrm{i.i.d.}}{\sim}\operatorname{Poi}(25),
\varepsilon_j\overset{\mathrm{i.i.d.}}{\sim}N(0,1),
$
independently of $U_j$, and define
\begin{equation*}
%\label{eq:size-informative-B}
B_j
=
\gamma
\left\{
\sqrt{1-\xi^2}\,\varepsilon_j
+
\xi\,\frac{M_j-\EE[M_j]}{\sigma(M_j)}
\right\},
\qquad
\xi\in(-1,1).
\end{equation*}
For the reference groups we set $N_j=M_j$, $j\in[K]$.
The auxiliary variable $M_{K+1}$ for the test group is latent and is used only to generate $B_{K+1}$. Conditional on $(U_j,B_j)$, the individual observations are generated
exactly as in Section~\ref{subsec:main-simulations},
$
Z_{j,i}\mid U_j,B_j
\overset{\mathrm{i.i.d.}}{\sim}
\mathcal N_d\left(
\mu(U_j)+B_je_d,\,
\Sigma(U_j)
\right).
$
The above construction preserves
$
\EE(B_j)=0,
\operatorname{Var}(B_j)=\gamma^2,
$
while, for each reference group,
$
\operatorname{Corr}(N_j,B_j)=\xi.
$
Thus, $\xi=0$ recovers the original Poisson-size simulation, whereas $\xi\neq 0$ makes the group size informative about the group-specific law through the latent response shift $B_j$. In particular, the group sizes remain exchangeable, while Assumption~\ref{assump:independence} no longer holds.

\begin{figure}[ht]
    \centering
    \includegraphics[width=0.6\linewidth]{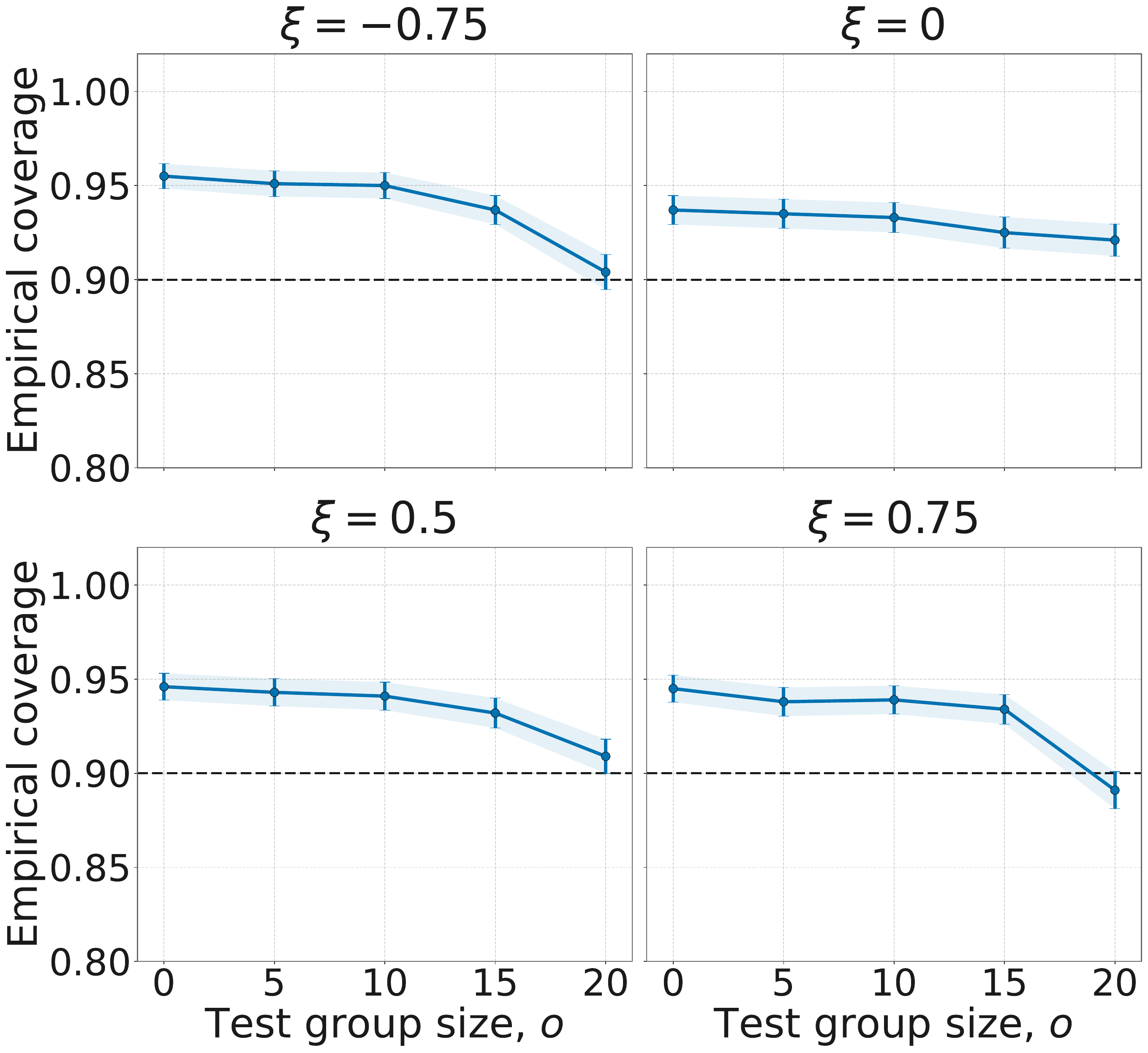}
    \caption{\footnotesize {Empirical coverage of GHCP under departures from
    group-size ignorability. The parameter $\xi$ controls the dependence
    between the reference-group size and the latent group effect, with
    $\operatorname{Corr}(N_j,B_j)=\xi$; thus $\xi=0$ corresponds to the
    size-ignorable setting and larger values of $\xi$ represent stronger
    departures from Assumption~\ref{assump:independence}. The dashed
    horizontal line denotes the nominal coverage level $1-\alpha=0.9$.
    Results are based on $B=1000$ repetitions.}}
    \label{fig:size-ignorability-coverage}
\end{figure}

\begin{figure}[ht]
    \centering
    \includegraphics[width=0.6\linewidth]{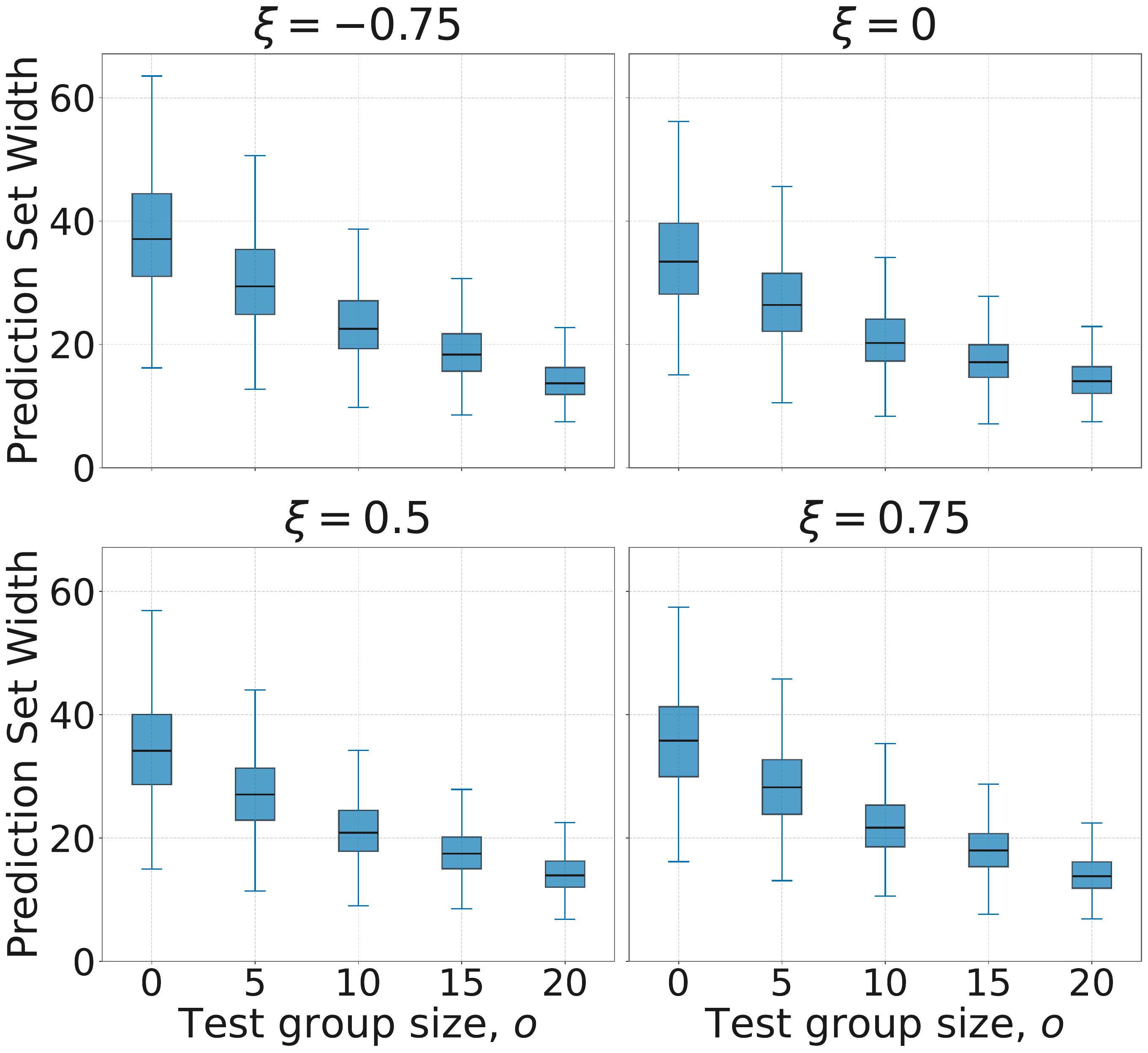}
    \caption{\footnotesize{Mean finite prediction-set width of GHCP under departures from group-size ignorability. The parameter $\xi$ satisfies $\operatorname{Corr}(N_j,B_j)=\xi$, with $\xi=0$ corresponding to the size-ignorable setting. Across all values of $\xi$, prediction sets become substantially shorter as the number $o$ of initially observed test-group samples increases. Results are based on
    $B=1000$ repetitions with $\alpha=0.1$.}}
    \label{fig:size-ignorability-width}
\end{figure}

Figures~\ref{fig:size-ignorability-coverage}, \ref{fig:size-ignorability-width} and
Table~\ref{tab:size-ignorability-sensitivity} show that the qualitative behavior of GHCP is stable under moderate departures from Assumption~\ref{assump:independence}. As $|\xi|$ increases, coverage remains close to or above the nominal $0.90$ level over most values of $o$. The departure is most visible when both the dependence is strongest
and the initial test-group sample is largest: at $\xi=0.75$ and $o=20$, the empirical coverage is $0.891$ with standard error $0.010$. The efficiency pattern is considerably less sensitive to $\xi$.
For every value of $\xi$, prediction-set width decreases substantially as more test-group observations become
available.

\begin{table}
\centering
\caption{\footnotesize Empirical coverage and mean interval width for GHCP
under size--intercept coupling $\xi$, Poisson DGP of Sec.\,3.1 except $B_j$,
$\gamma=5$, $\alpha=0.1$. Entries are means across $B=1000$ trials,
with standard errors in parentheses.}
\label{tab:size-ignorability-sensitivity}
\small
\setlength{\tabcolsep}{6pt}
\begin{tabular}{lccccc}
\toprule
& \multicolumn{5}{c}{Test-group sample size $o$} \\
\cmidrule(lr){2-6}
$\xi$ & 0 & 5 & 10 & 15 & 20 \\
\midrule
\multicolumn{6}{l}{\textit{Empirical coverage} ($1-\alpha=0.90$)} \\[2pt]
$\xi=-0.75$
& 0.955 (0.007) & 0.951 (0.007) & 0.950 (0.007) & 0.937 (0.008) & 0.904 (0.009) \\
$\xi=0$
& 0.937 (0.008) & 0.935 (0.008) & 0.933 (0.008) & 0.925 (0.008) & 0.921 (0.009) \\
$\xi=0.5$
& 0.946 (0.007) & 0.943 (0.007) & 0.941 (0.007) & 0.932 (0.008) & 0.909 (0.009) \\
$\xi=0.75$
& 0.945 (0.007) & 0.938 (0.008) & 0.939 (0.008) & 0.934 (0.008) & 0.891 (0.010) \\
\addlinespace[4pt]
\multicolumn{6}{l}{\textit{Mean interval width}} \\[2pt]
$\xi=-0.75$
& 38.027 (0.292) & 30.247 (0.236) & 23.329 (0.179) & 18.863 (0.146) & 14.184 (0.106) \\
$\xi=0$
& 34.274 (0.265) & 27.255 (0.218) & 21.037 (0.165) & 17.607 (0.136) & 14.478 (0.107) \\
$\xi=0.5$
& 34.785 (0.263) & 27.727 (0.211) & 21.498 (0.162) & 17.877 (0.132) & 14.374 (0.106) \\
$\xi=0.75$
& 36.149 (0.260) & 28.778 (0.210) & 22.275 (0.162) & 18.276 (0.131) & 14.169 (0.105) \\
\bottomrule
\end{tabular}
\end{table}

\subsection{Sensitivity to the merger weights}
\label{appsubsec:weight-sensitivity}

We next examine the sensitivity of GHCP to the weight assigned to the within-group predictor in \eqref{eq:weighted-merge} in Section \ref{subsec:G-HCP}. Recall that the merged predictor takes the form
$
\widetilde\mu_j
=
(1-\lambda_{\mathrm{local}})\widehat\mu^{\mathrm{global}}
+
\lambda_{\mathrm{local}}\overline Y_j,
$
where $\overline Y_j$ is computed from the first
$\tau=\lfloor o/2\rfloor$ observations in group $j$. We vary
$\lambda_{\mathrm{local}}$ while keeping all other components of GHCP fixed. In each repetition, all values of $\lambda_{\mathrm{local}}$ are
evaluated using the same generated data and donor selection. We use
$K=20$, $\alpha=0.1$, $\eta=0.5$, and $B=1000$ repetitions, with
$N_j\overset{\mathrm{i.i.d.}}{\sim}\operatorname{Poi}(25)$ and the same random-forest specification as in Section \ref{subsec:main-simulations}.

\begin{figure}[t]
    \centering
    \includegraphics[width=0.75\linewidth]{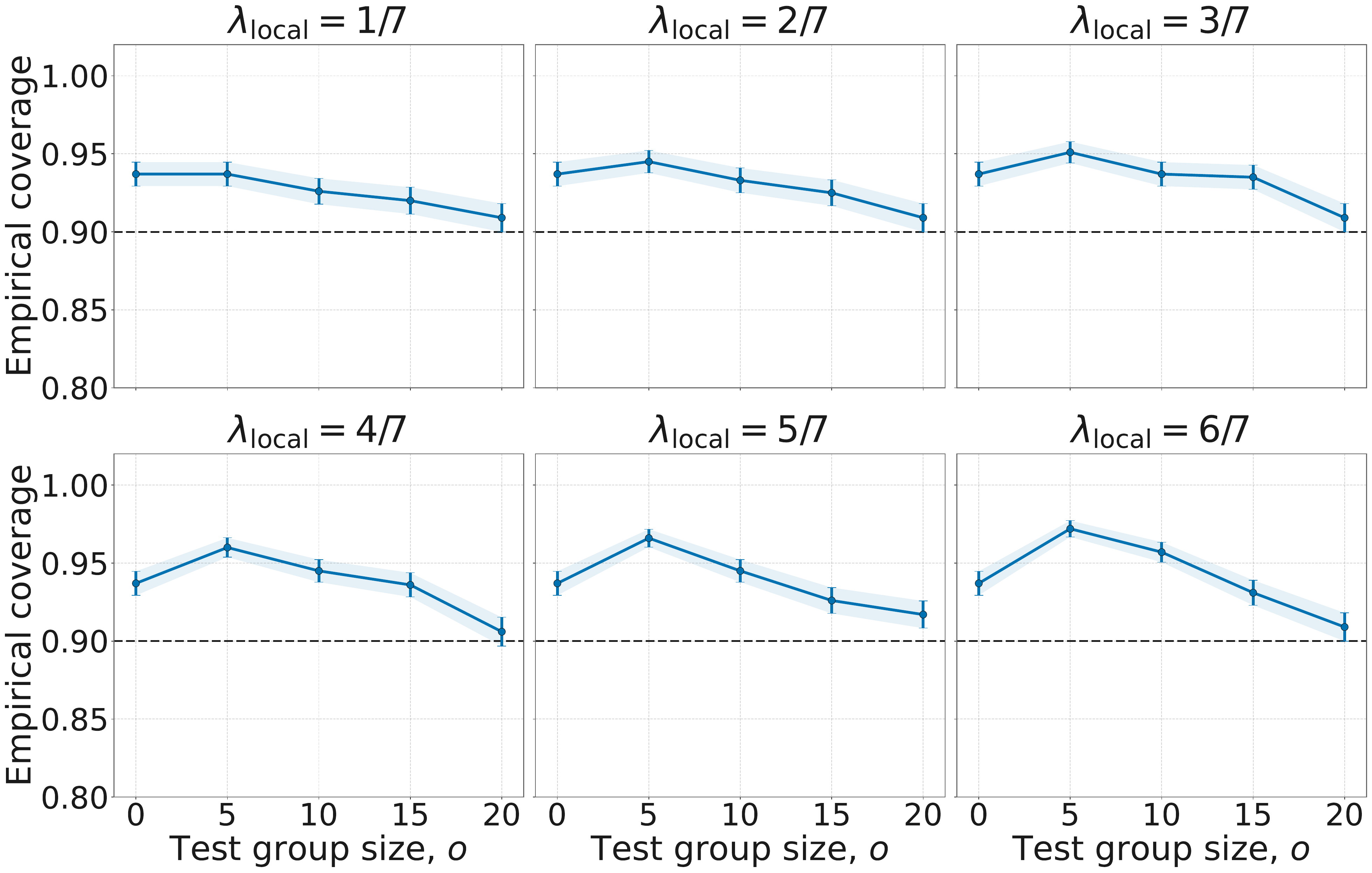}
    \caption{\footnotesize{Empirical coverage of GHCP as a function of the initial test-group sample size $o$ under the Poisson-size simulation with $\gamma=5$, for different values of the local merger weight $\lambda_{\mathrm{local}}$. The dashed horizontal line denotes the nominal coverage level $1-\alpha=0.9$. Results are based on $B=1000$ repetitions. At $o=0$, the local training
    sample is empty and all choices of $\lambda_{\mathrm{local}}$
    coincide.}}
    \label{fig:weight-sensitivity-original-coverage}
\end{figure}

\begin{figure}[t]
    \centering
    \includegraphics[width=0.75\linewidth]{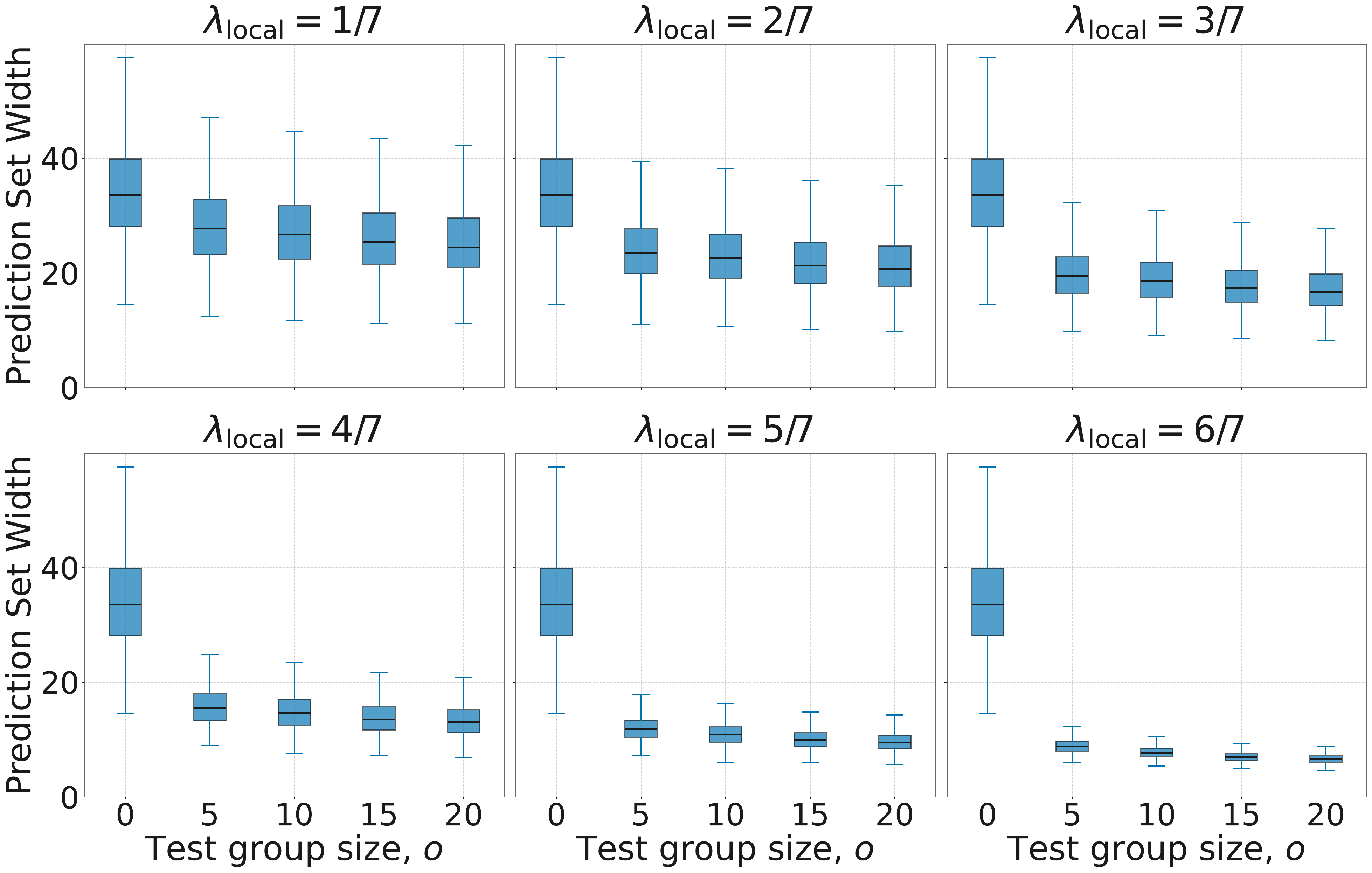}
    \caption{\footnotesize{Prediction-set widths of GHCP as a function of the initial test-group sample size $o$ under the Poisson-size simulation with $\gamma=5$, for different values of the local merger weight $\lambda_{\mathrm{local}}$. Larger local weights yield substantially shorter prediction sets in this setting.
    Results are based on $B=1000$ repetitions.}}
    \label{fig:weight-sensitivity-original-width}
\end{figure}

\paragraph{Original simulation setting.}
We first use the Poisson-size DGP of
Section~\ref{subsec:main-simulations}, with
$U_j\sim\operatorname{Unif}([1,5]^d)$ and $\gamma=5$. We consider
$
\lambda_{\mathrm{local}}
\in
\left\{1/7,2/7,\ldots,6/7\right\},
o\in\{0,5,10,15,20\}.
$
The results are shown in Figures~\ref{fig:weight-sensitivity-original-coverage}, \ref{fig:weight-sensitivity-original-width} and
Table~\ref{tab:weight-sensitivity-original}. Coverage remains close to the nominal level throughout. For every $o>0$, larger values of
$\lambda_{\mathrm{local}}$ lead to substantially shorter prediction
sets, while at $o=0$ all choices coincide because no within-group
training observations are available.

This behavior is consistent with the structure of the original DGP. The response contains both the latent group-specific shift $B_j$, with $\operatorname{Var}(B_j)=25$, and the group-level mean component $U_{jd}^2$.
Consequently, observations from the target group can provide substantial information about its response location, making the local mean especially useful in this setting.

\begin{table}[t]
\centering
\caption{\footnotesize {Sensitivity of GHCP to the merger weight
$\lambda_{\mathrm{local}}$ under the original Poisson-size simulation
with $\gamma=5$. Results are based on $B=1000$ repetitions with
$\alpha=0.1$ and $\eta=0.5$. Entries are Monte Carlo means with
standard errors in parentheses. At $o=0$, no within-group observations
are used for training, so all values of $\lambda_{\mathrm{local}}$
coincide.}}
\label{tab:weight-sensitivity-original}
\small
\setlength{\tabcolsep}{6pt}
\begin{tabular}{lccccc}
\toprule
& \multicolumn{5}{c}{Initial test-group sample size $o$} \\
\cmidrule(lr){2-6}
$\lambda_{\mathrm{local}}$
& $0$ & $5$ & $10$ & $15$ & $20$ \\
\midrule
\multicolumn{6}{l}{\textit{Empirical coverage ($1-\alpha=0.9$)}} \\[2pt]
$1/7$
& 0.937 (0.008)
& 0.937 (0.008)
& 0.926 (0.008)
& 0.920 (0.009)
& 0.909 (0.009) \\
$2/7$
& 0.937 (0.008)
& 0.945 (0.007)
& 0.933 (0.008)
& 0.925 (0.008)
& 0.909 (0.009) \\
$3/7$
& 0.937 (0.008)
& 0.951 (0.007)
& 0.937 (0.008)
& 0.935 (0.008)
& 0.909 (0.009) \\
$4/7$
& 0.937 (0.008)
& 0.960 (0.006)
& 0.945 (0.007)
& 0.936 (0.008)
& 0.906 (0.009) \\
$5/7$
& 0.937 (0.008)
& 0.966 (0.006)
& 0.945 (0.007)
& 0.926 (0.008)
& 0.917 (0.009) \\
$6/7$
& 0.937 (0.008)
& 0.972 (0.005)
& 0.957 (0.006)
& 0.931 (0.008)
& 0.909 (0.009) \\
\addlinespace[5pt]
\multicolumn{6}{l}{\textit{Mean prediction-set width}} \\[2pt]
$1/7$
& 34.58 (0.28)
& 28.57 (0.23)
& 27.62 (0.22)
& 26.40 (0.22)
& 25.39 (0.20) \\
$2/7$
& 34.58 (0.28)
& 24.27 (0.19)
& 23.34 (0.18)
& 22.15 (0.18)
& 21.28 (0.16) \\
$3/7$
& 34.58 (0.28)
& 20.04 (0.15)
& 19.11 (0.15)
& 17.99 (0.14)
& 17.23 (0.13) \\
$4/7$
& 34.58 (0.28)
& 15.93 (0.11)
& 14.96 (0.11)
& 13.97 (0.10)
& 13.34 (0.09) \\
$5/7$
& 34.58 (0.28)
& 12.08 (0.07)
& 11.07 (0.07)
& 10.17 (0.06)
& 9.69 (0.06) \\
$6/7$
& 34.58 (0.28)
& 8.96 (0.04)
& 7.82 (0.03)
& 7.04 (0.03)
& 6.65 (0.03) \\
\bottomrule
\end{tabular}
\end{table}

\paragraph{A setting favoring the global predictor.}
The preceding experiment does not imply that placing more weight on the local predictor is uniformly beneficial. To illustrate this, we consider a second DGP in which the advantage of the local mean is deliberately
reduced. We retain the same Gaussian model as in Section \ref{subsec:main-simulations}, but set
$
\gamma\equiv0,
$
and generate
$
U_{j1},\ldots,U_{j,d-1}
\overset{\mathrm{i.i.d.}}{\sim}\operatorname{Unif}(1,5),
U_{jd}\sim\operatorname{Unif}(1,2).
$
Setting $\gamma=0$ removes the latent group-specific intercept, while restricting the range of $U_{jd}$ substantially reduces the between-group variation in the response mean $U_{jd}^2$. Thus, in this setting there is less group-specific location variation for the local mean to recover.

\begin{figure}[ht]
    \centering
    \includegraphics[width=0.65\linewidth]{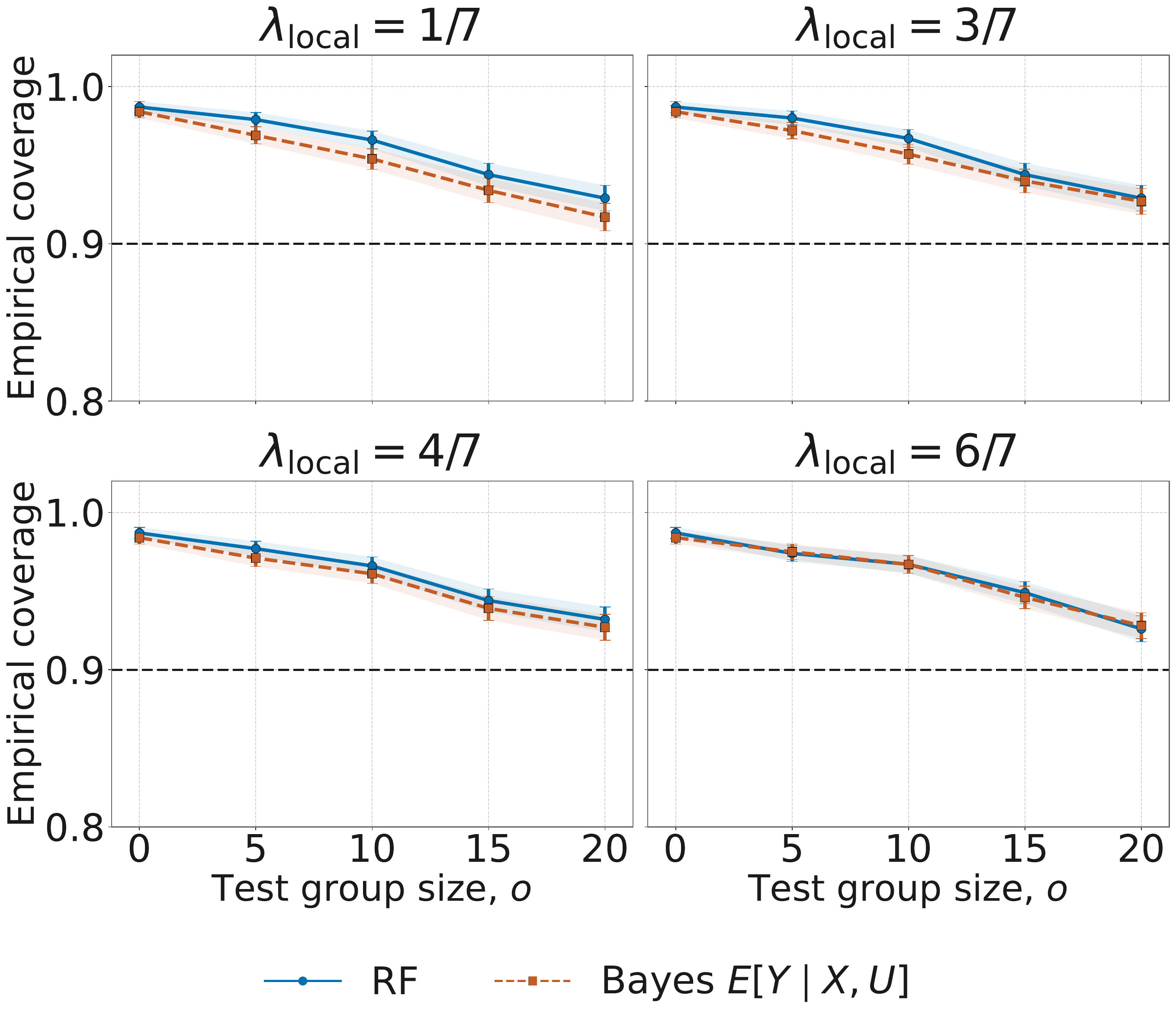}
    \caption{\footnotesize{Empirical coverage of GHCP in the modified simulation with $U_{jd}\sim\operatorname{Unif}(1,2)$ and $\gamma=0$, for different values of the local merger weight $\lambda_{\mathrm{local}}$.
    Results are shown for the random-forest and Bayes global predictors. The dashed horizontal line denotes the nominal coverage level $1-\alpha=0.9$. Results are based on $B=1000$ repetitions.}}
    \label{fig:weight-sensitivity-narrow-coverage}
\end{figure}

\begin{figure}[ht]
    \centering
    \includegraphics[width=0.65\linewidth]{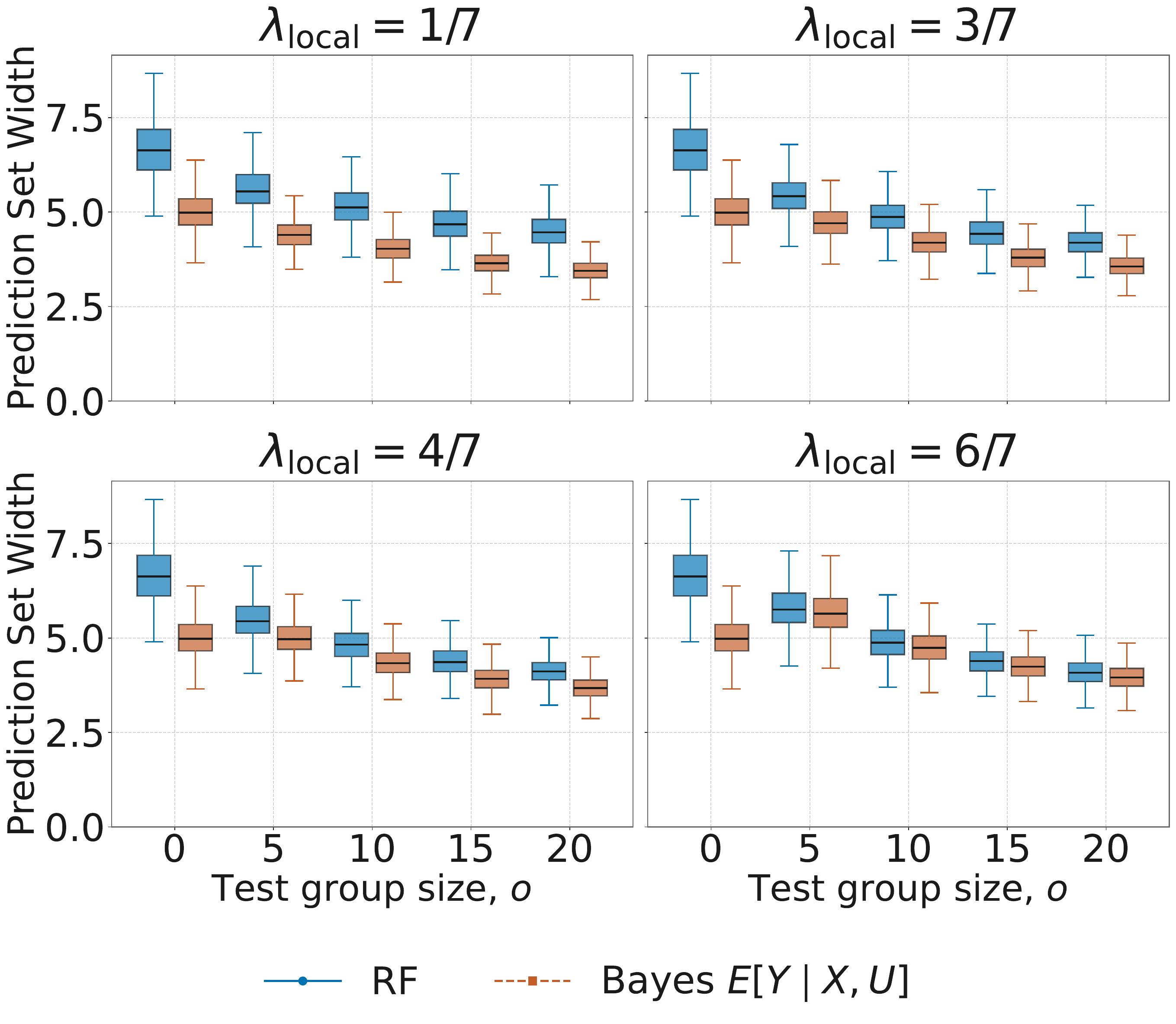}
    \caption{\footnotesize{Prediction-set widths of GHCP in the modified simulation with $U_{jd}\sim\operatorname{Unif}(1,2)$ and $\gamma=0$, for
    different values of the local merger weight
    $\lambda_{\mathrm{local}}$. Results are shown for the random-forest and Bayes global predictors. For the Bayes predictor, placing increasing weight on the local mean leads to wider prediction sets, illustrating that a larger local weight need not improve efficiency
    when the global predictor is informative.}}
    \label{fig:weight-sensitivity-narrow-width}
\end{figure}

To isolate the effect of the merger weight from variability in the fitted global predictor, we also consider the Bayes predictor $\EE(Y|X,U)$, as in Appendix \ref{appsubsec:simulation-bayes-error}, representing the best attainable global predictor under this DGP. We consider
$
\lambda_{\mathrm{local}}
\in
\left\{1/7,3/7,4/7,6/7\right\}.
$
Results are shown in Figure~\ref{fig:weight-sensitivity-narrow-coverage}, Figure~\ref{fig:weight-sensitivity-narrow-width} and
Table~\ref{tab:weight-sensitivity-narrow}. For the Bayes predictor, prediction-set width increases with $\lambda_{\mathrm{local}}$ for every $o>0$. Thus, this experiment shows that excessive weighting of the local mean is not bound to improve efficiency when $o$ is small. The corresponding pattern for the random-forest predictor is weaker, since it trades off the estimation error in the local mean against estimation error in the fitted global predictor.

\begin{table}[t]
\centering
\caption{\footnotesize{Sensitivity of GHCP to the merger weight
$\lambda_{\mathrm{local}}$ in the modified simulation with
$U_{jd}\sim\operatorname{Unif}(1,2)$ and $\gamma=0$. We show both the random-forest global predictor and the Bayes predictor $\EE(Y\mid X,U)$. Results are based on $B=1000$ repetitions with $\alpha=0.1$ and $\eta=0.5$. Entries are Monte Carlo means with standard errors in parentheses. At $o=0$, no within-group observations are used for training, so all values of $\lambda_{\mathrm{local}}$
coincide within each global-predictor specification.}}
\label{tab:weight-sensitivity-narrow}
\small
\setlength{\tabcolsep}{5pt}
\begin{tabular}{llccccc}
\toprule
& & \multicolumn{5}{c}{Initial test-group sample size $o$} \\
\cmidrule(lr){3-7}
Global predictor & $\lambda_{\mathrm{local}}$
& $0$ & $5$ & $10$ & $15$ & $20$ \\
\midrule
\multicolumn{7}{l}{\textit{Empirical coverage ($1-\alpha=0.9$)}} \\[2pt]
RF
& $1/7$ & 0.987 (0.004) & 0.979 (0.005) & 0.966 (0.006) & 0.944 (0.007) & 0.929 (0.008) \\
& $3/7$ & 0.987 (0.004) & 0.980 (0.004) & 0.967 (0.006) & 0.944 (0.007) & 0.929 (0.008) \\
& $4/7$ & 0.987 (0.004) & 0.977 (0.005) & 0.966 (0.006) & 0.944 (0.007) & 0.932 (0.008) \\
& $6/7$ & 0.987 (0.004) & 0.974 (0.005) & 0.967 (0.006) & 0.949 (0.007) & 0.926 (0.008) \\
\addlinespace[2pt]
Bayes
& $1/7$ & 0.984 (0.004) & 0.969 (0.005) & 0.954 (0.007) & 0.934 (0.008) & 0.917 (0.009) \\
& $3/7$ & 0.984 (0.004) & 0.972 (0.005) & 0.957 (0.006) & 0.940 (0.008) & 0.927 (0.008) \\
& $4/7$ & 0.984 (0.004) & 0.971 (0.005) & 0.961 (0.006) & 0.939 (0.008) & 0.927 (0.008) \\
& $6/7$ & 0.984 (0.004) & 0.975 (0.005) & 0.967 (0.006) & 0.946 (0.007) & 0.928 (0.008) \\
\addlinespace[5pt]
\multicolumn{7}{l}{\textit{Mean prediction-set width}} \\[2pt]
RF
& $1/7$ & 6.68 (0.03) & 5.64 (0.02) & 5.17 (0.02) & 4.72 (0.02) & 4.51 (0.01) \\
& $3/7$ & 6.68 (0.03) & 5.45 (0.02) & 4.89 (0.01) & 4.46 (0.01) & 4.22 (0.01) \\
& $4/7$ & 6.68 (0.03) & 5.49 (0.02) & 4.84 (0.01) & 4.39 (0.01) & 4.14 (0.01) \\
& $6/7$ & 6.68 (0.03) & 5.81 (0.02) & 4.90 (0.01) & 4.39 (0.01) & 4.10 (0.01) \\
\addlinespace[2pt]
Bayes
& $1/7$ & 5.03 (0.02) & 4.41 (0.01) & 4.04 (0.01) & 3.66 (0.01) & 3.46 (0.01) \\
& $3/7$ & 5.03 (0.02) & 4.74 (0.01) & 4.21 (0.01) & 3.80 (0.01) & 3.58 (0.01) \\
& $4/7$ & 5.03 (0.02) & 5.00 (0.01) & 4.35 (0.01) & 3.92 (0.01) & 3.69 (0.01) \\
& $6/7$ & 5.03 (0.02) & 5.68 (0.02) & 4.76 (0.01) & 4.25 (0.01) & 3.97 (0.01) \\
\bottomrule
\end{tabular}
\end{table}

\paragraph{Results}
Across both experiments and all weighting choices considered, incorporating sufficient observations from the test group consistently reduces prediction-set width. The magnitude of this gain, however, depends on how the global and local predictors are weighted, and hence on the relative information they provide under the underlying data-generating mechanism.

\FloatBarrier
\subsection{Additional results for ACS PUMS income data}
\label{appsubsec:acs-additional}

\begin{figure}[h]
    \centering
    \includegraphics[width=0.8\linewidth]{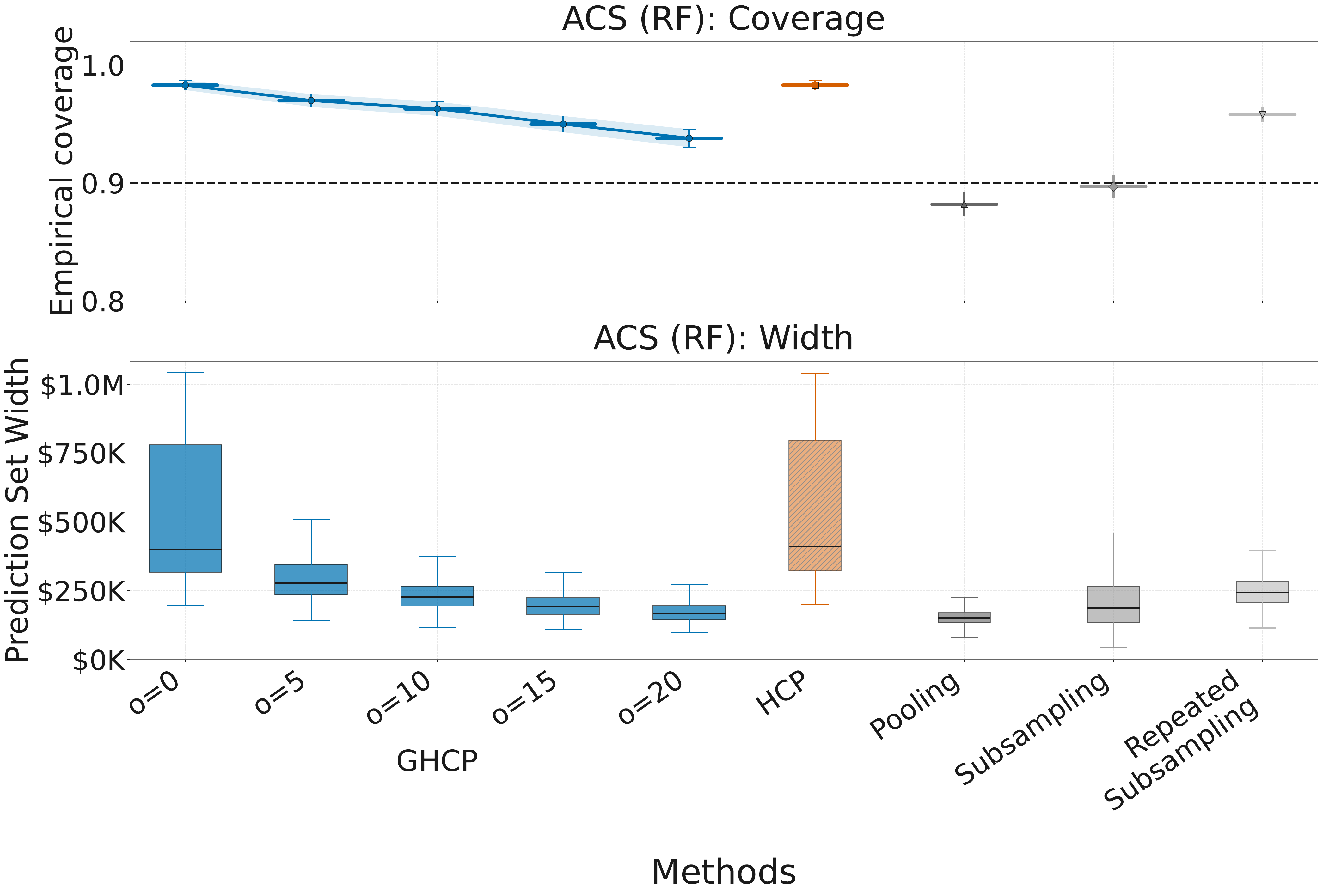}
    \caption{\footnotesize Empirical coverage and width on the ACS PUMS data for the full
    collection of methods.}
    \label{fig:app-acs-coverage}
\end{figure}

We reproduce the experiment of Section~\ref{subsec:acs-real} with
the full collection of competing methods as in Section \ref{appsubsec:added-baselines}. The data preprocessing, construction of reference and test PUMAs, individual-level covariates, and evaluation protocol are exactly as in Section~\ref{subsec:acs-real}. The results in Figure~\ref{fig:app-acs-coverage} and Table \ref{tab:acs-alpha01-extra} highlight the advantage of GHCP, which leverages the initial test group samples, over the other baselines.

\begin{table}
\centering
\caption{\footnotesize Empirical coverage and mean interval width for GHCP across initial test group sample sizes $o$, together with the baseline procedures, for the ACS experiment at $\alpha=0.1$. Values are averaged over $B=1000$ repetitions, with standard errors reported in parentheses. Widths are on the income scale.}
\label{tab:acs-alpha01-extra}
\small
\setlength{\tabcolsep}{6pt}
\begin{tabular}{lcc}
\toprule
Method & Coverage & Width \\
\midrule
GHCP, $o=0$  & 0.983 (0.004) & \$523{,}195 (7{,}685) \\
GHCP, $o=5$  & 0.970 (0.005) & \$315{,}773 (4{,}285) \\
GHCP, $o=10$ & 0.963 (0.006) & \$247{,}714 (3{,}027) \\
GHCP, $o=15$ & 0.950 (0.007) & \$200{,}557 (1{,}845) \\
GHCP, $o=20$ & 0.938 (0.008) & \$174{,}906 (1{,}595) \\
HCP & 0.983 (0.004) & \$531{,}277 (7{,}722) \\
Pooling CDF & 0.882 (0.010) & \$152{,}895 (852) \\
Subsampling Once & 0.897 (0.010) & \$248{,}823 (6{,}495) \\
Repeated Subsampling & 0.958 (0.006) & \$247{,}917 (1{,}820) \\
\bottomrule
\end{tabular}
\end{table}

\begin{figure}[h]
    \centering
    \includegraphics[width=0.9\linewidth]{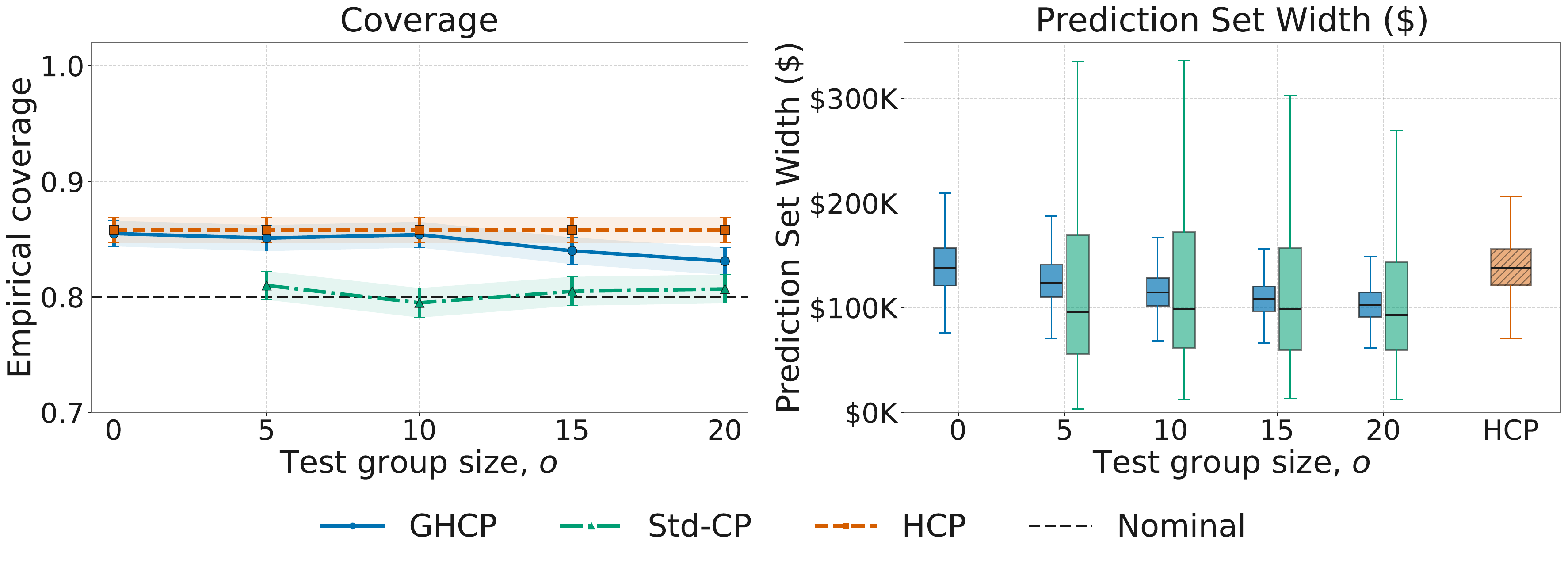}
    \caption{\footnotesize Empirical coverage and width on the ACS PUMS data for HCP, GHCP, and Std-CP ($\alpha=0.2$), with an RF global predictor for HCP/GHCP; GHCP additionally uses local within-group training. Std-CP is split conformal within the test group.}
    \label{fig:acs-main-coverage-2}
\end{figure}

The results in Figure~\ref{fig:acs-main-coverage-2} and Table~\ref{tab:acs-main-results-2} are for the same setup as above but for $\alpha=0.2$.
Std-CP uses split conformal prediction within the target PUMA with a studentized residual score function. At small $o$ this often yields infinite intervals. For $o\ge 10$, Std-CP produces non-trivial intervals but remains much more volatile than GHCP. At $o=0$, GHCP and HCP have similar mean widths. As $o$ increases, GHCP shrinks steadily, reaching a mean width of $103{,}543$ at $o=20$ (a $26.0\%$ reduction relative to HCP). At $o=20$, GHCP also beats Std-CP on mean width,
while maintaining nominal coverage.

\begin{table}
\centering
\caption{\footnotesize Empirical coverage and mean interval width for GHCP and randomized studentized Std-CP across initial test group sample sizes $o$, for the ACS experiment at $\alpha=0.2$. HCP does not use test group history, so its coverage and width are reported under $o=0$. Values are averaged over $B=1000$ repetitions, with standard errors in parentheses. Widths are on the income scale; $+\infty$ indicates at least one infinite replicate.}
\label{tab:acs-main-results-2}
\small
\setlength{\tabcolsep}{4pt}
\begin{tabular}{lccccc}
\toprule
 & $o=0$ & $o=5$ & $o=10$ & $o=15$ & $o=20$ \\
\midrule
\multicolumn{6}{l}{\emph{Coverage}} \\
GHCP   & 0.855 (0.011) & 0.851 (0.011) & 0.854 (0.011) & 0.840 (0.012) & 0.831 (0.012) \\
Std-CP & 1.000 (0.000) & 0.810 (0.012) & 0.795 (0.013) & 0.805 (0.013) & 0.807 (0.012) \\
HCP    & 0.858 (0.011) &  &  &  &  \\
\midrule
\multicolumn{6}{l}{\emph{Width}} \\
GHCP   & \$140{,}825 (828) & \$126{,}169 (697) & \$115{,}537 (596) & \$109{,}313 (565) & \$103{,}543 (531) \\
Std-CP & $+\infty$ & $+\infty$ & \$148{,}387 (5{,}123) & \$129{,}086 (3{,}770) & \$114{,}039 (2{,}615) \\
HCP    & \$139{,}890 (802) &  &  &  &  \\
\bottomrule
\end{tabular}
\end{table}

\FloatBarrier
\section{Proofs for Section~\ref{sec:donor-method}}
\label{appsec:proofs-main}

\subsection{Proof of Theorem~\ref{thm:donor-valid}}

We first introduce two lemmas that will be used in the proof of
Theorem~\ref{thm:donor-valid}. Lemma~\ref{lem:donation-restores} shows that the donation step transforms the retained reference groups and the test group into a hierarchically exchangeable set of groups. Lemma~\ref{lem:generic-hcp} then provides lower and upper coverage bounds when one of the groups is only partially observed. 
The proof of Theorem~\ref{thm:donor-valid} then follows by applying Lemma~\ref{lem:generic-hcp} to the collection obtained in Lemma~\ref{lem:donation-restores}.

First, recall from Section~\ref{subsec:restricted-selection} that a possibly randomized donor-selection rule
$S(N_{1:K},o)\subseteq[K]$ is permutation-equivariant if, for every deterministic size vector $n\in\mathbb N^K$ and every permutation $\pi$ of $[K]$,
\[
 S(\pi n,o)
\overset{d_R}{=}
\pi\{S(n,o)\},
\qquad
(\pi n)_j:=n_{\pi^{-1}(j)},
\]
where $\overset{d_R}{=}$ denotes equality in distribution with respect to the auxiliary randomization of the selection rule. We assume that this auxiliary randomness is independent of the data. The default donor pool $S(N_{1:K},o)=\{j\in[K]:N_j>o\}$ is a deterministic special case.

\begin{lemma}[Hierarchical exchangeability after donation]
\label{lem:donation-restores}
Suppose Assumptions~\ref{assump:group-exch}--\ref{assump:within-group-exch}
hold. Let $S\subseteq\{j\in[K]:N_j>o\}$ be selected by a permutation-equivariant rule whose auxiliary randomness is independent of the data. On $\{|S|>0\}$, suppose $J_0$ is drawn as $J_0\mid S\sim\Unif(S)$ and let $N_{J_0}$ be the (donated) test group size. Denote the multiset of reference group sizes and the training data as
\begin{equation*}
\mathcal N_S:=\{N_j:j\in S\},
\qquad
\mathcal D_{\mathrm{train}}:=\sigma(\widetilde W_j:j\in[K]\setminus S).
\end{equation*}
Then conditional on $(S,J_0,\mathcal N_S,\mathcal D_{\mathrm{train}})$, the retained reference groups indexed by $S\setminus\{J_0\}$ together with the test group with the donated length $N_{J_0}$ are hierarchically exchangeable. If $|S|=0$, the single test group of length $o+1$ is within-group exchangeable.
\end{lemma}

Fix $o\in\mathbb N_0$. Let $\widetilde W_1,\ldots,\widetilde W_M$ be hierarchically exchangeable
groups satisfying Assumptions~\ref{assump:HCP-group-exch} and
\ref{assump:HCP-within-group-exch}, where $\widetilde W_j=(W_{j,1},\ldots,W_{j,N_j}),
N_j>o$ almost surely.
Let $\tau:\mathbb N\times\mathbb N_0\to\mathbb N_0$ be the within-group training-size function defined in Appendix \ref{appsubsec:more-mergers}.
For $j\in[M]$, define
\(
m_j:=\tau(N_j,o),
L_j:=N_j-m_j~\text{and}~
r_j:=o-m_j.
\)
Any external training information used to construct the score rules is regarded as fixed; equivalently, all statements below may be read conditionally on that information. For each group, let $s_j$ be constructed using a procedure as in Appendix~\ref{appsubsec:more-mergers}. The held-out scores are
\(
s_{j,i}
:=
s_j(U_j,X_{j,m_j+i},Y_{j,m_j+i}),
i=1,\ldots,L_j.
\)
Define
\begin{equation}
\label{eq:generic-obs-measure}
\nu_{\mathrm{obs}}
:=
\sum_{j=1}^{M-1}\sum_{i=1}^{L_j}
\frac{1}{ML_j}\delta_{s_{j,i}}
+
\sum_{i=1}^{r_M}
\frac{1}{ML_M}\delta_{s_{M,i}}
+
\frac{N_M-o}{ML_M}\delta_{\infty},
\end{equation}
and let
$
\mathcal N:=\{N_1,\ldots,N_M\}
$
denote the unordered multiset of group sizes.

\begin{lemma}[Coverage for a partially observed group]
\label{lem:generic-hcp}
For every $\alpha\in(0,1)$,
\[\PP\left\{s_{M,r_M+1}\le\quant{1-\alpha}\nu_{\mathrm{obs}}\,\middle|\,\mathcal N\right\}\ge1-\alpha.\]
Consequently, $\PP\left\{
s_{M,r_M+1}
\le
\quant{1-\alpha}{\nu_{\mathrm{obs}}}
\right\}
\ge
1-\alpha.$
If, in addition, Assumption~\ref{assump:no-ties} holds, then
\[
\PP\left\{
s_{M,r_M+1}
\le
\quant{1-\alpha}{\nu_{\mathrm{obs}}}
\,\middle|\,
\mathcal N
\right\}
\le
1-\alpha
+
\frac1M
+
\frac1M
\max_{j\in[M]}
\frac{N_j-o}{N_j-\tau(N_j,o)}.
\]
In particular,
\(
\PP\left\{
s_{M,r_M+1}
\le
\quant{1-\alpha}{\nu_{\mathrm{obs}}}
\right\}
\le
1-\alpha+2/M.
\)
\end{lemma}
Now we prove Theorem \ref{thm:donor-valid}. Let
\(
\mathcal E
:=
\left\{
Y_{K+1,o+1}\in
\widehat C_{\mathrm{GHCP}}
(U_{K+1},X_{K+1,o+1})
\right\}.
\)

\medskip
\noindent\textbf{Case $|S|=0$.}
When no donor is available, the procedure uses only the test group, with the convention $N_{K+1}=o+1$. The first $\lfloor o/2\rfloor$ test observations are used for within-group training, while the remaining $L_{K+1}:=o+1-\lfloor o/2\rfloor$
held-out test positions consist of $L_{K+1}-1$ observed calibration scores and the unobserved test score. By Assumption~\ref{assump:within-group-exch}, conditional on the data used to construct the score function, these $L_{K+1}$
held-out test scores are exchangeable. The argument described in Appendix~\ref{appsec:background} therefore leads to
\(
\PP\{\mathcal E\mid |S|=0\}
\ge
1-\alpha.
\)
Under the additional Assumption~\ref{assump:no-ties}, a similar argument leads to
\(
\PP\{\mathcal E\mid |S|=0\}
\le
1-\alpha+1/L_{K+1}.
\)

\medskip
\noindent\textbf{Case $|S|>0$.}
Fix a nonempty set $A\subseteq[K]$, let $m:=|A|$, and fix $d\in A$. Let
\(
\mathcal N_A=\{N_j:j\in A\},~
\mathcal F_A
=
\sigma(\widetilde W_j:j\in [K]\setminus A).
\)
By Lemma~\ref{lem:donation-restores}, conditional on
$(S=A,J_0=d,\mathcal N_A,\mathcal F_A)$, the retained
reference groups and the test group with donated size $N_d$ form a
hierarchically exchangeable collection of $m$ groups. Moreover, the GHCP
construction corresponds to the setup of Lemma~\ref{lem:generic-hcp} with $\tau(n,o)=\lfloor o/2\rfloor$. For the test
group, $L_m=N_d-\lfloor o/2\rfloor, r_m=o-\lfloor o/2\rfloor$.
Consequently, the probability measure $\nu_{\mathrm{obs}}$ in
\eqref{eq:generic-obs-measure} coincides with
$\nu_{\mathrm{donor}}$ in \eqref{eq:donor-measure}. Applying the lower
bound in Lemma~\ref{lem:generic-hcp}, we have
\(
\PP\left\{
\mathcal E
\,\middle|\,
S=A,J_0=d,\mathcal N_A,\mathcal F_A
\right\}
\ge
1-\alpha.
\)
Taking conditional expectations over $\mathcal N_A$ and $\mathcal F_A$, we have
\(
\PP\{\mathcal E\mid S=A,J_0=d\}\ge1-\alpha.
\)
Averaging over $J_0$, then over all nonempty $A\subseteq[K]$, and
combining with the case $|S|=0$, we have
\(
\PP\{\mathcal E\}\ge1-\alpha.
\)
For the upper bound of coverage, under
Assumption~\ref{assump:no-ties}, we have from Lemma~\ref{lem:generic-hcp}
\[
\PP\left\{
\mathcal E
\,\middle|\,
S=A, |S|>0, J_0=d,\mathcal N_A,\mathcal F_A
\right\}
\le
1-\alpha
+
\frac{1}{|A|}
+
\frac{1}{|A|}
\max_{j\in A}
\frac{N_j-o}{N_j-\lfloor o/2\rfloor}.
\]
The right-hand side depends on the selected sizes but not on
$\mathcal F_A$ or the donor index $d$. Therefore, taking conditional
expectations over $\mathcal F_A$, $\mathcal N_A$, and $J_0$ given
$S=A$ yields
\[
\PP\{\mathcal E\mid S=A, |S|>0\}
\le
1-\alpha
+
\EE\left[
\frac{1}{|A|}
\left\{
1+
\max_{j\in A}
\frac{N_j-o}{N_j-\lfloor o/2\rfloor}
\right\}
\,\middle|\,
S=A, |S|>0
\right].
\]
Since $\rho_o(n)=(n-o)/(n-\lfloor o/2\rfloor)$
is nondecreasing in $n>o$, on $\{|S|>0\}$ we have
\[
\PP\{\mathcal E\mid S, |S|>0\}
\le
1-\alpha
+
\EE\left[
\frac1{|S|}
\left\{
1+
\rho_o\left(\max_{j\in S}N_j\right)
\right\}
\,\middle|\,
S, |S|>0
\right].
\]
Finally, since $\{|S|>0\}\in\sigma(S)$,
\begin{align*}
\PP\{\mathcal E\mid |S|>0\}
&=
\EE\left[
\PP\{\mathcal E\mid S, |S|>0\}
\,\middle|\,
|S|>0
\right] \\
&\le
1-\alpha
+
\EE\left[
\frac1{|S|}
\left\{
1+
\rho_o\left(\max_{j\in S}N_j\right)
\right\}
\,\middle|\,
|S|>0
\right].
\end{align*}

\subsection{Proof of Lemma~\ref{lem:donation-restores}}

On $\{M=0\}$, the calibration collection consists only of the test group
with length $o+1$. Across-group exchangeability is then vacuous, while
within-group exchangeability follows from
Assumption~\ref{assump:within-group-exch}. We therefore consider
$\{M>0\}$. Fix a nonempty set
$
A=\{a_1,\ldots,a_m\}\subseteq[K],
a_1<\cdots<a_m,
$
and write
$
E_A:=\{S=A\}.
$
Fix also $d\in A$. We prove the required stronger statement conditional on $E_A\cap\{J_0=d\}$, the multiset of selected sizes, and the training data.

It is convenient to work on an enlarged probability space on which,
for every group $j$, there is an infinite sequence
$Z_{j,1},Z_{j,2},\ldots$ conditionally i.i.d.\ from $\mu_j$, and define
\[
G_j
:=
\bigl(U_j, \mu_j ,Z_{j,1},Z_{j,2},\ldots\bigr),
\qquad j\in[K+1].
\]
The observed rows are obtained by truncating these sequences at their
observed lengths, so this augmentation does not change the joint law of
the observed data. Assumptions~\ref{assump:group-exch} and
\ref{assump:within-group-exch} imply that
$
(G_1,\ldots,G_{K+1})
$
is exchangeable. Moreover,
Assumptions~\ref{assump:independence} and
\ref{assump:within-group-exch} imply
\begin{equation}
\label{eq:size-complete-group-independence}
N_{1:K}
\perp\!\!\!\perp
G_{1:K+1}.
\end{equation}

Let
$
\mathcal N_A
:=
\{N_j:j\in A\},
\mathcal F_A
:=
\sigma(\widetilde W_j:j\in[K]\setminus A),
$
and write
$
\mathcal C_A:=\sigma(\mathcal N_A,\mathcal F_A).
$
For a deterministic size vector $n=(n_1,\ldots,n_K)$, define
$
q_A(n)
:=
\PP\{E_A\mid N_{1:K}=n\},
$
where the probability is over any auxiliary randomness used by the selection rule. Since the donor is drawn uniformly from $S$ using fresh randomness after $S$ has been selected, for every $d\in A$,
\begin{equation}
\label{eq:selection-donor-kernel}
\PP\{E_A,J_0=d\mid N_{1:K},G_{1:K+1}\}
=
\frac{q_A(N_{1:K})}{m}.
\end{equation}

\paragraph{Selected sizes.}
Let
$
N_A:=(N_{a_1},\ldots,N_{a_m}).
$
We first show that $N_A$ is exchangeable conditional on
$(E_A,J_0=d,\mathcal C_A)$. Fix $\pi\in\mathcal S_m$. Let $\rho$ be the permutation of $[K]$
satisfying
$
\rho(a_k)=a_{\pi(k)},
 k=1,\ldots,m,
$
and fixing every index outside $A$. For a vector $n$, write
$
n^\rho:=(n_{\rho(1)},\ldots,n_{\rho(K)}).
$
Since $\rho(A)=A$, by the permutation equivariance of the selection rule, $q_A(n^\rho)=q_A(n)$.
Moreover, because $\rho$ fixes $A^c$, both $\mathcal N_A$ and
$\mathcal F_A$ are unchanged by this permutation. Finally, let $\varphi:\mathbb R^m\to\mathbb R$ be bounded and measurable, and let $H$ be any bounded $\mathcal C_A$-measurable random variable.
Using \eqref{eq:selection-donor-kernel},
exchangeability of $N_{1:K}$, and
equivariance of the selection rule, we obtain
\begin{align*}
\EE\!\left[
\varphi(N_{a_1},\ldots,N_{a_m})
H
\1\{E_A,J_0=d\}
\right]
&=
\frac1m
\EE\!\left[
\varphi(N_{a_1},\ldots,N_{a_m})
Hq_A(N_{1:K})
\right]
\\
&=
\frac1m
\EE\!\left[
\varphi(N_{a_{\pi(1)}},\ldots,N_{a_{\pi(m)}})
Hq_A(N_{1:K}^{\rho})
\right]
\\
&=
\frac1m
\EE\!\left[
\varphi(N_{a_{\pi(1)}},\ldots,N_{a_{\pi(m)}})
Hq_A(N_{1:K})
\right]
\\
&=
\EE\!\left[
\varphi(N_{a_{\pi(1)}},\ldots,N_{a_{\pi(m)}})
H
\1\{E_A,J_0=d\}
\right].
\end{align*}
Here, the second equality also uses
\eqref{eq:size-complete-group-independence}: the joint law of
$(N_{1:K},G_{1:K+1})$ is unchanged when only the coordinates of
$N_{1:K}$ are permuted. Since the equality holds for every bounded
$\mathcal C_A$-measurable $H$,
\[
\EE\!\left[
\varphi(N_A)
\,\middle|\,
E_A,J_0=d,\mathcal C_A
\right]
=
\EE\!\left[
\varphi(N_A^\pi)
\,\middle|\,
E_A,J_0=d,\mathcal C_A
\right],
\]
where
$
N_A^\pi
:=
(N_{a_{\pi(1)}},\ldots,N_{a_{\pi(m)}}).
$ Thus $N_A$ is exchangeable under this conditioning. Now write
$
A\setminus\{d\}
=
\{b_1,\ldots,b_{m-1}\},
b_1<\cdots<b_{m-1},
$
and define the post-donation length vector
$
L^*
:=
(L_1^*,\ldots,L_m^*)
:=
(N_{b_1},\ldots,N_{b_{m-1}},N_d).
$
For fixed $(A,d)$, this is merely a deterministic reordering of $N_A$.
Therefore,
\begin{equation*}
%\label{eq:post-donation-length-exch}
(L_1^*,\ldots,L_m^*)
\overset d=
(L_{\pi(1)}^*,\ldots,L_{\pi(m)}^*)
\quad
\text{conditionally on }
(E_A,J_0=d,\mathcal C_A)
\end{equation*}
for every $\pi\in\mathcal S_m$.

\paragraph{Complete group--size pairs.}
Define
$G^*:=(G_1^*,\ldots,G_m^*):=
(G_{b_1},\ldots,G_{b_{m-1}},G_{K+1})
$
and
$P^*:=((G_1^*,L_1^*),\ldots,(G_m^*,L_m^*)).
$
We show directly that $P^*$ is exchangeable conditional on
$(E_A,J_0=d,\mathcal C_A)$. Fix $\pi\in\mathcal S_m$. Let $\rho$ permute the ordered list
$
(b_1,\ldots,b_{m-1},d)
$
according to $\pi$, and fixes its complement. Likewise, let $\eta$ permute
$
(b_1,\ldots,b_{m-1},K+1)
$
according to the same $\pi$, while fixing all remaining indices. Under the simultaneous transformation
$
(N_{1:K},G_{1:K+1})
\longmapsto
(N_{1:K}^{\rho},G_{1:K+1}^{\eta}),
$
the vector $P^*$ is transformed into
$
P_\pi^*
:=
\bigl(
(G_{\pi(1)}^*,L_{\pi(1)}^*),\ldots,
(G_{\pi(m)}^*,L_{\pi(m)}^*)
\bigr).
$
By exchangeability of $N_{1:K}$ and $G_{1:K+1}$ together with
\eqref{eq:size-complete-group-independence},
\begin{equation}
\label{eq:joint-independent-permutation}
(N_{1:K},G_{1:K+1})
\overset d=
(N_{1:K}^{\rho},G_{1:K+1}^{\eta}).
\end{equation}
Notice that the two permutations need not be the same: this is precisely where the independence in
\eqref{eq:size-complete-group-independence} is used. Both permutations leave $\mathcal C_A$ unchanged, and since
$\rho(A)=A$,
$
q_A(N_{1:K}^{\rho})=q_A(N_{1:K}).
$ Hence, for every bounded measurable function $\psi$ of $P^*$ and every
bounded $\mathcal C_A$-measurable random variable $H$,
\begin{align*}
\EE\!\left[
\psi(P^*)H\1\{E_A,J_0=d\}
\right]
&=
\frac1m
\EE\!\left[
\psi(P^*)Hq_A(N_{1:K})
\right]
\\
&=
\frac1m
\EE\!\left[
\psi(P_\pi^*)Hq_A(N_{1:K}^{\rho})
\right]
\\
&=
\frac1m
\EE\!\left[
\psi(P_\pi^*)Hq_A(N_{1:K})
\right]
\\
&=
\EE\!\left[
\psi(P_\pi^*)H\1\{E_A,J_0=d\}
\right],
\end{align*}
where the second equality follows from
\eqref{eq:joint-independent-permutation}. Since this holds for every
such $H$,
\[
\EE\!\left[
\psi(P^*)
\,\middle|\,
E_A,J_0=d,\mathcal C_A
\right]
=
\EE\!\left[
\psi(P_\pi^*)
\,\middle|\,
E_A,J_0=d,\mathcal C_A
\right].
\]
Thus, for every $\pi\in\mathcal S_m$,
$
P^*
\overset d=
P_\pi^*~
\text{conditionally on }
(E_A,J_0=d,\mathcal C_A).
$

\paragraph{Observed rows.}
For a complete group object $
g=(u,\mu,z_1,z_2,\ldots)
$
and a positive integer $\ell$, define
$
\Gamma(g,\ell)
:=
(
(u,z_1),\ldots,(u,z_\ell)
).
$
Applying this map coordinatewise to the exchangeable vector in $P^*$ gives
$
\widetilde W_a^*
:=
\Gamma(G_a^*,L_a^*),~
a=1,\ldots,m.
$
These rows are
$
\widetilde W_a^*
=
(W_{b_a,1},\ldots,W_{b_a,N_{b_a}})$ for $a=1,\ldots,m-1,
$
and
$
\widetilde W_m^*
=
(W_{K+1,1},\ldots,W_{K+1,N_d}).
$
Therefore,
\[
(\widetilde W_1^*,\ldots,\widetilde W_m^*)
\overset d=
(\widetilde W_{\pi(1)}^*,\ldots,
 \widetilde W_{\pi(m)}^*)
\]
conditionally on
$(E_A,J_0=d,\mathcal C_A)$, proving across-group
exchangeability assumption \ref{assump:HCP-group-exch}. 
It remains to verify the within-group condition. Fix $a\in[m]$ and
$\ell$ such that
$
\PP\!\left(
L_a^*=\ell
\,\middle|\,
E_A,J_0=d,\mathcal C_A
\right)>0.
$
Conditional on the latent group laws and the size vector,
Assumption~\ref{assump:within-group-exch} implies that the observations within each group are i.i.d. Moreover, the event
$\{E_A,J_0=d\}$ and $\mathcal N_A$ depend only on the size vector, while $\mathcal F_A$ contains only observations from groups outside $A$. Hence these conditioning variables are unchanged by a permutation of the observations within any one of the retained groups or within the test group. Therefore, for every $\sigma\in\mathcal S_\ell$,
\[
(\widetilde W_1^*,\ldots,\widetilde W_a^*,\ldots,
 \widetilde W_m^*)
\overset d=
(\widetilde W_1^*,\ldots,\widetilde W_a^{*,\sigma},\ldots,
 \widetilde W_m^*)
\]
conditionally on
$
(E_A,J_0=d,\mathcal N_A,\mathcal F_A,L_a^*=\ell).
$
This proves Assumption~\ref{assump:HCP-within-group-exch}. Thus the retained reference groups together with the test group assigned
the donated size $N_d$ are hierarchically exchangeable conditional on
$
(E_A,J_0=d,\mathcal N_A,\mathcal F_A).
$ 
This proves the claim for the case $\{|S| >0\}$.

\subsection{Proof of Lemma \ref{lem:generic-hcp}}

When $\tau\equiv0$ and $o=0$, the construction reduces to standard HCP. More generally, we adapt the weighted rank argument of \citet{lee2023distribution} to allow a within-group training block and a partially observed final group. For each $j\in[M]$, let
\(
\widetilde s_j=(s_{j,1},\ldots,s_{j,L_j})
\)
denote the vector of held-out scores. Since
$m_j=\tau(N_j,o)$, $L_j=N_j-m_j$, and $r_j=o-m_j$ are obtained from $N_j$ by the same deterministic rule in every group, these quantities are equivariant under permutations of the
groups. Moreover, $s_j$ is constructed from the first $m_j$
observations by the same measurable rule in every group (Section \ref{appsubsec:more-mergers}), the resulting
collection
\(
\bigl(N_j,m_j,L_j,r_j,\widetilde s_j\bigr), ~j\in[M],
\)
is exchangeable across $j$.

It remains to note the symmetry within each held-out score vector.
Conditional on the observations used to construct the score function $s_j(u,x,y)$, the remaining $L_j$ observations in group $j$ are exchangeable by Assumption~\ref{assump:HCP-within-group-exch}. Since the same score function $s_j$ is then applied to each of these observations, the coordinates of $\widetilde s_j$ are exchangeable as well.

\medskip
\noindent\textbf{Full data quantile.}
For $\beta\in(0,1)$, define
\begin{equation*}
%\label{eq:generic-full-q}
\mathsf Q_{1-\beta}
(\widetilde s_1,\ldots,\widetilde s_M)
:=\quant{1-\beta}{\nu_{\mathrm{full}}}=
\quant{1-\beta}{
\sum_{j=1}^{M}\sum_{i=1}^{L_j}
\frac{1}{ML_j}\delta_{s_{j,i}}
}.
\end{equation*}
For the realized score vectors $(\widetilde s_1,\ldots,\widetilde s_M)$, let us write $q_{1-\beta}:=\mathsf Q_{1-\beta}(\widetilde s_1,\ldots,\widetilde s_M)$.
By definition of quantile,
\begin{equation}
\label{eq:generic-lower-deterministic}
\frac1M
\sum_{j=1}^{M}\frac1{L_j}
\sum_{i=1}^{L_j}
\1\{s_{j,i}\le q_{1-\beta}\}
=
\nu_{\mathrm{full}}((-\infty,q_{1-\beta}]) \ge 1-\beta.
\end{equation}
Under Assumption~\ref{assump:no-ties}, all the scores are distinct and thus
$
\nu_{\mathrm{full}}(\{t\})\le 1/M$ for every finite $t$.
Moreover, by the definition of $q_{1-\beta}$,
$
\nu_{\mathrm{full}}((-\infty,q_{1-\beta})) < 1-\beta$.
It follows that
\begin{align}
\frac1M
\sum_{j=1}^{M}\frac1{L_j}
\sum_{i=1}^{L_j}
\1\{s_{j,i}\le q_{1-\beta}\}
&=
\nu_{\mathrm{full}}((-\infty,q_{1-\beta}))
+
\nu_{\mathrm{full}}(\{q_{1-\beta}\})
\le
1-\beta+\frac1M.
\label{eq:generic-upper-deterministic}
\end{align}
Note that $\nu_{\mathrm{full}}$ is invariant under permutations of the groups and permutations of the scores within each group.
Therefore, for every $\pi\in\mathcal S_M$,
\begin{equation*}
%\label{eq:generic-group-invariance}
\mathsf Q_{1-\beta}
(\widetilde s_1,\ldots,\widetilde s_M)
=
\mathsf Q_{1-\beta}
(\widetilde s_{\pi(1)},\ldots,\widetilde s_{\pi(M)}),
\end{equation*}
and, for every $j\in[M]$ and
$\sigma\in\mathcal S_{L_j}$,
\begin{equation*}
%\label{eq:generic-within-invariance}
\mathsf Q_{1-\beta}
(\widetilde s_1,\ldots,\widetilde s_j,\ldots,\widetilde s_M)
=
\mathsf Q_{1-\beta}
(\widetilde s_1,\ldots,\widetilde s_j^\sigma,\ldots,\widetilde s_M),
\end{equation*}
where $\widetilde s_j^\sigma:=
(s_{j,\sigma(1)},\ldots,s_{j,\sigma(L_j)})$.

\medskip
\noindent\textbf{Fixed-level rank identity.}
Throughout this argument, condition on the unordered multiset
$\mathcal N=\{N_1,\ldots,N_M\}$. This conditioning is invariant under
permutations of the groups and under within-group permutations. For the
realized full-data quantile $q_{1-\beta}$, define
\begin{equation*}
A_j(q_{1-\beta})
:=
\frac1{L_j}\sum_{i=1}^{L_j}
\1\{s_{j,i}\le q_{1-\beta}\}.
\end{equation*}

First condition further on the ordered length vector
$(N_1,\ldots,N_M)$. The index $r_M+1$ is then deterministic and belongs
to $[L_M]$. The held-out score vector in group $M$ is exchangeable, and
$q_{1-\beta}$ is invariant under permutations of this vector. Therefore,
\begin{equation*}
\PP\{s_{M,r_M+1}\le q_{1-\beta}\mid N_1,\ldots,N_M\}
=
\EE\left[A_M(q_{1-\beta})\mid N_1,\ldots,N_M\right].
\end{equation*}
Taking conditional expectations given $\mathcal N$ preserves this
identity. Next, conditional on the symmetric statistic $\mathcal N$, the
transformed group objects
$(N_j,m_j,L_j,r_j,\widetilde s_j)$ are exchangeable across $j$, while
$q_{1-\beta}$ is group-permutation invariant. Hence
\begin{equation*}
\EE\left[A_M(q_{1-\beta})\mid\mathcal N\right]
=
\EE\left[
\frac1M\sum_{j=1}^{M}A_j(q_{1-\beta})
\,\middle|\,\mathcal N
\right].
\end{equation*}
Therefore, we have
\begin{equation*}
%\label{eq:generic-group-average}
\PP\{s_{M,r_M+1}\le q_{1-\beta}\mid\mathcal N\}
=
\EE\left[
\frac1M\sum_{j=1}^{M}\frac1{L_j}
\sum_{i=1}^{L_j}
\1\{s_{j,i}\le q_{1-\beta}\}
\,\middle|\,\mathcal N
\right].
\end{equation*}
Combining the above equation with
\eqref{eq:generic-lower-deterministic} yields
\begin{equation}
\label{eq:generic-fixed-lower}
\PP\{s_{M,r_M+1}\le q_{1-\beta}\mid\mathcal N\}
\ge1-\beta.
\end{equation}
Under Assumption~\ref{assump:no-ties}, combining
the same equation with
\eqref{eq:generic-upper-deterministic} yields
\begin{equation}
\label{eq:generic-fixed-upper}
\PP\{s_{M,r_M+1}\le q_{1-\beta}\mid\mathcal N\}
\le1-\beta+\frac1M.
\end{equation}

\medskip
\noindent\textbf{Lower bound.}
For every finite $t$,
\begin{equation}
\label{eq:generic-cdf-identity}
\nu_{\mathrm{obs}}((-\infty,t])
=
\nu_{\mathrm{full}}((-\infty,t])
-
\frac1{ML_M}
\sum_{i=r_M+1}^{L_M}
\1\{s_{M,i}\le t\}.
\end{equation}
It follows that $\quant{1-\alpha}{\nu_{\mathrm{obs}}}
\ge
q_{1-\alpha}$.
Therefore, from~\eqref{eq:generic-fixed-lower}, we have
\(
\PP\left\{
s_{M,r_M+1}
\le
\quant{1-\alpha}{\nu_{\mathrm{obs}}}
\,\middle|\,
\mathcal N
\right\}
\ge
1-\alpha.
\)

\medskip
\noindent\textbf{Upper bound.}
For the upper bound, define
\[
\gamma_j
:=
\frac{L_j-r_j}{ML_j}
=
\frac{N_j-o}{M\{N_j-\tau(N_j,o)\}},
\qquad
\Gamma
:=
\max_{j\in[M]}\gamma_j.
\]
Since $m_j,L_j,r_j$ are functions of $N_j$, the value of $\Gamma$ is determined by $\mathcal N$. Equation~\eqref{eq:generic-cdf-identity} also implies
\begin{equation*}
%\label{eq:generic-cdf-lower}
\nu_{\mathrm{obs}}((-\infty,t])
\ge
\nu_{\mathrm{full}}((-\infty,t])-\gamma_M~\text{for every finite $t$}.
\end{equation*}
Since $0<\alpha-\1\{\Gamma<\alpha\}\cdot\Gamma<1$, the quantile $q_{1-\alpha+\1\{\Gamma<\alpha\}\cdot\Gamma}$ is well defined. By its definition,
\[
\nu_{\mathrm{full}}
((-\infty,q_{1-\alpha+\1\{\Gamma<\alpha\}\cdot\Gamma}])
\ge
1-\alpha+\1\{\Gamma<\alpha\}\cdot\Gamma.
\]
Using the above relation and $\gamma_M\le\Gamma$,
\begin{align*}
&\1\{\Gamma<\alpha\}
\nu_{\mathrm{obs}}
((-\infty,q_{1-\alpha+\1\{\Gamma<\alpha\}\cdot\Gamma}])
\ge
\1\{\Gamma<\alpha\}
\left\{
\nu_{\mathrm{full}}
((-\infty,q_{1-\alpha+\1\{\Gamma<\alpha\}\cdot\Gamma}])
-\gamma_M
\right\}
\\
&\ge
\1\{\Gamma<\alpha\}
\{1-\alpha+\Gamma-\gamma_M\}\ge
\1\{\Gamma<\alpha\}(1-\alpha).
\end{align*}
Consequently,
\[
\1\{\Gamma<\alpha\}
\1\left\{
s_{M,r_M+1}
\le
\quant{1-\alpha}{\nu_{\mathrm{obs}}}
\right\}
\le
\1\{\Gamma<\alpha\}
\1\left\{
s_{M,r_M+1}
\le
q_{1-\alpha+\1\{\Gamma<\alpha\}\Gamma}
\right\}.
\]
Since $\1\{\Gamma<\alpha\}$ and $\Gamma$ are $\mathcal N$-measurable,
\eqref{eq:generic-fixed-upper}, applied with $
\beta=\alpha-\1\{\Gamma<\alpha\}\Gamma$, implies
\begin{align*}
&\1\{\Gamma<\alpha\}
\PP\left\{
s_{M,r_M+1}
\le
\quant{1-\alpha}{\nu_{\mathrm{obs}}}
\,\middle|\,
\mathcal N
\right\}
\le
\1\{\Gamma<\alpha\}
\PP\left\{
s_{M,r_M+1}
\le
q_{1-\alpha+\1\{\Gamma<\alpha\}\Gamma}
\,\middle|\,
\mathcal N
\right\}
\\
&\le
\1\{\Gamma<\alpha\}
\left(
1-\alpha+\Gamma+\frac1M
\right).
\end{align*}
Also note,
\begin{align*}
\1\{\Gamma\geq \alpha\}
\PP\left\{
s_{M,r_M+1}
\le
\quant{1-\alpha}{\nu_{\mathrm{obs}}}
\,\middle|\,
\mathcal N
\right\}
&\le
\1\{\Gamma\geq \alpha\}
\le
\1\{\Gamma\geq \alpha\}
\left(
1-\alpha+\Gamma+\frac1M
\right),
\end{align*}

Summing the above two terms leads to
\(
\PP\left\{
s_{M,r_M+1}
\le
\quant{1-\alpha}{\nu_{\mathrm{obs}}}
\,\middle|\,
\mathcal N
\right\}
\le
1-\alpha+\frac1M+\Gamma.
\)
By the definition of $\Gamma$, 
\[
\PP\left\{
s_{M,r_M+1}
\le
\quant{1-\alpha}{\nu_{\mathrm{obs}}}
\,\middle|\,
\mathcal N
\right\}
\le
1-\alpha
+
\frac1M
+
\frac1M
\max_{j\in[M]}
\frac{N_j-o}{N_j-\tau(N_j,o)}.
\]
Finally, since $
0\le\tau(N_j,o)\le o<N_j$,
we have
\(
0
\le
\frac{N_j-o}{N_j-\tau(N_j,o)}
\le
1,
\)
and therefore $\Gamma\le1/M$. Marginalizing with respect to $\mathcal N$, we have
\(
\PP\left\{
s_{M,r_M+1}
\le
\quant{1-\alpha}{\nu_{\mathrm{obs}}}
\right\}
\le
1-\alpha+\frac2M.
\)

\subsection{Proof of Corollary~\ref{cor:all-donors}}

Under $\PP\{N_1>o\}=1$, exchangeability of the size vector implies $S=[K]$. Therefore, the upper-bound slack in Theorem~\ref{thm:donor-valid} can be written as
\begin{equation*}
\frac1K\left\{1+\EE\bigl[\rho_o(N_{\max})\bigr]\right\},
\qquad
N_{\max}:=\max_{1\le j\le K}N_j.
\end{equation*}
For every integer $n>o$,
\begin{equation*}
\rho_o(n)=\frac{n-o}{n-\lfloor o/2\rfloor}\le1.
\end{equation*}
Moreover, whenever $n>o+1$,
$\rho_{o+1}(n)<\rho_o(n)$. Indeed, the claim is immediate when $o$ is
even; when $o=2k+1$, it reduces to
\begin{equation*}
\frac{n-2k-2}{n-k-1}<\frac{n-2k-1}{n-k},
\end{equation*}
which follows from $n-k>n-2k-1$. Thus the slack decreases strictly
between consecutive values of $o$ for which all groups remain eligible.
Finally,
\begin{equation*}
\frac{1+\EE[\rho_o(N_{\max})]}{K}
\le\frac2K
\leq \frac2{K_1+1}
\qquad \text{for any}~K>2~\text{and}~K_1\leq K-1,
\end{equation*}

\subsection{Proof of Corollary \ref{cor:sufficient-donors}}
\label{app:proof-cor:sufficient-donors}
Let $p_o:=\PP\{N_1>o\}$ and $q_o:=1-p_o$. Since the group sizes are assumed to be i.i.d.,
\(
B:=|S|\sim\operatorname{Binomial}(K,p_o).
\)
By \eqref{eq:ghcp-donor-count-upper} and
$o+1-\lfloor o/2\rfloor\ge1$, it suffices to bound the GHCP coverage upper bound slack. This slack is bounded above by
\(
\PP\{B=0\}
+
2\EE\left[1\{B>0\}/B
\right].
\)
Since $1/b=\int_0^1 t^{b-1}\,\mathrm{d}t$,
we obtain
\begin{align*}
\EE\left[
\frac{\1\{B>0\}}{B}
\right]
&= \sum_{b=1}^{K}
\frac1b
\binom Kb
p_o^bq_o^{K-b} \\
&= 
\int_0^1
\frac{(q_o+p_ot)^K-q_o^K}{t}\,\mathrm{d}t=
\int_{q_o}^{1}
\frac{u^K-q_o^K}{u-q_o}\,\mathrm{d}u
=
\sum_{\ell=0}^{K-1}
\frac{q_o^\ell(1-q_o^{K-\ell})}{K-\ell}.
\end{align*}
Consequently,
\begin{align*}
\PP\{B=0\}
+
2\EE\left[
\frac{\1\{B>0\}}{B}
\right]
=
2\sum_{\ell=0}^{K-1}
\frac{q_o^\ell}{K-\ell}
+
q_o^K
\left(
1-2\sum_{\ell=0}^{K-1}\frac1{K-\ell}
\right).
\end{align*}
Since
\(
\sum_{\ell=0}^{K-1}1/(K-\ell)
=
\sum_{r=1}^{K}1/r
\ge1,
\)
the second term is nonpositive. Moreover, we have
$1/(K-\ell)
\le
(\ell+1)/K$ for $\ell=0,\ldots,K-1,~\text{as}~
(\ell+1)(K-\ell)\ge K$. Therefore,
\begin{align*}
\PP\{B=0\}
+
2\EE\left[
\frac{\1\{B>0\}}{B}
\right]
\le
\frac2K
\sum_{\ell=0}^{\infty}
(\ell+1)q_o^\ell=
\frac{2}{K(1-q_o)^2}
=
\frac{2}{Kp_o^2}.
\end{align*}
Finally, note that
$p_o
\ge
\sqrt{\frac{K_1}{K}+\frac1K}
$ implies $\frac{2}{Kp_o^2}
\le
\frac2{K_1+1}$, which completes the proof.

\subsection{Proof of Corollary~\ref{cor:restricted-valid}}

Let
\(
\mathcal E_\eta
:=
\left\{
Y_{K+1,o+1}\in
\widehat C_{\mathrm{GHCP},\eta}
(U_{K+1},X_{K+1,o+1})
\right\}.
\)
The restricted donor-selection rule in
Section~\ref{subsec:restricted-selection}, with the uniform tie-breaking
construction in Appendix~\ref{appsubsec:restriction}, is
permutation-equivariant. Hence Lemma~\ref{lem:donation-restores} applies
with $S_\eta$ in place of $S$.

First suppose $|S_\eta|>0$. Applying Lemma~\ref{lem:donation-restores} produces a hierarchically exchangeable collection consisting of the retained reference groups and the test group. Therefore, applying Lemma~\ref{lem:generic-hcp} with $\tau(n,o)=\lfloor o/2\rfloor$, we have
\(
\PP\{\mathcal E_\eta\mid |S_\eta|>0\}\ge 1-\alpha.
\)
If $|S_\eta|=0$, the procedure reduces to the same within-test group
construction used in the proof of Theorem~\ref{thm:donor-valid}. By
within-group exchangeability,
\(
\PP\{\mathcal E_\eta\mid |S_\eta|=0\}\ge 1-\alpha.
\)
Combining the two cases yields $\PP\{\mathcal E_\eta\}\ge 1-\alpha$.
Now suppose Assumption~\ref{assump:no-ties} holds. For the case
$\{|S_\eta|=0\}$, the corresponding upper bound from the proof of
Theorem~\ref{thm:donor-valid} provides
\(
\PP\{\mathcal E_\eta\mid |S_\eta|=0\}
\le
1-\alpha+
\frac{1}{o+1-\lfloor o/2\rfloor}.
\)
On the event $\{|S_\eta|>0\}$, the upper bound in
Lemma~\ref{lem:generic-hcp} leads to
\[
\PP\{\mathcal E_\eta\mid S_\eta,\{N_j:j\in S_\eta\},
\sigma(\widetilde W_j:j\notin S_\eta)\}
\le
1-\alpha
+
\frac1{|S_\eta|}
\left\{
1+
\max_{j\in S_\eta}
\frac{N_j-o}{N_j-\lfloor o/2\rfloor}
\right\}.
\]
Taking expectations over the two cases therefore yields
\begin{align*}
\PP\{\mathcal E_\eta\}
&\le
1-\alpha
+
\frac{\PP\{|S_\eta|=0\}}
{o+1-\lfloor o/2\rfloor}
\\
&\qquad
+
\EE\left[
\1\{|S_\eta|>0\}
\frac1{|S_\eta|}
\left\{
1+
\max_{j\in S_\eta}
\frac{N_j-o}{N_j-\lfloor o/2\rfloor}
\right\}
\right].
\end{align*}
Since every $j\in S_\eta$ is size-compatible, $N_j>o$, and hence
$
0\le
(N_j-o)/(N_j-\lfloor o/2\rfloor)
\le 1.
$
Consequently,
\[
\PP\{\mathcal E_\eta\}
\le
1-\alpha
+
\frac{\PP\{|S_\eta|=0\}}
{o+1-\lfloor o/2\rfloor}
+
2\EE\left[
\frac{\1\{|S_\eta|>0\}}{|S_\eta|}
\right],
\]
which proves the upper bound.

\section{Proofs for Section~\ref{appsec:derandomization}}
\label{appsec:proofs-alt}

We use the following standard arithmetic-mean merger result; see
\citet{vovk2020combining}.

\begin{lemma}
\label{lem:mean-merger}
Let $P_1,\ldots,P_B$ be super-uniform random variables. Define
\(
P_{\mathrm{merge}}
:=
\min\left\{
1,\,
(2/B)\sum_{b=1}^{B}P_b
\right\}.
\)
Then $P_{\mathrm{merge}}$ is also super-uniform, i.e., it satisfies
\(
\PP\{P_{\mathrm{merge}}\le u\}\le u, u\in(0,1).
\)
\end{lemma}

\begin{lemma}
\label{lem:fixed-order-valid}
Suppose Assumptions~\ref{assump:group-exch}--\ref{assump:within-group-exch} hold, $S$ is a donor pool selected by a permutation-equivariant rule as defined in Section~\ref{subsec:restricted-selection}, and the score function $s$ is constructed using only the reference groups outside $S$. Then, for every $\sigma\in\mathcal S_K$, every nonempty
$A\subseteq[K]$, and every $u\in(0,1)$,
\(
\PP\left\{\pi_\sigma(Y_{K+1,o+1})\le u
\,\middle|\, S=A \right\} \le u.
\)
\end{lemma}

\begin{proof}
Fix $\sigma\in\mathcal S_K$ and a nonempty $A\subseteq[K]$ with
$\PP\{S=A\}>0$, and let
\(
d
:=
d_\sigma(A)
=
\sigma\left(
\min\{\ell\in[K]:\sigma(\ell)\in A\}
\right).
\)
Thus, on the event $\{S=A\}$, the donor selected by the fixed ordering $\sigma$ is the fixed index $d$. Define
\[
\nu_d
:=
\sum_{m\in A\setminus\{d\}}
\sum_{i=1}^{N_m}
\frac{1}{|A|N_m}\delta_{s(W_{m,i})}
+
\sum_{i=1}^{o}
\frac{1}{|A|N_d}\delta_{r_i}
+
\frac{N_d-o}{|A|N_d}\delta_{\infty}.
\]
By the definition of $\pi_\sigma$ in \eqref{eq:order-pvalue}, on
$\{S=A\}$ we have
\begin{equation}
\label{eq:fixed-order-tail-mass}
\pi_\sigma(y)
=
\nu_d([t(y),\infty]).
\end{equation}
We next justify applying Lemma~\ref{lem:generic-hcp} with the fixed donor
$d$. Introduce an auxiliary donor
$J_0\mid S=A\sim\Unif(A)$, drawn independently of
$(N_{1:K},\widetilde W_{1:K+1})$. By
Lemma~\ref{lem:donation-restores}, conditional on
$\{S=A,J_0=d\}$, the reference groups indexed by
$A\setminus\{d\}$ together with the test group assigned size $N_d$
form a hierarchically exchangeable collection. Moreover,
\[
\left(
N_{1:K},\widetilde W_{1:K+1}
\,\middle|\,
S=A,J_0=d
\right)
\overset{d}{=}
\left(
N_{1:K},\widetilde W_{1:K+1}
\,\middle|\,
S=A
\right),
\]
since $J_0$ is conditionally independent of the data given $S=A$.
Consequently, conditional on $S=A$, the collection
\(
\left\{
\widetilde W_j:j\in A\setminus\{d\}
\right\}
\cup
\left\{
(W_{K+1,1},\ldots,W_{K+1,N_d})
\right\}
\)
is hierarchically exchangeable.

Next, let
\(
\mathcal N_A:=\{N_j:j\in A\},
\mathcal F_A:=\sigma(\widetilde W_j:j\in[K]\setminus A).
\)
On $\{S=A\}$, the score function $s$ is constructed from
$\mathcal F_A$ and is therefore fixed conditional on $\mathcal F_A$.
By the conditional statement in
Lemma~\ref{lem:donation-restores}, together with the fixed-donor
argument above, conditional on
$(S=A,\mathcal N_A,\mathcal F_A)$ the groups indexed by
$A\setminus\{d\}$ and the test group assigned size $N_d$ satisfy the
hierarchical exchangeability conditions of
Lemma~\ref{lem:generic-hcp}. There is no within-group training here. Hence Lemma~\ref{lem:generic-hcp} applies with
$
M=|A|,
m_j=0,
L_j=N_j,
r_j=o,
$
where the test group has the donated size $N_d$. Note that the corresponding distribution
$\nu_{\mathrm{obs}}$ becomes exactly $\nu_d$. Therefore, writing
$
q_d(u):=\quant{1-u}{\nu_d},
$
we have from Lemma~\ref{lem:generic-hcp}
\(
\PP\left\{
s(W_{K+1,o+1})\le q_d(u)
\,\middle|\,
S=A,\mathcal N_A,\mathcal F_A
\right\}
\ge 1-u.
\)
Marginalizing over $(\mathcal N_A,\mathcal F_A)$ yields
\begin{equation}
\label{eq:fixed-order-quantile-valid}
\PP\left\{
s(W_{K+1,o+1})>q_d(u)
\,\middle|\,
S=A
\right\}
\le u.
\end{equation}
By \eqref{eq:fixed-order-tail-mass},
\[
\pi_\sigma(Y_{K+1,o+1})
=
\nu_d([s(W_{K+1,o+1}),\infty])
=
1-\nu_d((-\infty,s(W_{K+1,o+1}))).
\]
This implies that if $\pi_\sigma(Y_{K+1,o+1})\le u$, then $\nu_d((-\infty,s(W_{K+1,o+1})))\ge 1-u.$ holds. Since $\nu_d$ has finite support on $\overline{\mathbb R}$ and the test score is finite, it also implies
\(
q_d(u)<s(W_{K+1,o+1}).
\)
Thus, on the event $\{S=A\}$,
\[
\left\{
\pi_\sigma(Y_{K+1,o+1})\le u
\right\}
\subseteq
\left\{
s(W_{K+1,o+1})>q_d(u)
\right\}.
\]
Combining this inclusion with
\eqref{eq:fixed-order-quantile-valid}, we have
\(
\PP\left\{
\pi_\sigma(Y_{K+1,o+1})\le u
\,\middle|\,
S=A
\right\}
\le u.
\)
\end{proof}

\subsection{Proof of Lemma~\ref{lem:donor-closed-form}}

Fix a nonempty $A\subseteq[K]$ and consider a realization for which
$S=A$. Let $M:=|A|$. For $m\in A$, define
\(
B_m(y)
:=
\sum_{i=1}^{N_m}
\1\{s(W_{m,i})\ge t(y)\},~
R(y)
:=
\sum_{i=1}^{o}
\1\{r_i\ge t(y)\}.
\)
For each $j\in A$, the number of permutations
$\sigma\in\mathcal S_K$ for which $j$ is the first member of $A$ is
$|\mathcal S_K|/M$.
Therefore,
\begin{align*}
&\frac1{|\mathcal S_K|}
\sum_{\sigma\in\mathcal S_K}\pi_\sigma(y)
=
\frac1M\sum_{j\in A}
\left\{
\sum_{m\in A\setminus\{j\}}
\frac{B_m(y)}{MN_m}
+
\frac{R(y)}{MN_j}
+
\frac{N_j-o}{MN_j}
\right\}
\nonumber\\
&=
\sum_{m\in A}
\frac{B_m(y)}{MN_m}
\left\{
\frac1M\sum_{j\in A}\1\{j\neq m\}
\right\}
+
\frac{R(y)}{M^2}
\sum_{j\in A}\frac1{N_j}
+
\frac1{M^2}
\sum_{j\in A}
\left(1-\frac{o}{N_j}\right).
%\label{eq:order-average-expansion}
\end{align*}
Note that,
\[
\frac1M\sum_{j\in A}\1\{j\neq m\}
=
\frac{M-1}{M}, \forall~m\in A,\qquad
\frac1{M^2}\sum_{j\in A}\frac1{N_j}
=
\bar w,
\qquad
\frac1{M^2}
\sum_{j\in A}
\left(1-\frac{o}{N_j}\right)
=
\frac1M-o\bar w.
\]
Therefore,
\begin{align*}
&\frac1{|\mathcal S_K|}
\sum_{\sigma\in\mathcal S_K}\pi_\sigma(y)
=
\frac{M-1}{M}
\sum_{m\in A}
\frac{B_m(y)}{MN_m}
+
\bar w R(y)
+
\frac1M-o\bar w
\\
&=
\frac{M-1}{M}
\sum_{m\in A}\sum_{i=1}^{N_m}
\frac{\1\{s(W_{m,i})\ge t(y)\}}{MN_m}
+
\bar w
\sum_{i=1}^{o}
\1\{r_i\ge t(y)\}
+
\left(\frac1M-o\bar w\right)
\1\{\infty\ge t(y)\}
\\
&=
\bar\nu_{\mathrm{donor}}
\left([t(y),\infty]\right).
\end{align*}
This proves the claim.

\subsection{Proof of Theorem~\ref{thm:donor-merge}}

First suppose that $S=A\neq\varnothing$, where $\PP\{S=A\}>0$. By Lemma~\ref{lem:fixed-order-valid},
\[
\PP\{\pi_\sigma(Y_{K+1,o+1})\le u\mid S=A\}\le u,
u\in(0,1),
\sigma\in\mathcal S_K.
\]
Therefore, by Lemma~\ref{lem:mean-merger},
\(
\PP\left\{
\pi_{\mathrm{merge}}(Y_{K+1,o+1})\le\alpha
\,\middle|\,
S=A
\right\}
\le\alpha.
\)
By \eqref{eq:ddhcp-pvalue-set},
\(
\widehat C_{\mathrm{D\text{-}GHCP}}
(U_{K+1},X_{K+1,o+1})
=
\left\{
y:\pi_{\mathrm{merge}}(y)>\alpha
\right\}.
\)
Hence, for every nonempty $A\subseteq[K]$ with $\PP\{S=A\}>0$,
\begin{equation*}
\PP\left\{
Y_{K+1,o+1}
\notin
\widehat C_{\mathrm{D\text{-}GHCP}}
(U_{K+1},X_{K+1,o+1})
\,\middle|\,
S=A
\right\}
\le\alpha.
\end{equation*}
When $S=\varnothing$, the construction reduces to the usual
split-conformal procedure based on the observed test group scores.
By the standard conformal prediction argument,
\begin{equation*}
\PP\left\{
Y_{K+1,o+1}
\notin
\widehat C_{\mathrm{D\text{-}GHCP}}
(U_{K+1},X_{K+1,o+1})
\,\middle|\,
S=\varnothing
\right\}
\le\alpha.
\end{equation*}
Combining the above two results proves the claim.

\end{document}